\documentclass[12pt,a4paper,twoside]{report}

\usepackage{duthesis}		
\usepackage{citesort}		
\usepackage{setspace}		
\usepackage{xspace}		
\usepackage{tarun,timesfonts}
\usepackage{latexsym,fancyhdr,psfig,epsfig}
\usepackage{amsmath}
\usepackage{amssymb}
\usepackage{latexsym} 
\usepackage{pifont}
  
\newcommand{\lapprox}{\lower.6ex\hbox{$\; \buildrel<\over\sim \;$}}

\newcommand{\gapprox}{\lower.6ex\hbox{$\; \buildrel>\over\sim \;$}}

\newcommand{\curly}{\lower1.ex\hbox{$\; \stackrel{\textstyle \wr}{\wr} \;$}}

\def\barr{\begin{array}}
\def\earr{\end{array}}
\def\berr{\begin{eqnarray}}
\def\err{\end{eqnarray}}
\def\berrno{\begin{eqnarray*}}
\def\errno{\end{eqnarray*}}
\def\be{\begin{equation}}
\def\fr{\frac}
\def\ra{\rangle}
\def\la{\langle}
\def\bs{\boldsymbol}
\def\te{\t_{\rm E}}
\def\dl{D_{\rm L}}
\def\ds{D_{\rm S}}
\def\dls{D_{\rm LS}}
\def\half{\frac{1}{2}}
    
\newcommand{\pder}[2]{\frac{\partial{#1}}{\partial{#2}}}
\newcommand{\dd}[1]{\frac{\partial}{\partial {#1}}}

\newcommand{\no}{\nonumber}
\newcommand{\q}{\;\;}
\newcommand{\qq}{\;\;\;\;}
\newcommand{\bbox}{\boldsymbol}

\renewcommand{\a}{\alpha}
\renewcommand{\b}{\beta}
\newcommand{\g}{\gamma}
\newcommand{\G}{\Gamma}
\renewcommand{\d}{\delta}
\newcommand{\D}{\Delta}
\newcommand{\e}{\epsilon}
\renewcommand{\k}{\kappa}
\renewcommand{\l}{\lambda}
\renewcommand{\L}{\Lambda}
\newcommand{\n}{\nabla}
\renewcommand{\o}{\omega}
\renewcommand{\O}{\Omega}
\newcommand{\r}{\rho}
\newcommand{\s}{\sigma}
\renewcommand{\S}{\Sigma}
\renewcommand{\t}{\theta}
\renewcommand{\v}{\varphi}
\renewcommand{\P}{\Phi}

\newcommand{\del}{\partial}
\renewcommand{\i}{\item}
\renewcommand{\SS}{\paragraph}
\newcommand{\Schw}{Schwarzschild\ }
\newcommand{\E}{Einstein\ }
\newcommand{\R}{Riemannian\ }
\newcommand{\Euc}{Euclidean\ }
\newcommand{\st}{spacetime\ }
\newcommand{\M}{Minkowskian\ }
\newcommand{\ass}{asymptotically\ }

\def\apj{{\it Astrophys.~J.}}
\def\aj{{\it Astronom.~J.}}
\def\mpla{{\it Mod. Phys. Lett. A}}
\def\apjl{{\it Astrophys.~J.~Lett.}}
\def\prd{{\it Phys.~Rev.~D.}}
\def\prl{{\it Phys.~Rev.~Lett.}}
\def\mnras{{\it Mon.~Not.~R.~Astr.~Soc.}}
\def\nature{{\it Nature}}
\def\science{{\it Science}}
\def\AnA{{\it Astr. Astrophys.}}
\def\grg{{\it Gen. Rel. Grav.}}

\begin{document}
\title{GRAVITATIONAL LENSING IN STANDARD AND ALTERNATIVE COSMOLOGIES}
\author{MARGARITA SAFONOVA \\ 
  DEPARTMENT OF PHYSICS AND ASTROPHYSICS \\
 UNIVERSITY OF DELHI \\
 DELHI $-$ 110 007 \\
 INDIA 
}

\supervisor{Prof. Daksh Lohiya}		
%
\coguide{Dr. Shobit Mahajan}	        
\hod{Prof. K.~C.~Tripathi}
\submitdate{April,~2002}
\copyrightfalse				
\figurespagetrue			
\tablespagetrue

\beforepreface				
\prefacesection{Acknowledgements} 

{\large
\emph{ 
I would like to acknowledge here all my friends, those who remain friends
in spite of large separations in space and time, and those here and now and
who, most probably, will remain friends in possible future large separations
in space and time. I would like to thank friends who became my
collaborators, and collaborators who became friends.}} 

\vskip 0.3in 
\begin{quote}
\large
\emph{
To begin with the very beginning, my deep thanks go to my M.Sc.
supervisor and my friend, Prof. Michael Vasil'evitch Sazhin. It was he who
introduced me to this beautiful field of science---Gravitational Lensing. He
taught me many things and methods I am using now, and he will forever remain
my teacher and my friend.}        
\end{quote}
\begin{quote}
\large
\emph{I would like to acknowledge here the hospitality of IUCAA and many people
who were helping me there to feel at home.}
\end{quote}  

\begin{quote}
\large
\emph{I would like to thank members of Department of Astrophysics, Delhi
University, most of all, Dr. Amitabha Mukherjee, and students of the
Department, who were always very nice and helpful to me. Beginning with my
first friend in DU, Harvinder, and continuing with Varsha, Abha, Deepak,
Abhinav and Namit.}  
\end{quote} 

\begin{quote}
\large
\emph{
\vskip 0.001in
I am also very grateful to the families of my friends;
especially to Abha's family, Amber and Gaurav, who were always so friendly and
attentive to me and my daughter that we felt like being a part of it.}
\end{quote}   

\begin{quote}
\large
\emph{
My special thanks go to my co-supervisor, Dr. Shobhit Mahajan, without whom
my thesis would have been unreadable.}  
\end{quote}

\begin{quote} 
\large
\emph{ I am infinitely grateful to Prof. A. Prasanna and to my collaborators,
Drs. Diego Torres, Gustavo Romero and Zafar Turakulov. Their ideas and work
exposed me to many new areas of science and helped me to shape my thesis.}
\end{quote} \begin{quote} \large
\emph{
I can't thank enough my husband, who was always there for any of my
questions and problems, whose ideas were sometimes better than mine, and who
helped me enourmously with my work.}
\end{quote} 

\vskip 0.1in 
{\large
\emph{
But this thesis would have been impossible without my guide, the person
who took me under his wing and nurtured me, who was patient (and sometimes not
so patient) with me, who was forgiving my bad moods and stood by me, who
encouraged my independence and didn't mind when I was venturing into new
territories. And thus, rephrasing the words of Ken Kesey in his ``One Flew
Over the Cuckoo's Nest", I dedicate this thesis}}

\normalsize
\newpage

\begin{flushright}
\Large

\emph{
\vskip 0.1in
To my supervisor, \\[0.1in]
who told me that there are no dragons\\
and then took me to their lairs.}
\end{flushright}

\prefacesection{List of publications}

\section*{Published Work}

\begin{enumerate}

\item {\bf M.~V. Safonova} \& D. Lohiya. \\
"Gravity balls in induced gravity model---`gravitational lens' effects.\\
{\it Grav. Cosmol.}, {\bf 6}, 327-334 (2000).

\item {\bf Margarita Safonova}, Diego F. Torres \& Gustavo E. Romero. \\
"Macrolensing signatures of large-scale violations of the weak
energy condition". \\
\mpla, {\bf 16}, 153-162 (2001).

\item {\bf Margarita Safonova}, Diego F. Torres \& Gustavo E. Romero. \\
"Microlensing by natural wormholes: theory and simulations".\\
\prd, {\bf 65}, 023001 (2002).

\end{enumerate}

\section*{Communicated Work}
\begin{enumerate}

\item Zafar Turakulov \& {\bf Margarita Safonova}.\\ ``Motion of a vector 
particle in a curved space-time. I. Lagrangian approach".
Submitted to \mpla. {\bf qr-qc/0110067}

\item Abha Dev, {\bf Margarita Safonova}, Deepak Jain, \& Daksh Lohiya \\
``Cosmological tests for a linear coasting cosmology". \\
Submitted to {\it Phys. Letters B}. {\bf astro-ph/0204150}

\item {\bf Margarita Safonova} \& Diego Torres, \\
``Degeneracy in exotic gravitational lensing"\\
Submitted to \prd, Brief Reports.

\end{enumerate}

\section*{Conference Presentations}

\begin{enumerate}

\item[i] \emph{Gravity ball as a possible gravitational lens}, 
Meeting of International Society on General Relativity and 
Gravitation ({\bf GR15}), IUCAA, Pune, December 1997.

\item[ii] \emph{Gravity balls in induced gravity model---
`gravitational lens' effects},
International Conference on Gravitation and Cosmology,
I.I.T. Kharagpur, January 2000.

\item[iii] \emph{Gravitational lensing as a tool to study alternative
cosmologies}, 12th Summer School on Astroparticle Physics and 
Cosmology, ICTP, Trieste, Italy, June 2000.

\item[iv] \emph{Macrolensing signatures of large-scale violations of
a weak energy condition}, Young Astronomers Meeting, IUCAA, Pune, 
January 2001. 

\item[iv] \emph{Energy conditions violations and negative masses in the
universe---gravitational lensing signatures}, Symposium on Cosmology and
Astrophysics, Jamia Milia Islamia, New Delhi, January 2002.

\end{enumerate}

\section*{Work in Preparation}

\begin{enumerate}

\item Zafar Turakulov \& {\bf Margarita Safonova}.\\ 
``Motion of a vector particle in a curved space-time. II. ``First order
approximation in \Schw\/ background".

\end{enumerate}
  

\afterpreface			
%
%
\chapter{General Overview of the Thesis}
This thesis contributes to the field of Gravitational lensing (GL)
and observational cosmology. We have investigated the possibility of 
detecting the existence of matter violating the weak energy conditions
through its lensing effect on background sources. In a different context 
we have investigated GL in an alternative cosmology with a linearly
evolving scale factor. We also considered gravitational lensing
statistics in such a cosmology and its compatibility with existing
observations. We have studied propagation of light in strong gravitational
fields and the equations of motion for a vector particle with spin. Using
geometrical optics approach we have shown that in \Schw and Kerr
geometries, massless particles deviate from geodesic motion.

Einstein's General Theory of Relativity (GR)
predicts that light rays are deviated from their straight path when they
pass close to a massive body. This prediction  was experimentally verified 
in 1919. Although the deflection is small, its effect can be enhanced by the
passage of light over long distances. The deflected rays have enough time to
intersect with one another to form caustics, and for objects which are compact
and massive or are at cosmological distances there is a possibility of
observing the effects of bending of light. Due to the bending of light,
background objects appear distorted and, in extreme cases, form multiple
images. This information can be used to obtain the distribution of mass in the
lens in a completely novel way. The images of background objects are magnified
by the action of lensing which makes them appear bigger (and therefore
brighter). Thus, a gravitational lens acts as a natural telescope providing us
information about the distant objects which are otherwise too dim to be
detected.

GL is a powerful tool for exploring the universe.  It can be used for the 
detection of exotic objects as well as for testing alternative theories of
gravity. Proposals have been made to discover cosmic strings, boson stars,
neutralino stars or wormholes through their gravitational lensing effects.
There is no compelling evidence that any of the observed GL systems are due to
these objects. However, it is essential to develop new lens models with objects
which are not forbidden on theoretical grounds.

The list of multiple-imaged gravitational lens systems has been growing
steadily since the discovery of the first lens system in 1979 (the famous `Old
Faithful' QSR 0957+561 A\&B). At present, more than 30 multiple-image
systems are confirmed, or are very likely to be, as gravitational lens
systems. These lens systems can provide us with information about the universe
as a whole. The global geometry of the universe, usually specified by its
matter density and a cosmological constant, remains a significant source of
uncertainty in modern cosmology. The possibility of using GL as a tool for the
determination of cosmological parameters, either by a detailed study of
specific lensing systems or through statistical analysis of samples of lenses,
has been long and frequently discussed. One of the results of the previous
works was that the mean image separations of lens systems have different
dependence on source redshift in different cosmologies and that it may
therefore be possible to measure the curvature of the universe directly.
Besides, the expected frequency of multiple image lensing events for high
redshift sources turned out to be quite sensitive to some cosmological
parameters. All this makes the gravitational lensing statistics an interesting
method to test different cosmological models.  

{\bf CHAPTER TWO} is an introduction to the gravitational lensing theory. We
give a brief introduction to the basic mathematical formalism for studying 
gravitational lensing and the background cosmology that we use to describe
spacetime in which light propagation takes place. We introduce the
terminology and concepts used in GL calculations, and review the basic
equations of gravitational lensing. The basic types of lenses are
presented and their properties are discussed. The second part of the chapter
reviews different astrophysical and cosmological applications of gravitational
lensing.

In {\bf CHAPTER THREE}, the substance of which has appeared as
Refs.~\cite{rita1,rita2}, we discuss the use of gravitational lensing as a
tool in search of exotic objects in the Universe. In the first section we
give an introduction to the energy conditions (EC) of classical GR and examine
the consequences of their possible violations. Typically, observed violations
are produced by small quantum systems and are of the order of $\hbar$. One
recent experimental study, however reported\ a violation, which could, in
principle, arise because of the existence of classically forbidden regions
carrying negative energy. It is currently far from clear whether there could
be macroscopic quantities of such an exotic, EC-violating matter.

Of all the systems which would require violations of the EC, wormholes are
the most intriguing. The salient feature of these objects is that an
embedding of one of their space-like sections in Euclidean space displays two
asymptotically flat regions joined by a throat. Since wormholes have to
violate the null EC in order to exist, the hypothesis underlying the positive
mass theorem no longer applies, and there is nothing, in principle, that can
prevent the occurrence of a negative total mass. In other words, we need to
have some negative mass near the throat to keep the wormholes open. If
wormholes exist, they could have formed naturally in the Big Bang, ``inflated"
from the ``quantum foam" that is thought to underlie spacetime. Alternatively,
they could have been constructed by an advanced extraterrestrial civilization
as terminuses for, say, pan-galactic subway system.

Discovery of any object with negative mass will not prove the existence of
wormholes for sure, although it will certainly enhance the possibilities for
wormholes to exist. In this chapter we analyze the gravitational effects light
experiences while traversing the regions with negative mass. 

In the second section we provide an in-depth study of the theoretical
peculiarities that arise in effective negative mass lensing, both for the
case of a point mass lens and source, and for extended source situations. We
describe novel observational signatures arising in the case of a source lensed
by a negative mass. We show that a negative mass lens produces total or
partial eclipse of the source in the umbra region and also show that the usual
Shapiro time delay is replaced with an equivalent time gain. We describe these
features both theoretically, as well as through numerical simulations. In the
third section we provide negative mass microlensing simulations for various
intensity profiles and discuss the differences between them. The light curves
for microlensing events are presented and contrasted with those due to lensing
produced by normal matter. Presence or absence of these features in the
observed microlensing events can shed light on the existence of natural
wormholes in the Universe.

In the last section we present a set of simulations of the macrolensing
effects produced by large-scale cosmological violations of the energy
conditions. These simulations show how the appearance of a background field of
galaxies is affected when lensed by a region with an energy density equivalent
to a negative mass ranging from $10^{12}$ to $10^{17}$ solar masses. We
compare with the macrolensing results of equal amounts of positive mass, and
show that, contrary to the usual case where tangential arc-like structures are
expected, there appear radial arcs---runaway filaments---and a central void.
These results make the cosmological macrolensing, produced by space-time
domains where the weak energy conditions is violated, observationally
distinguishable from standard regions. Whether large domains with negative
energy density indeed exist in the universe can now be decided by future
observations of deep fields.

In the {\bf FOURTH CHAPTER} we explore GL in an alternative cosmological
model and the concordance of this theory with the current gravitational lensing
observations. This chapter is divided into three sections. The first section
introduces the problems standard model experiences and ways of their
resolution. In the second section, part of which has appeared in
ref.~\cite{rita3}, we investigate the concordance of the gravitational lensing
statistics with the cosmology in which the scale factor is linearly evolving.
The use of gravitational lensing statistics as a tool for the determination of
cosmological parameters either by a detailed study of specific lens systems or
through a statistical analysis of a sample of lenses has been frequently
discussed. It has been pointed out that the expected frequency of multiple
imaging lensing events is sensitive to a cosmology. We use this test to
constrain the power index $\alpha$ of the scale factor. We calculate the
expected number of multiple image gravitational lens systems in a particular
quasar sample with a known distribution of redshifts. This is compared with the
observed frequency of lens systems found. Expected number of lens systems
depends upon the index $\alpha$ through the angular diameter distances. We
derive the expressions for angular diameter distances for this cosmology and
use them in the lensing probability functions. By varying $\alpha$, the number
of lenses changes, which on comparison with the observations gives us a
constraint on $\alpha$. We find that the value  $\alpha=1$ corresponding to
the coasting cosmology is in concordance with the number of observed lenses in
the considered sample. In the last section, substance of which has appeared as
ref~\cite{rita4}, we introduce the non-minimally coupled effective gravity
theory in which one can have non-topological soliton solutions. A typical
solution is a spherical region having $G_{\rm eff}=0$ outside, and the
canonical Newtonian value inside. Such a spherical domain (gravity-ball) is
characterised by an effective index of refraction which causes bending of
light incident on it. The gravity ball thus acts as a gravitational lens. We
consider the gravity ball to be of a size of a typical cluster of galaxies and
show that even empty (without matter) gravity ball can produce arc-like images
of the background source galaxy. In the case of background random field the
ball produces distortions (`shear') of that field. We also obtain
constraints on the size of the large gravity ball which can be inferred form
the existing observations of clusters with arcs.

The {\bf FIFTH CHAPTER} is dedicated to studies of the propagation of light in
strong gravitational fields. Most cosmological studies of lensing are
performed in the weak field approximation. However, there are interesting
astrophysical situations where light propagates in a strong gravitational
field. Weak field approximation becomes invalid in the vicinity of compact
objects like black holes and pulsars. Thus, gravitational lensing can have
additional effects. This chapter is divided into 5 sections. In the first two
sections we present introduction to the question and motivations for the
study. Third section describes the formulations of the theory, where
the equation of wave propagation coupled to the curvature of the spacetime is
derived. We also derive the modified geodesic equation and present a way to
solve the equation. Fourth section discusses the application of the results of
the third section to the velocity of photons propagation in the field of the
Kerr black hole. We conclude that the velocity remains subluminal contrary to
recent claims. Fifth section, an edited version of Ref.~\cite{rita5}, is
dedicated to another approach to the problem and presents the derivation of
the Papapetrou equation for massless particles from a simple Lagrangian.  

{\bf CHAPTER SIX} is the concluding chapter and presents a summary and remarks
with directions for the future work.

Some technical details are included in the appendices.

\chapter{Gravitational Lensing and Its Applications}
\section{Gravitational lensing as a cosmic telescope}
\label {sec:introduction}

\begin{flushright}
 "Do not Bodies act upon Light \\ at a distance, and 
by their action\\ bend its Rays; and is not this
action \\ strongest at the least distance?"\\
I. Newton, {\it Opticks}, 1704
\end{flushright}

Our Universe is controlled by gravity which doesn't limit its effects on
matter---light rays can be deflected and bent. In 1704 Newton proposed that
a light ray passing close to a massive body would be attracted and its path
bent. In 1911 Einstein obtained the so-called "Newtonian" value for the 
deflection angle from the Principle of Equivalence and the undisturbed
Euclidean metric. However with the full equations of General Theory
of Relativity (GR) Einstein obtained an angle twice the "Newtonian". The 1921
solar eclipse expedition confirmed the ``Einstein" value, making the Einstein
theory of General Relativity  a new paradigm and Einstein famous. 

In fact, during those early years many scientists contributed to the subject
of light deflection. It is worth mentioning such names as John Michell and H.
Cavendish, P. S. Laplace, A. Eddington, who appear to be the first to point
out that multiple (double) images can occur if two stars are sufficiently well
aligned, O. Chowlson, who is actually responsible for what we now call the 
"Einstein ring". The first to mention that action of gravity of a massive body
on light is similar to the refraction of light in an optical lens and called
it "gravitational lens", was a Czech engineer R. Mandl in his letter to
Einstein. In 1936 Einstein published a paper about this effect, where he
remarked that "there is no great chance of observing this phenomenon".
However, in 1937 the famous ``prophet" of astrophysics, Fritz Zwicky, published
a very optimistic paper about real possibility of discovery of gravitational
lens (GL) in case of a galaxy lensed by foreground galaxy. He was the first to
point out the usefulness of GL as an astrophysical tool which will allow a 
deeper look into the universe. He predicted many important applications of
GL, pointed out that magnification leads to a selection bias and estimated the
probability of detecting lensing to be very high.

In the mid 60's the discovery of quasars (QSOs) renewed interest in GL. The
subject was revived by S. Refsdal, who was later called by R. D. Blanford "the
most reliable prophet in gravitational lensing". In his first paper
\cite{refsdal1} Refsdal gave a full account of the properties of the point-mass
GL and calculated the time delay for the two images and mentioned a compact
object as a candidate for a lens. In a subsequent paper he considered the
application of GL for estimating the mass of the bending galaxy and the Hubble
constant through observable parameters of the source-lens system
\cite{Refsdal64}. In 1965 it was suggested by J. Barnothy that QSOs are in fact
Seyfert 1 galaxies, made to appear extra bright through other foreground
galaxy acting as a gravity lens. It is now believed that QSOs are indeed the
nuclei of galaxies, though not especially magnified by GL (with the possible
exception of BL Lac objects, that can, in fact, be magnified QSOs). However,
only after 1979 when the first GL system (QSO 0957+561~A~\&~B, ``The Old
Faithful") was discovered did a systematic search for lenses begin. By now
many GL systems have been identified which we can roughly  divide into three
classes: (i) more than 30 proposed multiple images of QSOs; (ii) several tens
of arcs and arclets; (iii) several radio and optical rings (and nearly rings)
with one possible candidate for an X-ray ring \cite{varsha}. One can now say
that gravitational lensing significantly affects our view and physical
understanding of the distant Universe and of its major constituents.

In spite of great theoretical and observational inspiration in GL field,
there exist certain problems, as in any other new field. One of the basic
problems is the amplification bias. The possible magnification of the source,
associated with the deflection, can fool the observer: some QSOs seen through
a foreground galaxy are much fainter than they appear to be. If one conducts
a flux-limited QSO search, some of these sources get boosted in flux above the
threshold of the sample. The net effect is that frequency of multiple imaging
appears much greater in flux-limited samples than in volume-limited ones.
Another direct consequence of the amplification bias is that one would
naturally expect to find an excess of amplifying galaxies near distant and
bright QSOs selected from a flux-limited sample. This has indeed been reported
\cite{vanDrom}.  

Another problem is the "verification", or "when is a lens a
lens". This became important after the discovery of several binary QSOs. There
exist several basic characteristics of images that are signatures of
gravitational lensing, though this list is by no means exclusive: (i) multiple
images of the same object; (ii) background image seen as a nearly complete
ring or as an extended arc; (iii) spectroscopic similarities of the images;
(iv) detection of the lensing galaxy or cluster in the right location and of
sufficient mass to create the image splitting; (iiv) images exhibiting
brightness variations characteristic of compact objects crossing the line of
sight. 

However, in spite of the problems and the uncertainties, the theoretical and
observational achievements of last few decades have made the GL one of the
most active and exciting fields of research. All possible applications of
gravitational lensing are surely impossible to list. From determination of the
Hubble constant and mass of the bender to using microlensing for dark
matter search to testing alternative gravitation theories and exotic objects
search---there is hardly an area of cosmology and astrophysics where GL has
not been applied. In this Introduction we will concentrate on the basic theory
underlying the gravitational lensing subject and the new progress of GL as
well as describe some of its interesting applications in astrophysics.

\newpage
 
\vskip 0.2in

\section{Elements of GR and cosmology and propagation of light}
\subsection{Basic notions of GR}
The basic principles of Einstein's GR are considered to be the Principle of
Relativity and Principle of Equivalence (EP). However, the decisive step
for the construction of the general relativity formalism was the suggestion by
Einstein and Grossman in 1912-1913 that the gravitational field must be
identified with the non-Minkowskian metric of the spacetime. Usually this
suggestion is deduced from the EP. However, some authors consider the metricity
of gravitation as the independent, if not, the main, principle of GR
\cite{sardanashwili,fock}. If we accept the metricity of the gravitational
field, then a pseudo-Riemannian \st must be chosen as the model for our
spacetime. (It is characterised by the structure of the manifold, pseudo-\R
metric, connection and curvature).

\vspace{-0.1in}
\SS{Differential manifold}
We will assume that the spacetime has the properties of a continuum, i.e.
it is a four-dimensional differential affine manifold $X_4$. Any point of it
can be labeled by real coordinates $x^{\mu}$ with $\mu =0,1,2,3$, where $0$
refers to the time coordinate, and 1,2,3 to the space coordinates. Any
affine space has a defined notion of the interval between its points. At
any point there exists an independent metric tensor field
$g_{\mu\nu}=g_{\mu\nu}(x)$ in order to allow for local measurements of
distances and angles. The square of the infinitesimal interval $ds$ between
$x^{\mu}$ and $x^{\mu}+dx^{\mu}$ is then determined by \be
ds^2=g_{\mu\nu}dx^{\mu}dx^{\nu}\,\,.
\label{eq:metrictensor} 
\end{equation} 
Thus, we can adopt the definition \cite{visser}:\\
{\bf Definition 1} {\it A spacetime is a four-dimensional manifold equipped
with a Lorentzian (pseudo-\R) metric with signature (-,+,+,+). (The manifold
should be Hausdorff and paracompact.)} This definition is important in the
studies of wormholes (see Chapter~\ref{sec:chapter3}). To admit the
construction of a wormhole, a spacetime shall be either ``almost-everywhere
Lorentzian" or be non-Hausdorff. \vspace{-0.1in}
 \SS{Connection}
In order to do physics in such a spacetime, we should have additional
structures on $X_4$. In an $X_4$ it does not make sense to say ``a
(nonzero) vector field is constant." To give such a statement a meaning, one
must introduce the notion of parallel transfer of vectors. Parallely
displaced from $x^{\mu}$ and $x^{\mu}+dx^{\mu}$, a vector $C^{\mu}$ changes
according to the prescription \be
d C^{\mu} = -\G^{\mu}_{\a\b}(x){C}^{\a}dx^{\b}\,\,.
\label{eq:transport}
\end{equation}
Here $dC^{\mu}$ is assumed to be bilinear in $ {C}^{\a}$ and $dx^{\b}$, the
set of the 64 coefficients $\G^{\mu}_{\a\b}$ is the affine connection. An $X_4$
equipped with a $\G$ is called a linearly connected space or $L_4$. The
parallel transport law (\ref{eq:transport}) can be extended to higher rank
tensor fields and densities, and it is possible to define their covariant
differentiation with respect to $\G$; for a vector,
$\n_{\nu}C^{\mu}=\del_{\nu} C^{\mu} + \G^{\mu}_{\a\nu} C^{\a}$. Now
it makes sense to state that ``a field is constant over spacetime"---its
covariant derivative has to vanish. In general, metric and connection are two
independent geometrical constructions on a manifold. But in GR, since the 
gravitational field is identified with the metric, a connection should satisfy
two additional conditions: 
\begin{itemize} 
\i Symmetry \hspace{2in}$\G^{\mu}_{\a\nu} = \G^{\mu}_{\nu\a}$,
\i Metricity \hspace{1.8in}$\qq\qq\n_{\nu}g_{\a\b}= 0$.

The last postulate guarantees that lengths, in particular the unit length,
and angles are preserved under parallel displacement. This enables a
choice of a coordinate system with connection coefficients vanishing at a
point $x$ and metric $g_{\mu\nu}(x)=\eta_{\mu\nu}$ (the \M metric) at this
point. Such a coordinate system is called the locally inertial system at $x$.

\end{itemize}
The above conditions allow us to express the components of such a connection
through the metric:
\be
\G^{\mu}_{\a\nu} = \half g^{\mu\s}\left(g_{\nu\s,\a} + g_{\a\s,\nu} -
g_{\nu\a,\s}\right)\,\,.\label{eq:christoffel}
\end{equation}
Components (\ref{eq:christoffel}) are called Christoffel symbols. 
\SS{Curvature}
Parallel transfer is a path-dependent concept. If we parallely transfer a
vector around an infinitesimal area back to its starting point, we find that
its components change. This change is proportional to the Riemann curvature
tensor \be
R_{\a\b\g}^{\qq \qq\d} := 2\del_{[\a}\G^{\d}_{\b]\g} + 
2 \G^{\d}_{[\a|\s}\G^{\s}_{|\b]\g}\,\,.
\end{equation}

\vspace{-0.1in}

\SS{Physical meanings}
Loosely speaking, the connection governs the ``acceleration" of a freely
falling particle in a gravitational field. A small test particle, free to move
under the unfluence of gravity alone, will follow a geodesic of the
metric.\footnote{A correction to this definition is in order (the meaning
will become more obvious in Chapter~5): geodesic equation is an equation for
motion of a moving free {\it spinless} test particle in an external
gravitational field (for ex. \cite{manoff}.} According to
(\ref{eq:metrictensor}), the length between two given points depends only on
the metric field. Therefore, the differential equation for the extremals can
be derived from $\d\int ds=  \d \int
\left(g_{\mu\nu}dx^{\mu}dx^{\nu}\right)^{1/2} = 0$ and results in 
\be
\fr{d^2x^{\mu}}{ds^2}+  \G^{\mu}_{\a\b} \fr{dx^{\a}}{ds}
 \fr{dx^{\b}}{ds } =0 \,\,,
\end{equation}  
the equation of geodesic. The Riemann tensor governs the difference in
acceleration of two freely falling particles that are near to each other. 
\vspace{-0.1in}
\SS{Einstein equations}
By contracting the Riemann tensor we obtain
\begin{itemize}
\i Ricci tensor
$$
R_{\mu\nu}\equiv R_{\a\mu\b\nu}g^{\a\b}\,\,,
$$
\i Ricci scalar 
$$
R \equiv R_{\a\b}g^{\a\b}\,\,,
$$
\i and \E tensor
$$
G_{\mu\nu}=R_{\mu\nu} - \half R g_{\mu\nu}\,\,.
$$
\end{itemize}
Finally, the \E field equations relate the curvature of \st (as measured by
the \E tensor $G_{\mu\nu}$) to the distribution of matter and energy (as
measured by the stress-energy tensor $T_{\mu\nu}$). Explicitly, 
\be
G_{\mu\nu}=\fr{8\pi G}{c^2} T_{\mu\nu}\,\,.
\end{equation}
The components $T^{00}$, $T^{0i}$, $T^{ij}$ are, correspondingly, energy
density, the energy flux and the stress. For most astrophysical and
cosmological purposes, one idealises bulk matter as a ``perfect fluid", for
which \be
T^{\a\b}=\left(\r c^2 +p\right)u^{\a}u^{\b} -p g^{\a\b}\,\,,
\end{equation}
where $\r$ denotes mass density and $p$ the pressure, both measured by a
comoving observer, and $u^{\a}$ is the 4-velocity, normilized to unity
\be
g_{\a\b}u^{\a}u^{\b}=1\,\,.
\end{equation}
The \E equations can be derived from the Einstein-Hilbert action principle
\be
S=\int\left( -\fr{c^3}{16 \pi G} R + L_{\rm M}\right) \sqrt{g} d^4x\,\,,
\end{equation}
where $L_{\rm M}$ is the matter Lanrangian.

\subsection{Weak fields}
\SS{Metric}
If the gravitational field is ``weak"---the metric of the \st differs little
from the flat, \M metric, one can write in approximately Cartesian coordinates
\be
g_{\mu\nu}\equiv \eta_{\mu\nu} + h_{\mu\nu}\,, \qq\qq |h_{\mu\nu}|<<1\,\,.
\label{eq:weak-field}
\end{equation}
GR then can be reduced to a ``linearized theory", where the metric for
any static distribution of matter can be given as \cite{The-book}
\be
ds^2\approx \left(1+\fr{2\P}{c^2}\right)c^2dt^2
-\left(1-\fr{2\P}{c^2}\right)d{\bf x}^2\,\,,
\label{eq:approxmetric}
\end{equation}
where $\P(x)$ is the Newtonian gravitational potential and $|d{\bf x}|= dl$
denotes the \E spatial line element. This ``post-\M" metric satisfies the
weak-field condition \ref{eq:weak-field}, if $|\P|<<c^2$ and matter moves
slowly $|{\bf v}|<<c$.

\SS{Effective refraction index}
From \ref{eq:approxmetric}, by putting
$ds=0$ and solving for $dl/dt$, we can obtain an effective speed of light,
$v_{\rm eff}$, to the first order
$$
v_{\rm eff}\equiv \fr{dl}{dt}=c\left(1+\fr{2\P}{c^2}\right)\,\,.
$$
The speed of light as measured in a locally inertial frame is $c$, but
since the coordinate system $(t,{\bf x})$ is not an inertial frame, the
{\it apparent coordinate speed} of light is different from its value
in vacuum. We can characterize the effect of light propagation in the
presence of gravitational potential by the {\it effective refractive index}
$n$ given by
\be
n(\mathbf{x}) = \frac{c}{v_{\rm eff}}= 1 - \frac{2 \Phi(\mathbf{x})}{c^2}\;.
\end{equation}	
The gravitational potential for a massive object is a negative quantity,
therefore, the apparent speed of light is slower in the presence of a
gravitational field. We assume that the lens is stationary, therefore its
gravitational field is a function of only space and is independent of time.
The light rays effectively move through a region of space with spatially
varying refractive index. This causes bending of light in analogy with the
usual optical phenomenon. We note that this effective refractive index is
independent of the wavelength of light, and, thus, to a very good
approximation, gravitational lensing is achromatic.  
\subsection{Strong fields}
The full \E equations are nonlinear. This is the principal difficulty in
extracting the exact solutions. Nevertheless, many solutions have been 
found; among them are the \Schw, the Friedmann-Robertson-Walker
(FRW) and the Kerr solutions.    

\SS{ADM split} In any well-behaved coordinate patch one can use the ``time"
coordinate to decompose the $(3+1)$-dimensional Lorentzian metric via the
Arnowitt-Deser-Misner (ADM) split \cite{MTW}. The ADM split yields: \be
g_{\mu\nu}(t,{\bf x}) \equiv 
\left[
\barr{ccc}
-(N^2-g^{ij}\b_i\b_j)  & \vdots  & \b_i \\
\ldots \ldots \ldots \ldots \ldots & . & \ldots \ldots\\
\b_i & \vdots  & g_{ij}
\earr 
\right]\,\,.
\end{equation}
The function $N(t,{\bf x})$ is known as the lapse function, while
$\bs{\b}(t,{\bf x})$ is known as the shift function. The three-metric
$g_{ij}(t,{\bf x})$ describes the geometry of ``space", while the lapse and
the shift functions describe how the space slices are assembled to form a \st.
This ADM split allows one to adopt quick and dirty definition of a horizon. In
every asymptotically flat region $N\rightarrow 1$, $\b \rightarrow 0$ and
$g_{ij} \rightarrow \d_{ij}$ asymptotically as one approaches spatial
infinity. Associated with each \ass flat region one may define a putative
horizon by vanishing of the lapse function. Roughly speaking, when $N=0$ time
has slowed to a stop. ADM split is essential to define the ADM mass (see App.
A). 

\subsection{Standard cosmological model}\label{sec:cosmology}
\SS{FRW Universe}\label{sec:FRW}
Our universe is believed to be homogeneous and isotropic on large scales.
This is borne out by the observations of the distribution of galaxies and the
remarkable isotropy of the Cosmic Microwave Background Radiation. These
assumptions, together with the field equations of GR, give solutions for the
geometry of the universe. The space-time metric for such a universe is given
by the  FRW line element (for example, \cite{weinberg}) 
\be
ds^2 = c^2dt^2 - a^2(t) \left[ \frac{dr^2}{1-kr^2} +r^2
(\sin^2\theta d\theta^2 + d\phi^2)\right]\;,
\label{eq:frw}
\end{equation}
where $(r,\theta,\phi)$ are the spatial {\it comoving} spherical polar
coordinates of a space-time point and $t$ is the cosmic time, $a(t)$ denotes
the scale factor of the universe at time $t$ and $k$ is the curvature index
of the spatial hypersurfaces $t=\mbox{constant}$. In this form of the metric we
can rescale the coordinates in such a way that constant $k$ is $+1,\,-1$ or
$0$, corresponding to spatial sections of constant positive, negative or
vanishing curvature, respectively. With such a rescaling, the coordinate $r$
in the metric is dimensionless and $a(t)$ has dimensions of length. The
dynamics of the universe is obtained from the field equations of GR. These
equations are presented later in this chapter. But many properties of the
universe, which are kinematic in nature, can be obtained solely from the FRW
metric. 

Light from distant objects appears redshifted due to the expansion of
the universe. The observed redshift is defined as $ z \equiv (\lambda_{\rm o} -
\lambda_{\rm e})/\lambda_{\rm e}$, where $\lambda_{\rm e}$ and 
$\lambda_{\rm o}$ are the emitted and the observed wavelengths,
respectively. It is  related to the expansion parameter by
\be
\frac{a_0}{a} = 1 + z\;,
\end{equation}
where $a_0$ is the present value of the scale factor and $a$ is
the value of the scale factor when the light ray was emitted from the
source. We define the Hubble parameter $H(t)$, which measures the rate of
change of the scale factor at any time $t$ as $H(t) =\dot{a}(t)/a(t)$. The
present value of the Hubble parameter is denoted as $H_0$. Its latest
numerical value is ascertained to be $H_0=70 \pm 7 \,\,{\rm kms}^{-1} {\rm
Mpc}^{-1}$ \cite{Freedman2000}. The dynamics of the universe depends on the
matter content of the universe. This can be specified by the energy density
$\r(t)$ and the pressure $p(t)$, which are often related by an equation of
state of the form $p = w \, \rho$; the classic examples are 
\begin{itemize}
\item{\bf Non-relativistic matter: ($w=0$) }~ Galaxies are the tracers of  the
expansion of the universe in the sense that they follow the general expansion
of the universe. Treated as a fluid, they exert negligible pressure, therefore,
to an excellent approximation, they can be treated as pressureless dust.
\item {\bf Radiation: ($w=1/3$)}~A major component of the early universe was
in the form of the radiation. It is believed that after the radiation era, the
universe has undergone a long period of matter domination, with the radiation
providing only  $10^{-4}$ of the closure density today.

\item {\bf Cosmological Constant ($\Lambda$): $(w=-1)$}~ Though originally 
introduced as an arbitrary constant by Einstein \footnote{The original
motivation was that the universe was believed to be static and
therefore a cosmic repulsive force was needed to balance the
attractive force of gravity. This was later on abandoned by \E when
it was discovered that the universe is, in fact, expanding.}, it has
made a comeback in the recent times \cite{sahni00}. The current
observations suggest that about $2/3$ of the present day density is in
the form of cosmological constant $\Lambda$, and only $1/3$ is in
the form of matter.
\end{itemize}
In general though there need not be a simple equation of state. There maybe
more than one type of material, such as combination of radiation and
non-relativistic matter. Certain types of matter, such as a scalar field,
cannot be described by an equation of state at all.  

\SS{Dynamical Equations}
The crucial equations governing the evolution of the scale factor of the
universe and the matter-energy content of the universe, are the Friedmann
equations: 
\begin{eqnarray}
\label{eq:friedmann1}
&& \frac{2\ddot{a}}{a} + \frac{\dot{a}^2+kc^2}{a^2}- \Lambda c^2=
-\frac{8\pi G}{c^2} p\\ \label{eqn:friedmann}
&& \left(\frac{\dot{a}}{a}\right)^2 + \frac{kc^2}{a^2} =  \frac{8\pi{\rm
G}}{3} \rho   + \frac{\Lambda c^2}{3}, 
\end{eqnarray}
where we have also included a constant $\Lambda$ term. The spatial geoometry
is flat if $k=0$. For a given $H$, this requires that the density equals the
critical density 
$$
\rho_c = 3H_0^2/8\pi{\rm G}\,\,,
$$
Densities are often measured as fractions of $\r_{\rm c}$:
$$
\O(t)\equiv \fr{\r}{\r_{\rm c}}\,\,,
$$
the dimensionless density parameter as $\O_{\rm M} = \rho_0/\rho_{\rm c}$ and
the Lambda parameter as $\O_{\rm Lambda} = \L c^2/3H_0^2$.

\SS{Distance Measures}
\label{sec:distance}
In an expanding, curved space-time, the measure of distance is not
uniquely defined. The distance between two points can be defined by:
\begin{itemize}
\item The light travel distance.
\item The flux received from a standard candle. 
\item The angle subtended by a standard ruler.
\end{itemize}
In a Euclidean space all these three measures would coincide. However,
in a non-Euclidean space-time the three measures are all different. We
need to define all the distances separately in complete analogy with
their Euclidean counterparts.
\begin{itemize}
\item [(a)]{\bf Comoving Coordinate Distance:} 
A quantity of interest is the coordinate distance $r(z)$ up to a 
redshift $z$.
Since light rays travel along the null geodesics 
of the space-time, $ds=0$. For a FRW metric we obtain
\begin{equation}
\int_{t_o}^t \frac{cdt'}{a(t')} = \int_r^0 \frac{dr}{\sqrt{1-kr^2}}\;.
\end{equation}
We can convert the integrals over $t$ into integrals over $z$ by
differentiating $1+z=a_0/a$ to obtain $dz/dt = -(1+z)H(z)$ and
substituting in the previous equation. This gives
\begin{equation}
\frac{c}{H_0 a_0} \int_0^z \frac{dz'}{h(z')} = \int^r_0
\frac{dr}{\sqrt{1-kr^2}}\;, 
\end{equation}
where we defined the dimensionless Hubble parameter $h(z)=H(z)/H_0$. For a
flat space-time, where $k=0$, we obtain \begin{equation}
r(z) = \frac{c}{a_0 H_0} \int_0^z \frac{dz'}{h(z')}\;.
\end{equation}

\item [(b)]{\bf Luminosity Distance:} The Luminosity distance 
$D_{{\rm L}}$ is defined in such a way as to preserve the \Euc inverse-square
law of diminishing of light with distance from a point source. This gives
\begin{equation} D_{{\rm L}} = a_0 r (1+z)\;.
\end{equation}

\item [(c)] {\bf Angular Diameter Distance:} The Angular Diameter Distance
$D_{\rm A}$ is defined in such a way as to preserve a geometrical property of
\Euc space, namely, that the angular size subtended by an object should fall
off inversely with $d_{\rm A}$. This gives 
\begin{equation}
D_{\rm A} = \frac{a_0 r}{1+z}\;.
\end{equation}
In later sections we will use the angular diameter distance measured
by an observer at $z_1$ up to $z_2$. This is given by
\begin{equation}
D_{\rm A}(z_1,z_2) = \frac{a_0 r_{12}}{1+z_2}\;,
\end{equation}
which gives
\begin{equation}
D_A(z_1,z_2) = \frac{c}{(1+z_2)H_0} \int_{z_1}^{z_2} \frac{dz'}{h(z')}\,.
\end{equation}
\end{itemize}
 
\section{Basic concepts of gravitational lensing}
 
\subsection{Approximations}\label{sec:approximations}

In the following section we describe results on gravitational lensing based on
the Refs.~\cite{The-book,Narayan,Zakharov,fort-mellier,wu,sazhin-review}.
The formal description of GL is based on several approximations.  In the
previous section we have already made an assumption of weak gravitational
fields, thus, justified the use of linearized field equations of GR. Indeed,
even in clusters of galaxies, the deflection angles are well below $1'$ and
the maximum image separation in multiple-imaged systems are not more than
$10''$. Thus, we can express the approximations used as
\begin{itemize}
\i We assume that the gravitational field can be described by
the linearized metric:
$$
ds^2 = \left(1+  \frac{2 \Phi}{c^2}\right)~c^2dt^2 - \left(1- 
\frac{2\Phi}{c^2}\right)~ dl^2\;,
$$
where $\Phi$ is the Newtonian potential due to the gravitational field.

\item {\it Geometrical optics approximation}---the scale over which
the gravitational field changes is much larger than the wavelength
of the light being deflected.

\item {\it Small-angle approximation}---the total deflection angle is
small. The typical bending angles involved in gravitational lensing of
cosmological interest are $ < 1^{\prime}$; therefore we describe the lens
optics in the paraxial approximation.

\item {\it Geometrically-thin lens approximation}---the maximum
deviation of the ray is small compared to the length scale on which the
gravitational field changes. Although the scattering takes place continuously
over the trajectory of the photon,   appreciable bending occurs only within a
distance of the order of the impact parameter.
\end{itemize}
\begin{figure}[ht]
\centerline{  
\psfig{figure=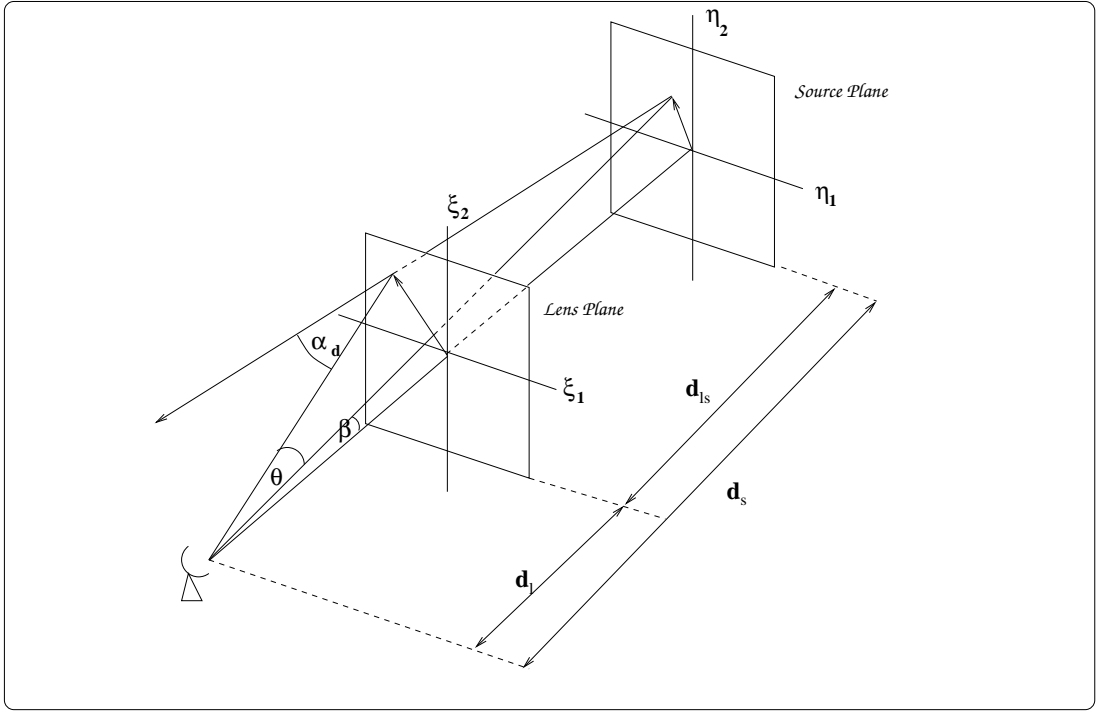,width=0.6\hsize,angle=-90}}
\medskip
\linespread{0.8}
{\caption[Two-dimensional picture of gravitational lensing situation]
{Schematic diagram of gravitational lensing. The two hypothetical planes are
defined to erect a convenient coordinate system. } 
\label{fig:lensgeom}}
\end{figure}
We begin the discussion of gravitational lensing by defining two
planes, the source and the lens plane. Fig.~\ref{fig:lensgeom} describes a
typical lensing situation. A convenient origin, passing through the lens is
chosen on the sky. The planes, described by Cartesian coordinate systems,
pass through the source and deflecting mass and are perpendicular to the
optical axis (the straight line extended from the source plane through the
deflecting mass to the observer). These planes are hypothetical and are
solely for the purpose of visualization. The coordinates of the image with
respect to the origin are $(\xi_1,\xi_2)$ and that of the source are $(\eta_1,
\eta_2)$, respectively. Since the components of the image and the
source positions are much smaller in comparison to the distances to lens and
source planes, we can write the coordinates in terms of the observed angles.
Therefore, the image coordinates can be written as $(\theta_1,\theta_2$) and
those of the source as $(\beta_1,\beta_2)$.

\vspace{-0.3in}
\subsection{Deflection angle}

To calculate the bending angle produced by a gravitational field, we use the
{\it ray equation} which describes the path of a light ray through a spatially
varying refractive index $n(\bf x)$. To derive the ray equation we start from
Fermat's principle, which states that the light travel time from the source
to the observer is an extremum, \be
\delta\int_{t_{{\rm s}}}^{t_{{\rm o}}} dt = 0\;,  
\label{eq:fermat}
\end{equation} 
where the subscript ``$s$'' stands for the source and ``$o$'' stands for the
observer. The integral is evaluated along the trajectory of the light ray. To
obtain the ray equation we parameterize the ray path by $s$, therefore
\be
\frac{d {\bf x}}{dt} = \frac{d{\bf x}}{ds}\frac{ds}{dt}\;.
\end{equation}
Substituting $ |d{\bf x}/dt| = c/n({\bf x})$ in Eq.~\ref{eq:fermat}, we obtain
\be
\delta \int_{s_{em}}^{s_{obs}} n({\bf x})\left(\sum_i (dx_i/ds)^2 \right) ^{1/2} ds=0 \;.
\end{equation}
Using the Euler-Lagrange equation and choosing the parameter $s$ to be the
path length $l$ we obtain
\be
\frac{d}{dl} \left [ n({\bf x})\,\,\widehat{{\bf x}} \right] = 
\bs{\nabla} n\;,
\label{eq:rayeq}
\end{equation} 
where $\widehat{{\bf x}} = d{\bf x}/dl$ is the unit tangent vector along the
path of the ray ${\bf x}(l)$. In astrophysical applications of GL the
bending angles are small, therefore, to obtain the deflection angle using this
formula we can integrate equation (\ref{eq:rayeq}) along the unperturbed path
of the ray $\gamma$ to obtain
\be
{\mbox {\boldmath $\alpha$}}_d \equiv (\widehat{{\bf x}}_{em} - 
\widehat{{\bf x}}_{obs})_{\perp} = 
-\int_{\gamma} \bs{\nabla}_{\perp} n \,dl \;,
\label{eq:angle}
\end{equation}
where the component perpendicular to the unperturbed ray is used in
the calculation. Using this formula we can obtain the deflection angle
angle due to a point mass. We set our coordinate system such that the
source and the lens lie along the $z$-axis, and the origin is chosen at the
position of the lens, the component of the position vector perpendicular to
the $z$-axis being denoted as ${\mathbf \xi}$. The impact parameter $\xi_0$ is
the distance of the unperturbed ray from the centre orthogonal to the direction
of propagation. From (\ref{eq:angle}) we obtain the deflection angle
$\alpha_{\rm d}$ as \be 
\alpha_{\rm d} =- 2GM\,\int_{\gamma_0} \bs{\nabla}_{\xi}
\left (\frac{1}{\sqrt{\xi^2 + z^2}}\right) \,dz =
\frac{2GM}{\xi_0} \,\int_{-\infty}^{\infty} \left( 1 +
\left(\frac{z}{\xi_0}\right)^2 \right)^{-3/2} \,\frac{dz}{\xi_0}\,\,,
\end{equation}
where, since the main contribution to the integral comes from the range
$-\xi_0 < z<\xi_0$, we have put the limits of the integral as minus and plus
infinity. Thus, the Einstein deflection angle of a light ray passing near a
compact mass $M$ at a distance $\xi$ is 
\begin{equation}
\alpha_{\rm d}(\xi)=\frac{4 G}{c^2}\frac{M}{\xi}\,\,.
\label{eq:alpha-einstein}
\end{equation}
This bending angle is twice the value of what would be expected from the
Newtonian theory. Einstein's General Theory of Relativity was vindicated when
this angle was measured for the case of the Sun, where the predicted value
($\sim 1.75^{\prime \prime}$) was confirmed by observations during a total
solar eclipse in 1919.\\
\vspace{0.01in}
For an extended mass we can obtain this angle by integrating
individual deflections due to all mass elements constituting the lens. In
{\it thin-lens approximation} the deflection angle $\alpha_{\rm d}$ can be
obtained by projecting the volume mass density of the deflector onto the lens
plane $\bs{\xi}=(\xi_1,\xi_2)$, which results in a surface mass density
$\Sigma(\bs{\xi})=\int\rho(\bs{\xi},z)dz$. The deflection angle is a
superposition of Einstein angles for mass elements $dm=\Sigma(\xi)\,d^2\xi$.
Considering all the deflecting mass to be concentrated only in the lens plane
and the deflection taking place only in the lens plane, a deflection angle can
be expressed as a two dimensional vector \begin{equation} {\bs \a}_{\rm
d}({\bs \xi})=\frac{4G}{c^2}\int\!\!\int \fr{\S({\bs \xi'})\,({\bs
\xi}-\bs{\xi'})} {|{\bs \xi}-\bs{\xi'}|^2}\,d^2\xi' \,\,.
\end{equation}
From here we obtain
\begin{equation}
{\bs \alpha} =
\fr{4G}{c^2}\fr{D_{\rm L}D_{{\rm LS}}}{D_{\rm S}} \int\!\!\! \int 
\Sigma(\bs{\t}^{\prime}) \, \fr{\bs{\t} -\bs{\t}^{\prime}} 
{|\bs{\theta} -\bs{\theta}^{\prime}|^2} \, d^2\theta^{\prime} \,,
\label{eq:alpha-thru-Sigma}
\end{equation} 
where we have used the fact that $ \bs{\xi}=D_{\rm L} \bs{\t}$ and
defined displacement vector $\bs{\a}$ as 
\begin{equation}
\bs{\alpha}=\fr{\dls}{\ds}{\bs \a}_{\rm d}\,\,.
\end{equation}
There is no unique definition of distances in a curved spacetime. Distances
which should be used in this equation are the angular diameter distances to
ensure that the equation remains valid for a more general spacetime
(see~\ref{sec:cosmology}). Defining the critical density as $ \S_{\rm cr} =
(c^2/4 \pi G)(D_{\rm S}/D_{\rm L} D_{{\rm LS}})$ and the dimensionless
quantity $\kappa( \bs {\t}) = \S (\bs{\theta})/  \S_{\rm cr}$,  we can write
equation (\ref{eq:alpha-thru-Sigma}) as \be  \bs{\alpha} = \fr{1}{\pi}
\int\!\!\! \int \kappa({\bs \t}^{\prime}) \, \fr{{\bs \t}-{\bs \t}^{\prime}}
{|{\bs \t} - {\bs \t}^{\prime}|^2} \,d^2\t^{\prime}\;.
\label{eq:alpha-thru-kappa}
\end{equation} 

\subsection{Lens equation and the lensing potential}

Considering the projection of the light ray on the two planes, we can
derive a relation between the source coordinates and the image
coordinates in terms of the bending angle $ \bs{\a}_{\rm d}$  
\begin{equation}
\bs{\b} = \bs{\t} - \fr{D_{\rm LS}}{D_{\rm S}}{\bs \a}_{\rm d}\,,  
\label{eq:lensequation}
\end{equation}
Equation (\ref{eq:alpha-thru-kappa}) can be written as 
\begin{equation} 
{\bs \a} = {\bs \nabla} \psi\,, \qquad \psi =  \frac{1}{\pi} \int\!\!\!
\int \kappa({\bs \theta}^{\prime})  \ln |{\bs \theta}
-{\bs \t}^{\prime}|\, d^2\theta^{\prime} \;. 
\label{eq:potential}
\end{equation} 
Using the identity $ {\nabla}^2 \ln |{\bf x}-{\bf x^{\prime}}| = 2 \pi
\delta^2({\bf x}-{\bf x^{\prime}} )$ we obtain the equation which the
dimensionless relativistic lens potential $\psi$ satisfies:
\be
\nabla ^2 \psi ( \bs{\t}) = 2\kappa( \bs{\t})\;.
\label{eq:poisson}
\end{equation}
In terms of this potential the lens equation (\ref{eq:lensequation}) can be
written as \be
\bs{\beta } = \bs{\theta } - \bs{\nabla } \psi ( \bs{ \theta })\;.
\label{eq:basicleq}
\end{equation}
In general, angles $\beta,\,\theta,\,\alpha$ and $\alpha_{\rm d}$ may not be
coplanar and so, Eq.~\ref{eq:basicleq} is a vector equation. Given the matter
distribution of the lens and the position of the source the lens equation may
have more than one solution, which means that the same source can be seen at
several positions in the sky. The lens equation describes a mapping $\bs{\t}
\mapsto \bs{\b}$ from the lens plane to the source plane.  

\vspace{-0.3in}
\subsection{Magnification}\label{sec:magnifications}

Besides multiple imaging, the differential deflection across a light bundle
affects the properties of the images. In particular, the cross-sectional area
of the bundle gets distorted and the flux of the images is influenced. The
source subtends a solid angle $\Delta\o_{\rm s}$ at the observer in the
absence of lensing. In the case of lensing $\Delta\o_{\rm i}$ is a solid angle
subtended by the image. Gravitational lensing preserves the surface brightness
of the source (we assume that during deflection no absorption or
emission of light is taking place and that deflection by a nearly static
deflector introduces no additional frequency shift between the source and
observer, except a cosmological redshift). The flux is $S=I\cdot\Delta\o$, $I$
being specific intensity. For an infinitesimally small source the ratio
between the solid angles gives the flux amplification due to lensing $$
|\mu|=\frac{S_i}{S_s}=\frac{d\o_{\rm i}}{d\o_{\rm s}}\,\,. 
$$
Local properties of the lens mapping are described by its Jacobian matrix
$\cal A$ 
\be
{\cal A} \equiv \pder{\bs \b}{\bs \theta}=\left(\d_{ij} -
\fr{\del^2 \psi_i({\bs \t})}{\del \t_i \del \t_j }\right) \equiv \d_{ij} -
\psi_{ij}\,\,. 
\label{eq:Jacobian}
\end{equation}
A solid-angle element $\delta\beta^2$ of the source is mapped to the solid-angle 
element $\delta\theta^2$ of the image, and the magnification is given by
$$
\frac{\delta\theta^2}{\delta\beta^2} = \mbox{det} \,\mu\,\,.
$$
The Jacobian matrix $\cal A$ is thus the inverse of the magnification factor
\be
\mu(\bs{\t}) = \left | \mbox{det}\pder{\bs \b}{\bs \t}\right |^{-1} = 
\fr{1}{\mbox{det}{\cal A}(\bs{\t})}\,\,. \label{eq:mu}
\end{equation}
Eq.~\ref{eq:Jacobian} shows that the matrix $\psi_{ij}$ describes deviation
of the lens mapping from the identity mapping. From (\ref{eq:poisson}) we
have
\be
\mbox{Tr} \,\psi_{ij} = 2 \k\,\,.
\label{eq:trace}
\end{equation}
Two additional combinations of $\psi_{ij}$ are important:
\begin{align}
& \g_1=\fr{1}{2}\left(\psi_{11}-\psi_{22}\right);\no \\
& \g_2 = \psi_{12} = \psi_{21} \,\,.\no
\end{align}
With these definitions we can write the Jacobian matrix $\cal A$ as
\be
{\cal A} = 
\left(
\barr{cc}
1-\k-\g_1 & -\g_2\\
-\g_2 & 1-\k +\g_1
\earr
\right)\,\,;\
\end{equation}
and the magnification factor
\be
\mu = \fr{1}{(1-\k^2)-\g^2}\,\,.
\label{eq:finalmu}
\end{equation}
The eigenvalues of $\cal A$ are $a_{1,2} = 1-\k \mp \g$, where $\g =
\sqrt{\g_1^2 +\g_2^2}$ and the determinant is det${\cal A}=(1-\k^2)-\g^2$.
When the line of sight competely misses the deflector, $\S({\bs \t})=0$ and the
$\k$ term vanishes in Eq.~\ref{eq:poisson}. So, $\k$ represents the
amplitude of the convergence due to the matter within the light-ray (also
called Ricci focusing), while the $\g$ term is the amplitude of the shear due
to the matter outside the beam (also called Weyl focusing). Eigenvalues of
$\cal A$ describe the image distortion in the radial and tangential
directions for a circular source, resulting in an ellipse.

The zeroes of the Jacobian of the lens mapping are called the singular points
of the lens mapping. For the isolated lenses the lens mapping would go to
identity at large distances from the lens mapping. For such lenses the zeroes
of the Jacobian are either points or closed curves in the image plane. These
curves are called the critical curves. Their images in the source plane are
calles the caustics. Caustics separate the regions of different image
multiplicities. When a source crosses a caustic the number of images changes by
two.

\subsection{Lens models}\label{sec:point-mass}

The mass distribution inside galaxies and clusters of galaxies is in general
quite complicated and may not have any symmetry. However, since the circular
mass distributions are easier to handle analytically they are very convenient
to use in gravitational lensing. Symmetry allows the lens equation to be
separated in the polar coordinates making the equations analytically
tractable. Besides, for many celestial bodies, like ``Jupiters", stars, black
holes, and even galaxies when the light rays pass outside the deflector, a
point mass approximation is valid. Galaxies and even clusters of galaxies are
also well approximated by the singular isothermal sphere. Below we give a
description of these two models with corresponding lensing equations.

\SS{Point Mass (Schwarzschild) Lens\hspace{-0.16in}}\footnote{In
lens theory the term ``point mass" is used whenever one is concerned with light
rays deflected with the impact parameters greater than the Schwarzschild radius
of a static spherical object; the exterior of such an object is always
described by the Schwarzschild metric, hence the term \emph{Schwarzschild
lens}.}

\begin{figure}[ht]
\centerline{
\epsfig{figure=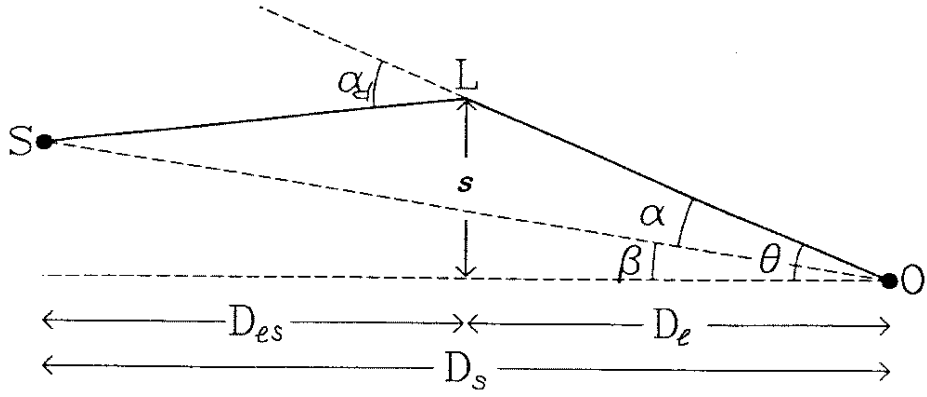,width=0.8\textwidth}}
\caption{Basic geometry of gravitational lensing}
\label{fig:basic}
\end{figure}

\noindent ({\it i}) {\bf Lens equation}\\ Due to axial symmetry,
the propagation of light reduces to one dimensional problem. Let us consider
the situation described in Figure~\ref{fig:basic}.

From the Figure follows the geometric relation
\be
\t D_{\rm S} = \b D_{\rm S} + \a_{\rm d} D_{\rm LS}\,\,.
\end{equation}
Substituting for $\a$ from (\ref{eq:alpha-einstein}), we rewrite it as
\be
\b = \t - \fr{D_{\rm LS}}{D_{\rm L} D_{\rm S}} \fr{4 GM}{c^2 \t}\equiv
\t - \fr{\t_{\rm E}^2}{\t}\,\,,
\label{eq:pointleq}
\end{equation}
where
\be
\te^2 =\fr{4 GMD_{\rm LS}}{c^2 D_{\rm L} D_{\rm S}}\,\,.
\label{eq:theta_e}
\end{equation}
Angle $\t_{\rm E}$ denotes the angular radius, called the {\it Einstein
radius}. It provides a natural angular scale to describe the lensing geometry.
Sources which are closer than about $\theta_{E}$ to the optic axis are
significantly magnified, whereas sources which are located well outside the
Einstein ring are magnified very little. Besides, it is the radius of a {\it
tangential critical curve}. In the given case of, the caustic is a point on tye
optical axis. If a source is displaced slightly off the axis, two bright images
are created on opposite sides of the lens centre, one just inside, and the
other just outside the critical radius. The equation (\ref{eq:pointleq}) has
two real roots: \begin{equation} \theta_{1,2} = \frac{1}{2} \b \pm\frac{1}{2}
\sqrt{\b^2 + 4 \theta_{\rm E}^2} \,\,, \end{equation}
which correspond to two physical images of the source $S$. The
angular separation between the images is
\begin{equation}
\Delta\theta = \theta_1 - \theta_2 = \sqrt{\b^2+ 4
\theta_{\rm E}^2}\geq 2\theta_{\rm E}\,\,.
\label{eq:deltatheta} 
\end{equation}
The separation between the source and the deflector is related to the
image position by
\begin{equation}
\theta_1 + \theta_2 = \b\,\,.
\end{equation}
Thus, the lens equation has two solutions of the opposite sign. The source
has an image on each side of the lens, one inside Einstein radius, one
outside. If the source is a disk of radius $\varphi_{\rm s}$, the images will 
represent ellipses, squeezed along the axis connecting them and stretched in
perpendicular direction. For example, if $a, b$---semi-major
axes, the area of ellipse is $\Omega =  \pi ab$. The relations with the
radius of the source we can write as  
\begin{equation}
\fr{a}{\varphi_{\rm s}} = \fr{\t_1}{\b};\qq
\fr{b}{\varphi_{\rm s}} = \fr{d\t_1}{d\b}\,\,.
\end{equation}
\noindent
In the case of perfect alignment between source, lens and observer, ($\b=0$),
an  observer will see a ring with radius $\theta_{\rm E}$ and thickness
\begin{equation}
\left. 2\,\varphi_{\rm s}\,\fr{d\t_1}{d\b} \right|_{\b\to\ 0} =
2\varphi_s \left[ \frac{1}{2} + \frac{1}{2}\,\frac{\b}{\sqrt{\b^2
+ 4\,\theta_{\rm E}^2}} \right]  = \frac{1}{2}\,2\varphi_{\rm s} =
\varphi_{\rm s}\,\,,
\end{equation}
equal to the source radius. The solid angle which it subtends on the
sky is then $2\pi\theta_{\rm E}\varphi_{\rm s}$.

\noindent
({\it ii}) {\bf Magnifications}

For a circular symmetric lens, the magnification factor $\mu$
(Eq.~\ref{eq:mu}) is reduced to 
\be
\mu=\fr{\t}{\b}\fr{d\t}{d\b}\,\,.
\label{eq:pointmag}
\end{equation}
For a point mass lens, which is a special case of a circular symmetric lens,
we substitute for $\b$ using the lens equation (\ref{eq:pointleq}) to obtain
the magnifications of the two images,
\be
\mu_{\pm} = \left[1-\left(\fr{\t_{\rm}}{\t_{\pm}}\right)^4\right]^{-1} = 
\fr{u^2+2}{2u\sqrt{u^2+4}}\pm\fr{1}{2}\,\,,
\end{equation}
where $u$ is the angular separation of the source in units of Eistein angle.
The total magnification of the two images is
\be
\mu=|\mu_{+}| +|\mu_{-}| = \fr{u^2+2}{u\sqrt{u^2+4}}\,\,.
\end{equation}
$u=1$, or $\b=\te$, is often taken to be a typical case
that characterizes the efficiency of the lens. This corresponds to $\mu=1.34$
or $\D\mu=0.32$ in apparent magnitude. Point-like masses play an important role
in the study of {\it microlensing}, which arises when the separation of the
images is too small to be resolved and the lensing effect can only be observed
through the lensing-induced time variability of the source.

\vspace{-0.13in}
\subsubsection{Singular Isothermal Sphere}\label{sec:SIS}

When we consider galaxies as lenses we need to allow for the distributed nature
of the matter. The studies of the flat rotation curves of galaxies and the
galaxy/gas distributions in clusters of galaxies suggest that the total matter
profiles in these systems follow the singular isothermal sphere (SIS) model
very well \be \r(r) = \fr{\s_v^2}{2\pi Gr^2}\,\,,
\label{eq:rho}
\end{equation}
where $\s_v$ measures the line-of-sight velocity dispersion. In this model it
is assumed that the mass components behave like particles of an ideal gas,
confined by their combined spherically symmetric gravitational potential.
It is assumed also that the gas is isothermal, so that $\s_v$ is constant
across the galaxy $m\s_v^2= kT$. Upon projecting along the line-of-sight, we
obtain the surface mass density
\be
\S(\xi) = \fr{\s_v^2}{2G\xi}\,\,,
\end{equation}
where $\xi$ is the distance from the centre of the two-dimensional profile.
Referring to Eq.~\ref{eq:alpha-einstein}, we find 
\be
\a_{\rm d} = 4\pi\fr{\s_v^2}{c^2}\,\,,
\end{equation}
which is independent of the impact parameter. The Einstein angle in this case
is \be
\te=4\pi \fr{\s_v^2}{c^2}\fr{D_{\rm LS}}{D_{\rm S}}=
\a_{\rm d}\fr{D_{\rm LS}}{D_{\rm S}}\,\,.
\end{equation}
The solution to the lens equation $\b=\t-\te$ is
\be
\t_{\pm} = \t_{\rm E} \pm \b\qq\mbox{for } \b<\te\,\,.
\end{equation}
Thus, the lens has two images on the oposite sides of the lens centre. For
$\b>\te$ only one image appears at $\t=\t_{+}=\b+\te$. The image separation
is just the diameter of the Einstein ring: $\D \t= \t_{+} +\t_{-} \equiv
2\te$. Although the surface mass density is infinite at $\xi=0$, the
behaviour of the model for larger values of $\xi$ seems to approximate the
matter distribution of galaxies fairly well. Real galaxies, however, cannot
follow the density law (\ref{eq:rho}) due to an infinite density at the
centre and an infinite mass. Other models exist and are frequently employed.
For example, if the singularity is removed from the centre, the model is
called a softened SIS---a SIS with a finite core radius $r_{\rm c}$ (ISC).
In this case a lens is capable of producing either one, or three images.
However, the usual absence of the third image implies that even if there is a
core, it must be small, $\lapprox 200$ parsecs \cite{Wallington93}. \\
The magnifications of the two images follow from Eq.~\ref{eq:pointmag} and
are \be \mu= \fr{|\t|}{|\t|-\te}  \,\,, \end{equation} the circle $|\t|=\te$
is a tangential critical curve. Images are stretched in the tangential
direction by a factor $|\mu|$, whereas the distortion factor in the radial
direction is unity (see Section~\ref{sec:magnifications})     

\section{Astrophysical applications of gravitational lensing}

Listing already existing and possible future astrophysical and
cosmological applications of gravitational lensing is nearly an impossible
task, so vast has become this field in the last years. Zwicky's idea of the
gravitational lensing as a cosmic telescope is proving itself with each
observational discovery. We can see the magnified distorted images of galaxies
which otherwise are far too dim to be observed, not to say, studied.
Gravitational lensing effect allows us to test the General Relativity Theory,
to probe the nature of the lensing object, the source and the intermediate
space, and to test the large-scale structure of the universe. We will describe
several interesting applications of gravitational lensing.

\subsection{Determination of Hubble parameter and mass of the deflector}

\SS{Hubble parameter}

One of the first applications of gravitational lensing, suggested by
Refsdal \cite{Refsdal64}, is the determination of the Hubble constant via the
direct measurement of the time delay $\Delta t$ between the observed light 
curves of multiply imaged quasars. For axially symmetric lens this method
can be described using the wavefront picture. Wavefront characterizes the locus
of all points with equal light-time-travel from the source. Light rays in
vacuum are  perpendicular to the wavefronts, which are spherical close to the
source. However, they become deformed by the gravitational field of deflector,
may intersect themselves and cross the observer several times, producing
multiple imaging. Every passage past the observer corresponds to the image of
the source in the direction normal to the wavefront. The time delay for pair
of images is the time between two crossings of the wavefront.

The wavefronts from a distant, doubly imaged quasar cross each other at the
symmetry point. They represent the same light propagation time and for an
observer, located at a distance $x$ from the symmetry axis, the time delay must
be equal to the distance between the wavefronts at the observer divided by the
velocity of light. The deflection law can be written as \cite{Refsdal64}
\begin{equation}
\a_{\rm d} \propto |\xi|^{\varepsilon -1}\,\,,
\end{equation}
with $\varepsilon =0$ for a point mass lens. Using Hubble relation for
small redshifts \begin{equation}
c z = D H_0\,\,,
\end{equation} one obtains the expression for the Hubble parameter $H_0$ in
terms of observable quantities \begin{equation}
H_0 \D t = \fr{1}{2}\fr{z_{\rm s} z_{\rm d}}{(z_{\rm s}-z_{\rm d})}
\D \t^2 (2-\varepsilon)\,\,.
\label{eq:hubbledelay}
\end{equation}

\SS{Determination of mass}

One more direct application of gravitational lensing is the determination of
the mass of the deflector. The simplest  situation here is when the lens is
a spherically symmetric object and the source lies exactly behind the lens
centre. The lens can then form an Einstein ring. In this case, the bending
angle is \begin{equation} 
\alpha_{\rm d}(\t)=\frac{4GM(<\theta)}{c^2D_{\rm L}\theta}\,\,,
\end{equation} and lensing equation with the source at the origin becomes
\begin{equation}
\theta=\alpha_{\rm d}(\theta)\frac{D_{\rm LS}}{D_{\rm S}}\,\,.
\end{equation} Combining these two equations we obtain \begin{equation}
M(<\theta) = \pi (D_{\rm L} \theta)^2 \Sigma_{\rm cr}\,\,.
\label{eq:condition}
\end{equation} The mass inside the Einstein ring can be determined, once its
angular diameter and redshifts of the lens and the image are known. Even if
the alignment of the source, deflector and the observer is not perfect, and the
ring is not observed, this mass estimate may be very useful and rather
accurate. For example, the mass inside the inner $0''.9$ of the lensing galaxy
in the quadruple quasar QSO $2237+0305$ ("Einstein cross") has been determined
with an accuracy  of a few percent \cite{The-book}, with the largest
uncertainty being due to the estimate of the Hubble constant. This method
doesn't depend on the nature or state of matter, but it measures only the
projected mass and only in the inner part of the lensing galaxy.

Another method of determining the mass of the galaxy-deflector is from 
making use of the Eq.~\ref{eq:hubbledelay}. From the observed image
separation  $\D \t \simeq 2\te$ (Eq.~\ref{eq:deltatheta}) we find
\begin{eqnarray*}
2\te & \simeq & \theta_1 + \theta_2  \\
\te^2 &\simeq &\frac{1}{4} \theta_1 \theta_2
\end{eqnarray*}
So that,
\begin{equation}
\left(\frac{\D \t}{2}\right)^2 \simeq \frac{4 G M D_{\rm LS}}
{c^2 D_{\rm L} D_{\rm S}}  \sim \frac{M H_0 (z_{\rm s}-z_{\rm l})}
{c z_{\rm s} z_{\rm l}}
\end{equation}
Thus, the observed image separation $\D \t \propto M H_0$, if the
redshifts of the  lens and the source are known. Using the relation
between $\Delta t$ and  $H_0$ (\ref{eq:hubbledelay}) we obtain $M \propto
\Delta t$. Thus, from the direct measurement of the $\Delta t$ one can
determine the mass $M$ of the galaxy-lens located within an angular radius $\D
\t /2$.

Determination of the mass and mass distribution of the cluster of galaxies has 
become posible since the discovery of {\it arcs} and {\it arclets}. {\it Arcs}
are the result of very strong distortion of background galaxies (when a
part of an extended source covers different parts of the diamond-shaped
caustic, associated with cluster of galaxies as a deflector). Assuming the
cluster mass distribution to be axially-symmetric, we can have a rough mass
estimate from (\ref{eq:condition}), where $\theta$ now is the distance of the
arc from the cluster centre, and $\Sigma_{\rm cr}$ can be determined, if the
redshift of the arc can be determined. Also, since the arc roughly traces the
Einstein radius, we can use the Eq.~\ref{eq:theta_e}. This method loses
accuracy if the cluster is highly asymmetric or has significant substructure
(clumps of dark matter, for example). However, it is believed that the
assumption, describing clusters as isothermal spheres with finite cores, works
well \cite{fort-mellier}.

Discovery of {\it arclets} (less elongated images of background galaxies than
arcs) and weakly distorted images of background galaxies opened up the
possibility of studying the mass distribution in the outer parts of the
clusters. Shape of a galaxy image is affected by the tidal gravitational field
along its corresponding light bundle. This distortion is small and since
galaxies have intrinsically different shapes, the effect cannot be determined
in the individual galaxy image. However, with the sky densely covered by
randomly oriented faint galaxy images, a statistical study of the distortions
of these far-away sources is possible. From the coherent alignment of images of
an ensemble of galaxies one can draw the distortion pattern which traces the
gravitational field of the foreground cluster. By reconstruction techniques one
can measure the tidal field related to the gravitational potential of the
cluster and obtain the surface mass density.

\subsection{A candidate string-lensing field}

\begin{figure}
\centerline{
\vspace{-1.5 cm}
\hspace{-1.0 cm}
\makebox
{
\epsfig{figure=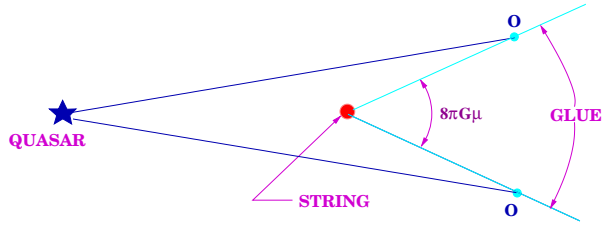,width=0.5\textwidth}}}
\vskip 0.6in
\caption[Lensing by a straight cosmic string]{The conical space around a
straight string can be obtained from Eucledean space by cutting out the wedge
of angular size $8\pi G\mu$ and identifying the exposed surfaces. Light rays
emitted by quasar intersect behind the string and observer $O$ sees a double
image of quasar $Q$. If $l$ and $d$ are the distances from the string to the
quasar and to the observer, respectively, then the angular separation between
the images is $\delta=8 \pi G \mu l(d+l)^{-1}$.}\label{fig:string}
\end{figure}

One interesting aspect of the lensing phenomena is lensing by a straight
cosmic string. Gravitational interaction of strings is characterized by the
parameter $G \mu$, where $\mu$ is mass per unit length of the string $\mu
\sim \eta^2$. Here $\eta$ is the energy scale of symmetry breaking $\sim
10^{16}$ GeV  for grand unification scale. Thus,  \begin{equation}
G \mu \sim (\eta/m_{\rm pl})^2\,\,,
\end{equation} $m_{\rm pl}$---Planck mass $\sim 10^{19}$ GeV and $G=m_{\rm
pl}^{-2}$. For grand unification strings $G \mu \sim 10^{-6}$ \cite{Vilenkin}.
The metric around the straight string is flat, so it cannot be detected by
gravitational interaction. However, the space around the string is actually a
conical space, that can be made out of a Euclidean space by cutting out a wedge
of angular size $\delta$ and by identifying the opposite sides of the wedge
(Fig.~\ref{fig:string}).

The deficit angle $\delta$ is $\delta = 8 \pi G \mu$. As seen from
the Figure~\ref{fig:string}, the conical nature of space around the string can
give rise to lensing phenomena. If there is an intervening string between us
and background galaxy, such a string should produce an identical twin pair of
images of the galaxy over a strip of space owing to an angle deficit
$\delta$ around the string. The discovery of the field with a peculiar group of
4 ``twin" galaxies was actually reported in a field near the quasar UM 679
\cite{Hu87}.

These 4 pairs are remarkably twinlike with characteristic separations $\sim
2.4''$. The separation of the twins (and the width of the strip over which
the splitting occurs) is determined by $\mu$ and a value $\sim 2.4''$ roughly
corresponds to a string mass of $10^{-6}$ in dimensionless units. Further
investigation of the candidate  string-lensing field \cite{Hu90}
revealed that in an area of $\pm 1'$ around the original twins there are 7
twinlike galaxies, which satisfy the magnitude and color difference criteria
for being lensing pairs. These twins vary wildly in magnitudes and colors but
the distribution of separations is strongly peaked at $2''-2.5''$.

One more interesting phenomena can occur if the background source lies
partially out of the wake since the galaxies are comparable in sizes with the
lensing strip. Whereas a fully lensed galaxy should have the same color and
spectrum in both images, in a partially lensed one the presence of strong color
gradients, such as in a UV excess nucleus, can produce color differentiation
between the images. Such an event was observed in the reported case. Here one
member is bluer in continuum light, while the pair is identical in images of
emission lines. Thefore, this object can be a partially lensed galaxy.

Strings can exist also in loops, though the exact metric of a long-lived loop
string is unknown today. Its lensing properties in the linearised gravitational
approximation were studied by Gott \cite{gott85} and Wu \cite{wu89}. It was
shown that the loop can produce three images if the source is inside the loop;
one is the original source seen through the loop, and two images on opposite
sides from the light rays which passed outside the loop.    

\subsection{Detection of gravitational waves by lensing}

Gravitational lenses can be used to detect gravitational waves, as a
gravitational wave affects the travel-time of a light ray. In a gravitational
lens, this effect produces time delays between the different images. Such
"detectors" are 22 orders of magnitude larger than any of the existing or
contemplated detectors, and they are sensitive to much lower frequences
\cite{allen90}. Time delay, produced by the wave, can be measured if
the source (e.g., quasar) has variable brightness, thus producing images whith
brightness variations correlated to a time-shift. For this purpose, the
most useful systems would be those which are highly symmetric, so that the time
delay due to the difference in the path length is small.

A gravitational wave affects the time delay because it perturbes the metric 
tensor, and therefore, modifies the path lenght of the two light rays. The
metric is given by  \begin{equation}
g_{ab} = \eta_{ab} + h_{ab}
\end{equation}
where $\eta_{ab}$ is the Minkowski metric, and $h_{ab}$ is a small perturbation.
One can calculate the time delay by examining the influence of the metric
perturbation on the equation of the null geodesic. The measured time delay in
the lens system can then be used to put an upper limit on the amplitudes of
stochastic background of the gravitational waves at low frequencies.
Sensitivity of such a "detector" is greatest at wavelengths comparable to
the overall size of the lens system. Allen \cite{allen90} made such
an estimation for the lens $0957+561$ for the frequency range $\omega  > 2\pi
\times 10^{-18}$ Hz. The amplitude of the gravitational waves must be less than
\begin{equation}
h < 2\times 10^{-5} \left[ \frac{\omega}
{2 \times 10^{-17}}\right]\frac{sec}{radian}
\end{equation}
or the expected time delay $\Delta t_{rms}$ would exceed the
measured value of 420 days.

Though it may be difficult to separate the "intrinsic", geometrical time delay,
from the delay caused by the gravitational wave, this idea may be still useful
if the gravitational wave amplitude $h$ is larger than $\eta$, the angle
between the images. The spatial motion of the geodesics that form the two
images, induced by the gravitational wave, becomes significant. One effect of
the gravitational perturbation is to change the angle between the images,
usually by increasing it. The relative intensities of the two images also
change.

\subsection{Determination of the lens parameters from gravitationally lensed 
gamma-ray bursts}

Recent results \cite{Meegan92} from the Burst and Transient Source Experiment
(BATSE) on the $Compton\ Gamma\ Ray\ Observatory$ show an isotropic
distrubution of gamma-ray bursts across the sky and rule out a population of
sources within the Galaxy. The most natural explanation is that the bursts
have a cosmological origin. If $\g$-ray bursts occur at high redshifts, then
some bursts are likely to undergo  gravitational lensing by foreground matter.
This would lead to the detection of multiple bursts with  identical profiles
but with different time delays and magnifications from a single event. Given a
set of a multiple bursts (two or four), produced by a gravitational lens, what
one one deduce about the nature of the lensing mass? Narayan and Wallington
\cite{Narayan92} showed that, if the lens is compact and pointlike, the
quantity $(1+z_{\rm l})M$ can be determined directly from the observations
without any information about the angular diameter distance to the lens or the
source, and without knowledge of the source redshift. Here $M$ is mass of the
lens and $z_{\rm l}$ is its redshift. What does a determination of  
$(1+z_{\rm l})M$ mean? First, it gives an upper bound on the lens mass 
\begin{equation} M_{\rm max} =  (1+z_{\rm l})M = \left(
\frac{c^3}{4G}\right)p_1 \,\,, 
\end{equation} 
where 
\begin{equation}
p_1 = \frac{(1+z_{\rm l})4GM}{c^3}\,\,.
\end{equation}
This bound is independent of the size or geometry of the universe and of the
redshift of the source or the lens. Second, if one can obtain an upper bound
on $z_{\rm l}$, then one will also have a lower bound on $M$. Using the
$V/V_{\rm max}$ data from BATSE, most cosmological models of gamma-ray
bursts currently estimate the redshifts of the faintest observed sources to
be $z_{\rm s} <1.5$. Accepting this estimate, we have $z_{\rm l} \leq 1.5$, 
and obtain the lower bound  
\begin{equation}
M_{\rm min} = M_{\rm max}/ (1+z_{\rm l, max}) = M_{\rm max}/2.5\,\,.
\end{equation}
The ability to bound the mass of the lens from both sides to within a factor 
of $\sim 2.5$ is an impressive accomplishment. For a point mass lens the
measurable parameter $p_1$ has a magnitude given by 
\begin{equation}
p_1 = 0.019(1+z_{\rm l}) \left(\frac{M}{M_{\odot}} \right) \mbox{msec}\,\,,
\end{equation}
which means that we can hope to detect point lenses with masses $> 10^2
M_{\odot}$. If a  significant fraction of the mass of the universe is in the
form of compact objects with masses of up to $\sim 10^6 M_{\odot}$ in the dark
halos of galaxies or in the intergalactic medium, then $\g$-ray bursts will
reveal their presence through lensing and will provide accurate mass
determinations. Moreover, once a sufficient number of lensed bursts has been
detected, statistical techniques may be used to determine the fractional mass
density $\Omega_{\rm lens}$ of the lenses \cite{Mao92}. 

If the lens is not pointlike but has an extended mass distribution, the
observations can be  used to obtain the velocity dispersion of the lens. If
galaxy lenses are not singular, but have finite cores, then the image
configuration will consist of three or five images. The extra burst will have
the longest time delay and will generally be significantly weaker than the
rest. The relative magnification of this burst will give useful information on
the core radius of the lens. 

\subsection{Discovering planetary systems through gravitational microlensing}

Traditional methods for the search of planetary systems involve either indirect 
observations (e.g. astrometry measurements) or direct infrared observations
(search for  dust lanes around the stars). A new method, based on the
gravitational lensing of the  bulge star by an intervening disk star was
suggested in 1992 \cite{Gould92}.  Planetary systems of galactic disk stars
can be detected through microlensing of stars in the Galactic bulge. Planets
in a solar-like system located half-way to the Galactic Centre should leave a
noticeable signature on the light curve of a gravitationally lensed bulge star.

The gravitational lensing of distant sources by intervening individual
compact masses,  typically in the mass range $10^6 \, M_{\odot}>M>10^{-4}\,
M_{\odot}$, is called "microlensing". This term originates from the fact that
the undetectable separation between images is of the order of microarcseconds
for a solar mass located at a  cosmological distance. For Galactic stars,
however, the separation is of the order of a milliarcsecond. This angular
separation is too small to observe. However, the resulting  magnification
can change the integrated light from the images for the time duration of the
microlensing event. The brightness of the lensed star increases, peaks, and
then decreases. The resulting light curve is smooth and completely described
by three parameters: the temporal width, the maximum magnification and the
time of  maximum magnification. The duration of such an event is from several
weeks to several months. It is symmetric in time about its maximum
magnification and is achromatic, which allows one to distinguish it from
variable stars. 

If there is a planet around the lensing star, the light curve may be
significantly altered. The planet of mass $m$ will typically affect the image
magnification only for a fraction of $(m/M)^{1/2}$ of the duration of the
entire event. With the typical stellar velocities ($\sim 200$ km/sec), this is
a day or so for a Jupiter-mass planet. The observed light curve will look
almost exactly like a light curve of an isolated star, except for a sharp
spike during the fraction of time when a source moves inside a planetary
Einstein radius. The planet affects appreciably the microlensed image only if
the planet and the unperturbed image are separated by a distance of the order
of the planet's own Einstein radius,  $(4 GMD_{\rm eff}/c^2)^{1/2}$, where
$D_{\rm eff}= D_{\rm L} D_{\rm S}/D_{\rm LS}$.

The probability of detecting such events in the total number
of microlensing events can be estimated. If a solar-like planetary system lay
at a random position along the line of sight to the Galactic bulge, and if the
bulge source came within one Einstein radius of the central star of this
system, then the system could be detected $\sim 20$\% of the time (assuming
minimum detectable perturbation $\delta_{min} = 5$\%). The largest contributor
will be a Jupiter-like planet, $\sim 17$\%, Saturn-like will give $\sim 3$\%,
and all the other planets $\ll 1$\%. 

From the light curve the ratio of planetary to stellar mass, $\epsilon=m/M$,
can be determined. If the lensing star is a G dwarf or earlier, its spectrum
can be taken. From the spectral type and the luminosity one may determine
the mass and distance and, thereby, infer the mass of the planet and its
projected distance from the star. The typical planetary signal lasts for a day
or less. Thus, to detect a Jupiter-mass planet, observations should be taken
every 4 hours, detection of a Neptune-mass planet will require hourly
observations.

\subsection{Light deflection in strong gravitational fields}

Up to now we were considering weak gravitational fields, where the essential
assumption for lensing formalism is that the bending angles of light rays are
very small. All tests of GTR within the Solar System, including the bending and
delay of light rays passing the Sun, have examined the gravitational
interaction only in connection with weakly self-gravitating objects (for
example, the Sun has a surface gravitational potential $G M/c^2 R \approx
2\times 10^{-6}$). The measured relativistic effects are but small
perturbations to Newtonian mechanics and these tests say nothing about the
strong field situation. There are, however, astrophysical systems where
gravitational fields are strong and light bending leads to new interesting
effects. Observing such systems and measurement of their parameters can yield
tests of GTR to a greater precision.

\subsubsection{GL Effects in Accreting Systems}

\SS{Relativistic "looks" of a neutron star}

General relativistic effects are quite substantial for neutron stars (NS)
of radii smaller than about $2r_{\rm S}$, where $r_{\rm S}=2 G M_{\rm
NS}/c^2$ is the Schwarzschild radius of NS. One expects these
effects to play a major role in the interpretation of the spectrum and light
curves of such stars. The characteristic quantity here is the ratio
$\rho=r_{*}/r_{\rm S}$, where $r_{*}$ is NS's radius. For stars with $\rho\leq
2$ radiation emitted deep inside the strong gravitational field of such a star
will be significantly modified as seen by the distant observer. Light ray,
emitted on or near the surface will be redshifted and may be deflected by more
than $45^0$ \cite{Nollert89}.

It was shown that rays reach the observer with larger impact parameters
than they would in flat space and, in addition, photons from the "back" half
of the star can reach the observer. As a result, the part of the star which
would be visible in flat space now appears larger and we may see parts of
the star which would otherwise be hidden from view. While in flat
spacetime exactly half of the surface of the star is visible, for $\rho>1.7$
the whole star becomes visible, the point at $180^0$ appears as the circular
boundary of the disk.

For the NS with emission from polar regions the visibility of the hot spots
is markedly different from the flat space case. Ref.~\cite{Nollert89} shows the
simulated picture of a pair of antipodal hot spots at 12 different phases of
rotation with the angle between rotation axis and a hot spot to be $45^\circ$.
 
\paragraph{Light curves from relativistic neutron stars}

Pechenick {\it et al} \cite{pechenik} have investigated the influence of GL
effects on the beaming of radiation from a hot spot on the surface of a slowly
rotating star in the limit where \Schw metric is applicable. It was
demonstrated that the deflection of light in the vicinity of a NS produces
large deviations of model light curves from those expected in the absence of
gravitational effects. For thermal emission, gravity tends to flatten the
light curve for the NS with $\rho\geq 2$, giving very little observed
variation for nominal pulsar values. For NS with $\rho\leq 2$ a new feature
appears in the form of a spike at $\theta=0^\circ$, whose width is about the
angular diameter of a polar cap and whose relative height increases with
decreasing of a polar cap radius. For the most relativistic case this feature
represents a jump of $1.2\,mag$ over the essentially flat continuum. The
sharp peak is actually due to the gravitational lensing effect of the star on
the cap at $\theta=180^\circ$ to the line of sight. It arises from photons
emitted near the tangential plane and bent through large angles
(up to $227^\circ$). The region near $\theta=180^\circ$ appears as a ring
whose apparent brightness exceeds that of a comparable region at
$\theta=0^\circ$. Thus, gravity alone is capable of producing strongly beamed
radiation even for isotropically radiating polar caps. 
 
These results are applicable to the structure of $X$-ray pulsars in binary
systems where accretion is assumed to give rise to hot polar regions. That
the $X$-ray light curve arises simply from rotational eclipsing of these caps
seems unlikely. One interesting possibility is that emission originates above
the NS surface. It was demonstrated that for the NS with $\rho =2.5$ any
emitting region higher than $0.19 R_{\rm NS}$, regardless of shape and size,
will always be in view. For a typical NS no point that is more than 2 or 3 km
above the surface is ever out of sight. For accreting pulsars it would be
relevant for the case when the infalling matter decelerates at some distance
above the surface due to shocks or radiation pressure. In this case, one has an
accretion "column" rather than a polar cap or hot spot. In the general
relativistic treatment of the emission from the column, there appear two new
qualitatively different effects \cite{riffert}. First, the frequency redshift
is different for radiation arising from different heights. Second, the star
and the accretion column will both produce some shadowing of the light rays.
This effect depends on the emission height and on the direction of
observation. In general, one sees that the column beam is strongly backwardly
bent for the most relativistic cases.

\paragraph{Emission from accretion disk}

The spectrum of X-rays produced by an accretion disk around a black hole 
is influenced markedly by GL effects \cite{Cunningham75}. Due to the forward
"peaking" of the emitted radiation by the rapidly moving gas in the inner disk
and to the gravitational focusing, the radiation is concentrated toward the
equatorial plane. The concentration toward the plane from the inner disk is
more severe for large values of $a/M$, where $a$ is angular momentum per
unit mass and $M$ is mass of the black hole. As the disk thickens in its outer
regions, a distant observer directly in equatorial plane is in a shadow and in
Newtonian case (disk accretion model neglecting the relativistic effects)
receives no radiation. In actuality, he sees some blueshifted radiation from
every radius outside the radius of marginal stability $r_{\rm ms}$ (radius at
which gas begins to plunge into the hole). Radiation seen by other observers
is less blueshifted, and that seen by the axial observer is always redshifted.
For a Schwarzschild black hole, the effects of redshift and focusing are minor,
since a disk around such black hole has a large inner radius ($r_{\rm
ms}=6M)$. Only the equatorial observer sees a spectrum different from
Newtonian spectrum (since this observer sees only focused radiation he always
sees a spectrum dominated by high-energy radiation from the inner disk). For
Kerr black hole with $a\leq 0.9 M$ redshift and focusing effects on the
observed spectrum are striking. Even though all observers receive practically
the same integrated flux, the average photon energy differs by approximately
an order of magnitude between the equatorial and axial observer. The axial
observer sees radiation from the cold, outer regions of the disk primarily;
radiation from the inner regions is redshifted and defocused. Consequently, he
sees a spectrum, attenuated at high energies in comparison to the Newtonian
spectrum. The equatorial observer sees radiation from the hot, inner region to
be blueshifted and strongly focused, consequently, spectrum is enhanced at
high energies, compared with the Newtonian. If emission is not isotropic, for
the axial observer $n_{\rm e}\approx0$ ($n_{\rm e}$---the angle of the emitted
radiation with the surface normal) for radiation from large radii. This angle
increases as $r_{\rm e}$ decreases. For other observers, the emission angle
decreases as $r_{\rm e}$ increases, reaching a minimum for $r_{\rm e}=3-4M$.
The radius for min $n_{\rm e}$ is smallest for the equatorial observer. Thus,
radiation reaching the observer from very small radii has large emission
angles. Near the horizon radiation must be emitted parallel to the disk surface
to escape being trapped by the disk or the hole.

\subsubsection{GL Effects in Compact Binaries}

Compact binaries present a unique laboratory for testing GR in the strong-field
limit. This has become possible due to increasing discoveries of binaries,
where one compact object is a neutron star (pulsar, beaming at us) and the 
other is a white dwarf (WD) (for example, 1855+09), a neutron star
(NS) (for example, 1913+16 system) or a black hole (BH) (for example, a binary
PSR B0042-73 in SMC is argued to have a massive $10--30\, M_\odot$ BH
companion \cite{lipunov95}).

To date, there exists one test of relativistic gravity \cite{Damour83}, the 
$\dot{\omega}-\gamma-\dot{P_{b}} $ test. It is obtained by combining the five
timing parameters of the binary:  eccentricity---$e$, orbital period---$P$ and
three "post-keplerian" (PK) ones: advance of periastron $\dot{\omega}$, time
dilation and gravitational shift parameter $\gamma$ and the orbital period
decay $\dot{P}$. All these are linked by the theory-dependent constraint,
which can be defined as $ \dot{P}/f^{\rm
theory}(e,\,P,\,\gamma,\,\dot{\omega})=1$. For PSR 1913+16 Damour
\cite{Damour83} finds \begin{equation} \frac{\dot{P^{obs}}}{f^{\rm
GR}(e^{obs},\,P^{obs},\,\gamma^{obs},\,\dot{\omega^{ o b s }})} =0.995\pm0.021
\end{equation} By contrast, in other theories of gravity the influence of
strong-field effects on the function
$f^{theory}(e,\,P,\,\gamma,\,\dot{\omega})$ is enormous and it changes
drastically both in sign and in magnitude \cite{Damour83}. Therefore, this
result constitutes good (better than $0.5\%$ accuracy \cite{Taylor92})
confirmation of GR and, above all, a very selective confirmation of the
ability of GR to describe the strong (and$/$or rapidly varying) gravitational
fields. The $\dot{\omega}-\gamma-\dot{P_{b}} $ test is a mixed test, which
combines strong-field effects (related to $\dot{\omega}$ and $\gamma$) with
radiative effects (related to $\dot{P}$). Thus, one cannot logically conclude,
when it is satisfied, that both the specific strong-field and radiative
predictions of GR have been independently confirmed \cite{Damour92}. In the
case of a binary's orbit inclination to the sky plane approaching $90^\circ$
the pulsar's two more PK parameters, characterizing the range $r\equiv
Gm_c/c^3$ and the shape $s\equiv \sin{i}$ of the Shapiro time delay, may be
measured. This has already allowed the determination of the neutron star
mass with a precision better than $18\%$ (B1855+09 \cite{Ryba91}).

\paragraph{Self-lensing by binaries}

Eclipsing binaries present a remarkable situation where, if the orbital
plane inclination to the plane of the sky is close to $90^\circ$, the
gravitational amplification of the companion by the compact object is possible.
The observable effect here is periodic brightness enhancement which would
depend on the geometry of the system. Schneider \cite{jscneider89} gives for
the binary pulsar PSR $1957+20$ the periodic brightness amplification between
$10^{-4}$ and $10^{-2}$, with a probability that amplification really exists
at about $10\%$,  if an adopted radius of optical companion is
$R_c=0.1R_\odot$. For an impact parameter $b=0$, the duration of amplification
is of $\,T_{\rm A}\approx 400$ sec. Though the variability in brightness of an
optical companion was observed in the case of PSR1957+20, Schneider warns that
it is still difficult to make the presently predicted amplification
observable. The star can have intrinsic variability due to the geometry of the
"evaporation" mechanism driven by the pulsar. Gould \cite{gould95} suggested
monitoring millisecond binaries to search for lensing events. He showed that
self-lensing in binary pulsars could be used to probe the structure of the
emitting region of pulsars, in particular, to check whether the emission
originates from the light cylinder or from a smaller locus. For pulsars of
period $P$, the light cylinder has a radius $Pc/2\pi\approx10^{8}\, [P/(20\,
ms)]\, cm$. If the lensing magnification curve follows the classic form for a
point source, this would show that the source is small on scales of the
Einstein ring, $R_{\rm E} \approx3 \times10^{8}\, (a/10^{11})^{1/2}$ cm.

\paragraph{Deflection of the light beam by companion}

If the inclination of the sky plane of the orbit of binary pulsar
$i$ is close to $90^{\circ}$, the pulsar beam during a fraction of the orbital
revolution will be gravitationally deviated by the companion and this effect
will manifest itself in the timing formulae. Bending of the pulsar beam by a
companion can lead to the deviation of the apparent position, to brightness
amplification of the source and to time delays. As in the Shapiro effect,
the deflection of pulsar's beam shows in the timing as a rapid, sharp growth
of the magnitude of the post-fit residuals of times-of-arrival (TOAs) on a
short time interval during the superior conjunction of the pulsar and its
companion. This effect is superimposed on the Shapiro effect \cite{Doroshenko}.

Pulsar's TOAs are determined by measuring the phase offset between
each observed profile and a long-term average one. In addition, a set of
post-fit residuals are determined---the differences between measured TOAs and
those calculated using the classical spin-down model of pulsar rotation.
When many TOAs are available, spaced over months or years, it generally
follows that, at least the pulsar's celestial coordinates, spin parameters and
Keplerian orbital elements will be measurable with high precision. The binary
systems most likely to yield measurable PK parameters are those with large
masses, high eccentricities and astrophysically "clean", so that orbits are
mostly dominated by gravitational interactions between two compact masses.
Apart from mass determination, measurement of the light deviation effect in
near edge-on compact binaries would help to get  further constraints on the
position of the pulsar's spin. In this  effect the shape of TOAs residuals
strongly depends on the spatial orientation of the pulsar's rotational axis
\cite{Doroshenko}.

Epstein \cite{epstein} was, perhaps, the first who paid attention to the 
phenomenon of pulsar beam deviation in the gravitational field of it's
companion in the case of $\cos i\leq 10^{-3}$. To be detected, the PSR beam
must, after each revolution, point in the direction defined by the null
geodesic connecting the orbital location of the pulsar to the observer. Due to
the light bending produced by the gravitational field, this direction depends
on the orbital phase. The result of this gravitational shift is a delay due
to the excess angle swept out by the pulsar in each spin period. The
influence of two relativistic effects (Shapiro and gravitational shift) should
be compared with the special relativistic effect---abberation, which also
causes shifts of the pulsar beam and delays. The discussion on the importance
of the different delays appears in \cite{jschneider90}. Author discussed the
importance of the different delays which must be included in the timing
formulae of binary system with orbit inclination close to  $90^{\circ}$ to the
line of sight. He studies the gravitational shift delay for compact binaries
at orbital phases close to superior conjunction. For PSR 0655+64, Schneider
derives delays of $1-10 \mu$s (due to the gravitational shift effect) during a
small fraction of the orbital period. Observation of this delay in a binary
system must be considered as evidence that pulsars are indeed rotating
beacons, instead of stars with periodic isotropic flares \cite{Goicoechea95}.
All derivations are done in standard Schwarzschild formalism, neglecting the
changes imposed by the mass of the pulsar in the system, which bears some
justification. When the pulsar mass is assumed to be negligible, there
exists only one geodesic, radial, joining the centres of the two stars. More
than one such path would imply that a photon emitted radially from the centre
of the companion star can  aquire a transverse motion, which is not possible
in the Schwarzschild solution. However, once the mass of the pulsar is
"switched on", there exists a family of geodesics between the two stars, all
of which have geometric path lengths greater or equal to the initial radial
one. The addition of the pulsar mass must increase all geodesic path lengths
and those passing near to both stars the most. Thus, it appears that the
corrections due to the pulsar mass can only increase the magnitude of the
predicted effects \cite{gorham}.

Doroshenko and Kopeikin \cite{Doroshenko} reduced the favourable inclination
angle to cos$i \leq 0.003$ and showed that this estimate becomes less
restrictive as the pulsar's spin axis approaches the line of sight. For the
first time they showed the time behaviour of the residuals of the TOA in the
vicinity to the moment of superior conjunction of the pulsar and its
companion. Their numerical estimates showed that beam deviation effect is too
small to be detected in the presently the best available example of PSR
B1855+09 with cos$i=0.04$.    

\chapter{Gravitational Lensing as a Tool in Search of
Natural Wormholes and Negative Matter in the Universe.}
\section{Introduction}
\label{sec:chapter3}
\vspace{-0.1in}
\paragraph{Energy conditions and their violations} 
The energy conditions (EC) of classical GR for a case of FRW spacetime are a
set of simple constraints on various linear combinations of the energy density
and pressure. They are the null (NEC), the weak (WEC), the strong (SEC), and
the dominant (DEC) energy conditions. For a FRW spacetime and a diagonal
stress-energy tensor $T_{\mu\nu}=(\rho,-p,-p,-p)$ with $\rho$ the energy
density and $p$ the pressure of the fluid, they read: 
\begin{eqnarray}
\hbox{NEC} & \iff &  \quad (\rho + p \geq 0 ), \nonumber \\ \hbox{WEC} & \iff
& \quad (\rho \geq 0 ) \hbox{ and } (\rho + p \geq 0),  \nonumber \\
\hbox{SEC} & \iff & \quad (\rho + 3 p \geq 0 ) \hbox{ and } (\rho + p \geq 0),
\nonumber \\ \hbox{DEC} & \iff & \quad (\rho \geq 0 ) \hbox{ and } (\rho \pm p
\geq 0). 
\end{eqnarray}
The EC are, then, just simple constraints on various linear combinations of
the energy density and the pressure of the matter generating the spacetime
curvature. Since normal matter has both positive energy density and positive
pressure, it automatically satisfies the null, weak and strong energy
conditions. If we impose the condition that speed of sound in normal matter
is less than the speed of light, we would have in addition $\,\del \r/\del p
<1$. Assuming there is no cosmological constant, integrating this gives
$p<\r$, so that dominant energy condition is satisfied as well. Roughly
speaking, the violation of the DEC is usually associated with either a large
and negative cosmological constant, or with superluminal acoustic modes. 
Quite recently, it was shown \cite{VISSER-HUBBLE} that the presently favoured
value of Hubble parameter [$H_0=(65,85)$] implies that SEC must be violated
somehere between the epoch of galaxy formation and the present. Since normal
matter satisfies SEC, this would require the introduction of ``abnormal
matter" (we would need large quantities of abnormal matter, sufficient to
overwhelm the gravitational effects of normal matter). Violating the SEC is
usually associated with a positive cosmological constant or cosmological
inflationary epoch. 

In the 60's and 70's these energy coonditions were widely used as key
foundations for a number of important theorems. For example, for ``the positive
mass theorem" (which states that objects made of the matter that satisfies the
DEC can never antigravitate), for a variety of theorems that predict the
creation of singularities in stellar collapse and in cosmological scenarios.
The discovery of Hawking that nonrotating black holes can evaporate and,
correspondingly, their surface area can shrink (in direct violation of the
area increase theorem---``second law of black hole mechanics", based on the
SEC) forced physicists to face the fact that though in classical physics the
EC are perfectly reasonable assumptions, quantum fields can violate them. Many
physical systems, both theoretical and experimental, are known to violate one
or more EC. Perhaps, the most quoted is the Casimir effect (for more
information on this and other examples, see \cite{visser}). Typically,
observed violations are of quantum nature and are of the order of $\hbar$. It
is currently far from clear whether there could be macroscopic violations of
EC. If they do exist, macroscopic negative masses could be part of the
ontology of the universe. 

\paragraph{Negative Masses}

The violations of the EC, in particular the weak one, would admit the
existence of negative mass. The possible existence of negative
gravitational masses has been investigated since the end of the
nineteenth century \cite{JAMMER}. From a Newtonian point of view, we can
differentiate four possible situations \cite{BONDI}: ({\it i}) both inertial
and gravitational masses are positive, ({\it ii}) inertial mass is positive
and gravitational mass is negative, ({\it iii}) inertial mass is negative and
the gravitational mass is positive, and ({\it iv}) both inertial and
gravitational masses are negative. Most of the Nineteenth Century literature
on negative masses is devoted to case (ii). From a relativistic point of view,
however, the situation is quite different: the Equivalence Principle
(EP)\footnote{Definitions: Strong equivalence principle---Spacetime is
everywhere a Lorentzian manifold. Freely falling particles follow geoesics of
the metric. Weak equivalence principle---All freely falling particles follow
the same trajectories independent of their internal composition.} requires
that gravitational and inertial masses cannot be considered distinct any
longer. Test particles move along geodesic lines in accordance with the
initial conditions, independently of the fact that their energy density is
positive or negative. Negative energy densities or negative masses exert a
repulsive force not only for ordinary matter, but for exotic matter as well.
This might conceivably pose a stability problem for large amounts of exotic
material, requiring large values of tension in order to keep an equilibrium
configuration \cite{motho}. In more complicated astrophysical systems
stability could be achieved through electromagnetic forces.

\paragraph{Wormholes}

\begin{figure}[t]
\begin{center}
\epsfxsize=5cm 
\epsffile{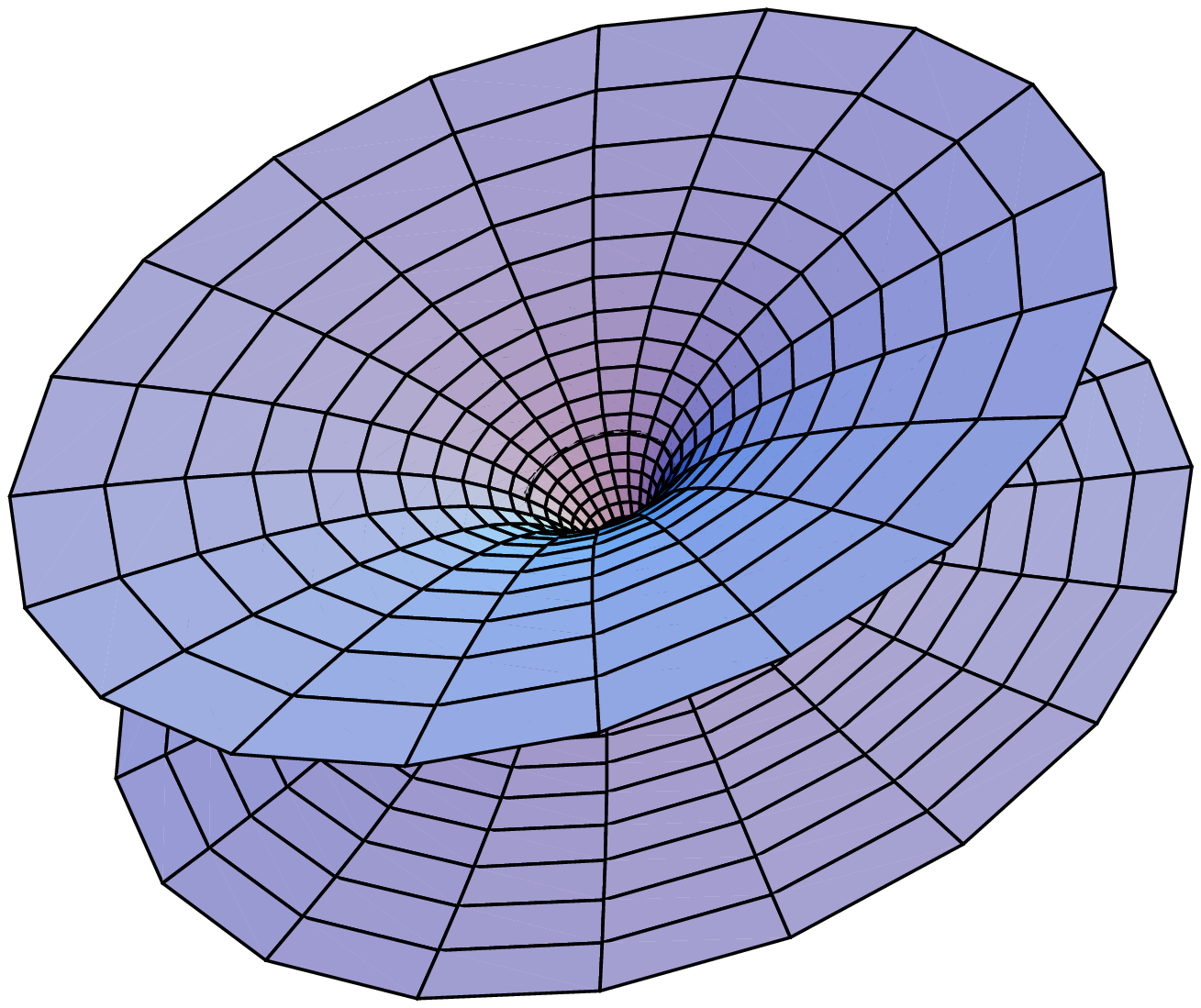} \hspace{.8cm}
\epsfxsize=5cm 
\epsffile{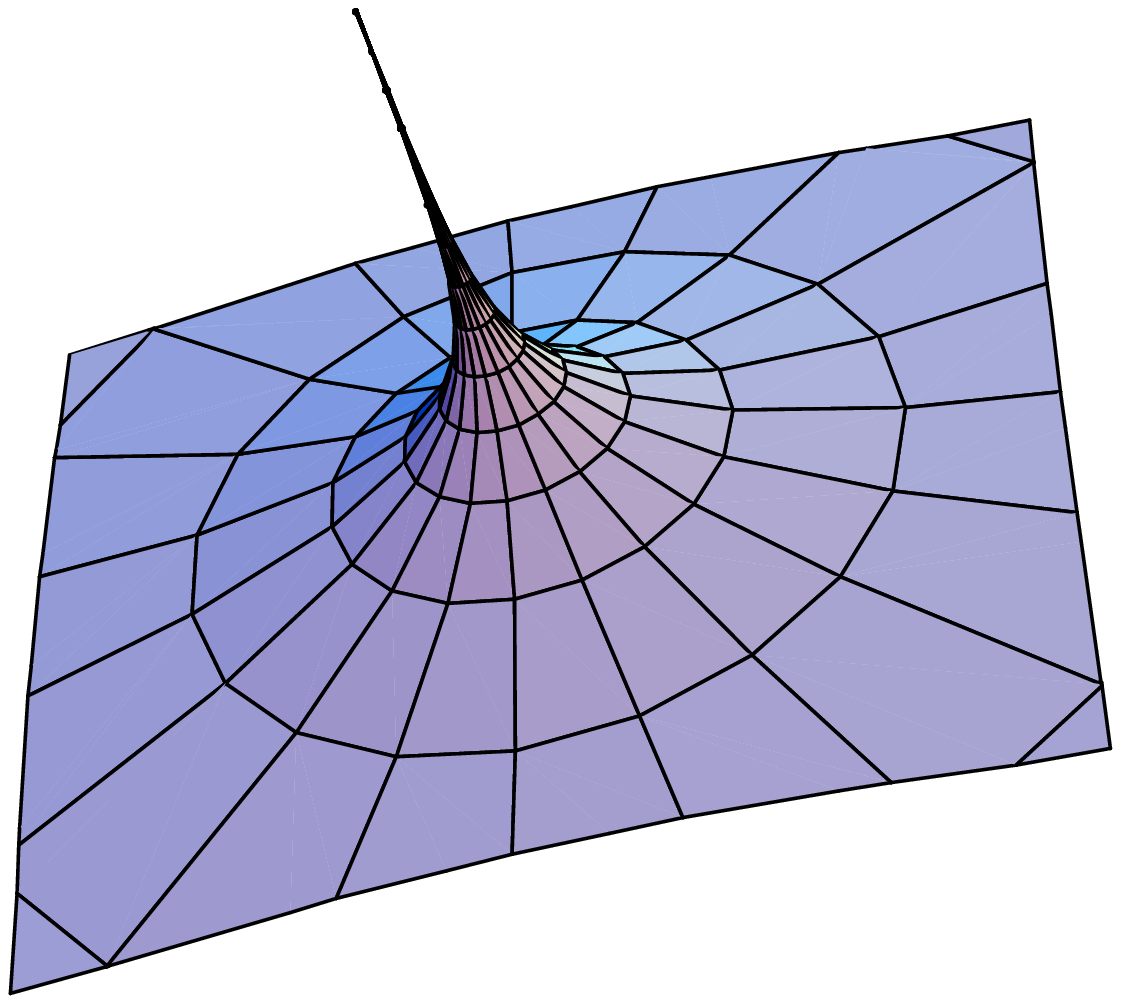}
\end{center} \vspace{.25cm} \caption[Embedding diagramme for a wormhole]{Left:
Embedding diagramme for a wormhole. Two mouths, joined by a tunnel, can connect
regions otherwise separated or disconnected (here the normal space should fold
as a sheet of paper, whereas the wormhole would be a tunnel from one side of
the sheet to the other). Right: Embedding diagram for a black hole. The
singularity here is represented as a pinch off of the wormhole tunnel.}
\label{fig:wh_bh}
\end{figure}

Of all the systems which would require violations of the EC in order to
exist, wormholes are the most intriguing. The salient feature of these
objects is that an embedding of one of their spacelike sections in Euclidean
space displays two asymptotically flat regions joined by a throat. Wormholes
basically represent {\it bridges} between otherwise separated regions of the
spacetime (Fig.~\ref{fig:wh_bh}) and need a special kind of matter in order to
exist. This matter, known as exotic, violates the EC.

Wormhole solutions to the Einstein field equations have been extensively
studied in the last decade (for the detailed description of the wormhole
solution see Appendix A). A simple search in PRD online resources yields 57
papers in the last three years including the word "wormhole" in its full
record. Msot of these are works deal with different trials to find
analytical solutions representing Lorentzian wormholes in space-time. There
are known wormhole solutions representing rotating structures, charged,
uncharged, static, evolving wormholes, wormholes within alternative gravity
theories, etc. But, clearly, this is only one aspect of the problem. Once one
has an analytical solution, it is not at all clear that the physical
constructs it represents can, or do, indeed, exist. This second aspect of
the problem is not commonly treated in the literature: very little work has
been done to develope astrophysical, or other kind of tests, to see
whether wormholes really are part of the ontology of the universe (see
\cite{visser,wh} and references cited therein.

\paragraph{Can a wormhole have total negative mass?}

Since wormholes have to violate the null energy condition in order to exist,
the hypothesis underlying the positive mass theorem no longer applies. There
is nothing in principle that can then prevent the occurrence of a negative
total mass. In other words, we need to have some negative mass near the throat
to keep the wormholes open \cite{h-v}. Examples of wormhole solutions, both
with positive and negative mass, followed the pioneering work by Morris and
Thorne \cite{motho} (see \cite{visser} for a review). Visser, for instance,
suggested a particular class of solutions, lacking spherical symmetry,
configurations in which there is a flat-space wormhole, framed by struts of an
exotic material \cite{cube}. These kinds of wormholes and many others,
including the simplest one known as absurdly benign wormhole \cite{motho},
would have a negative mass density. Typically near the throat the following
relationship would hold: \be {\rm mass\;at\;throat}=-\frac{{\rm radius} \cdot
c^2}{G}\,\,.  \end{equation} For a radius equal to 1 meter the mass of the
mouth will be equal to $-1$ Jupiter mass. The total mass seen at
infinity\footnote{In any \ass flat spacetime it is possible to define the
so-called ``imprint at infinity". Traversable wormholes, by definition, reside
in \ass flat spacetimes. To deduce wormholes conservation laws, the ADM mass is
invoked. It can be defined in terms of a suitable limit of surface integrals at
spatial infinity. The most important feature of ADM mass is that it is
conserved (see App.~A).} will depend on the details of the model, such as the
neighboring matter, and can be positive, negative or zero according to the
specific case. Whether or not mass separation is possible is not clear yet.

Some speculations have been made about how inflation can be
responsible to enlarge a microscopic wormhole---believed to exist
in the Planck foam---out to macroscopic dimensions \cite{ROMAN}. This
mechanism could result in a population of natural wormholes. 

\paragraph{Large-scale violations and observational strategy}

The result of the already mentioned work \cite{VISSER-HUBBLE} on the
violation of SEC in the recent epoch could imply the existence of a massive
scalar field or a positive cosmological constant. These are favoured by 
current observations. Even if the global energy density of the universe is
WEC-respecting, it is still not clear whether there exist spacetime domains
where large-scale violations of the EC occur, allowing the formation of
physical systems with an energy density equivalent to a total negative mass of
the size of a galaxy or even a cluster of galaxies.

Fifteen years after the seminal paper by Morris and Thorne \cite{motho}, we
face the following situation: there is no observational evidence supporting
the existence of natural wormholes or serious theoretical reasons for their
impossibility. Blackholes shared such a status during years until the
discovery of galactic x-ray sources and quasars in the 1960s. Clearly, we
have no better way than devising observational tests for deciding the
existence of negative masses. For instance, if natural wormholes with
negative masses or spacetime domains having large-scale violations of the EC
actually exist in the universe (e.g. if the original topology after the
Big-Bang was multiply connected), then there could be some observable
electromagnetic signatures that might lead to their identification. The
approach of the present work points out in this direction: we do not know
whether there exist an astrophysical population of objects with negative mass,
but {\it if} they indeed exit, then we provide specific observational
signatures that we can expect.

\paragraph{Previous Studies on Lensing by Negative Masses}

The idea that wormholes can act as gravitational lenses and induce
a microlensing signature on a background source was first suggested by Kim
and Sung \cite{swkim}. Unfortunately, their geometry was that of a perfect
alignment of a source, both wormhole's mouths and an observer, which is quite
unlikely. They also considered both mouths to be of positive mass. Cramer et
al. \cite{cramer} carried out more detailed analysis of negative mass
wormholes and considered the effect they can produce on background point
sources at non-cosmological distances. The generalization to a cosmological
scenario was carried out by Torres et al. \cite{diego-grbwh}, although lensing
of point sources was still used. As far as we are aware, the first and only
bound on the possible existence of negative masses, imposed using
astrophysical databases, was given by Torres et al. \cite{diego-grbwh}.
Recently, Anchordoqui et al. \cite{doqui} searched in existent gamma-ray
bursts databases for signatures of wormhole microlensing. Although they
detected some interesting candidates, no conclusive results could be obtained.
Peculiarly asymmetric gamma-ray bursts \cite{rom}, although highly uncommon,
might be probably explained by more conventional hypothesis, like precessing
fireballs (see, for instance, Ref. \cite{Zwart}). 
 
In the following sections we describe microlensing by natural wormholes of
stellar and sub-stellar masses. We provide an in-depth study of the
theoretical peculiarities that arise in negative mass microlensing, both for a
point mass lens and source, and for extended source situations. We present
negative mass microlensing simulations, showing the resulting shapes of the
images, the intensity profiles, the time gain function, the radial and
tangential magnifications, and other features. Our work extends and deepens
previous papers in several ways, and gives a method of analyzing
observational predictions quantitatively.   

In the last section of this chapter we present the results of a set of
simulations showing macrolensing effects we could observe if such a large
amount of negative energy density existed in our universe.

\section{Negative mass lensing formalism and basic equations for a point lens}
In this chapter we consider lensing only by a point negative mass
lens, and thus we can use all the assumptions concurrent with the
treatment of the Schwarzschild lens (see Section~\ref{sec:point-mass}).
The image coordinates can be written as $(\theta_1,\theta_2$) and
those of the source as $(\beta_1,\beta_2)$.

\subsection{Effective refractive index of the gravitational field
of a negative mass and the deflection angle}

The `Newtonian' potential of a negative point mass lens is given by \be
\Phi(\xi,z) = \frac{G |M|}{(b^2 +z^2)^{1/2}}\,\,, \end{equation} where $b$ is
the impact parameter of the unperturbed light ray and $z$ is the distance
along the unperturbed light ray from the point of closest approach. We have
used the term Newtonian in quotation marks since it is, in principle,
different from the usual one. Here the potential is positively defined and
approaches zero at infinity \cite{visser}. In view of the assumptions stated
above, we can describe light propagation close to the lens in a locally
Minkowskian spacetime perturbed by the {\em positive} gravitational potential
of the lens to first post-Newtonian order. In this weak field limit, we
describe the metric of the negative mass body in orthonormal coordinates
$x^0=ct$, ${\bf x}=(x^i)$ by \be ds^2 \approx \left(1 +  \frac{2
\Phi}{c^2}\right)c^2dt^2 - \left(1 - \frac{2\Phi}{c^2}\right) dl^2,
\end{equation} where $dl=|{\bf x}|$ denotes the Euclidean arc length. The
effect of the space-time curvature on the propagation of light can be
expressed in terms of an effective index of refraction $n_{\rm eff}$ given by
\be n_{\rm eff}=1 - \frac{2}{c^2} \Phi\,. \end{equation} Thus, the effective
speed of light in the field of a negative mass is \be v_{\rm eff}
= c/n_{\rm eff} \approx c + \frac{2}{c} \Phi\,. \end{equation}
Because of the increase in the effective speed of light in the gravitational
field of a negative mass, light rays would arrive faster than those following
a similar path in vacuum. This leads to a very interesting effect when
compared with the propagation of a light signal in the gravitational field of
a positive mass. In that case, light rays are delayed relative to propagation
in vacuum---the well known {\it Shapiro time delay}. In the case of a
negative mass lensing, this effect is replaced by a new one, which
we shall call {\it time gain}. We will describe this effect in more detail in
the following subsections. 
Defining the deflection angle as the differenceof the initial and final ray
direction \be \bbox{\alpha} : = \hat{\bf e}_{\rm in} -\hat{\bf e}_{\rm out},
\end{equation} where $\hat{\bf e}:=d {\bf x}/dl$ is the unit tangent vector of
a ray ${\bf x}(l)$, we obtain the deflection angle as the integral along the
light path of the gradient of the gravitational potential 
\be 
\bbox{\alpha} = \frac{2}{c^2} \int \bbox{\nabla}_{\bot}\Phi dl\,\,,
\label{eq:nablaphi}
\end{equation} 
where $\bbox{\nabla}_{\bot}\Phi$ denotes the projection of
$\bbox{\nabla}\Phi$ onto the plane orthogonal to the direction
$\hat{\bf e}$ of the ray. We find
\be
\bbox{\nabla}_{\bot}\Phi(b,z) = -\frac{G |M|{\bf b}}{(b^2
+z^2)^{3/2}}\,.
\end{equation}
Eq.~\ref{eq:nablaphi} then yields the deflection angle
\be
\bbox{\alpha} = -\frac{4 G |M| {\bf b}}{c^2 b^2}\,\,.
\label{eq:deflection-angle}
\end{equation}
It is interesting to point out that in the case of the negative
mass lensing, the term `deflection' has its rightful meaning---the
light is deflected away from the mass, unlike in the positive mass
lensing, where it is bent towards the mass.

\subsection{Lensing geometry and lens equation}

\begin{figure}[ht!]
\centerline{\epsfig{figure=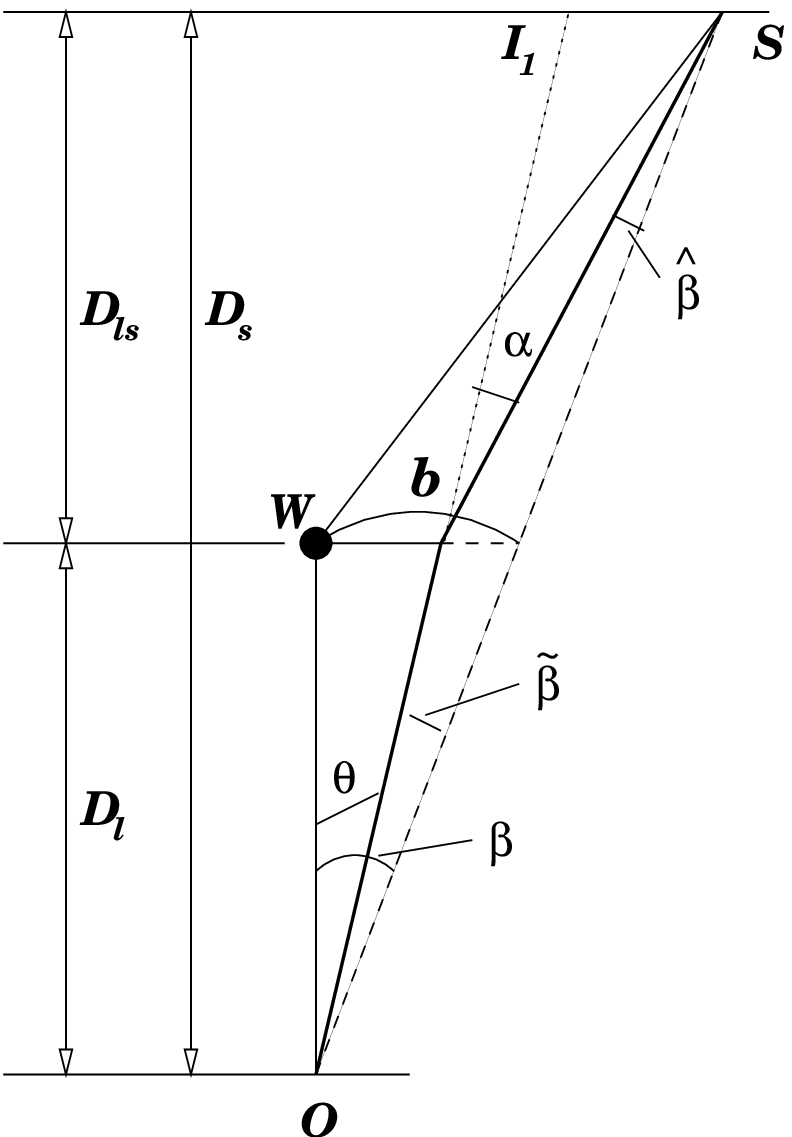,width=0.4\textwidth}}
\vspace*{.25cm} \caption[Lensing geometry of a negative mass]{Lensing
geometry of a negative mass. $O$ is the observer, $S$ is the source, $W$ is
the negative mass lens, $I_1$ is one of the images. $\beta$ is the angle
between the source and the lens---position of the source, $ \theta$ is the
angle between the source and the image---position of the image,
and $\alpha$ is the deflection angle. $b$ is the impact parameter
and $D_l$, $D_s$ and $D_{ls}$ are angular diameter distances.
Other quantities are auxiliary.} \label{fig:nlens}
\end{figure}

In Fig.~\ref{fig:nlens} we present the lensing geometry for a point-like
negative mass. From this figure and the definition of the deflection angle
(Eq.~\ref{eq:deflection-angle}), we can obtain the relation between the
positions of the source and the image. We only need to relate the radial
distance of the source and the image from the center, since due to circular
symmetry, the azimuthal angle $\phi$ is not affected by lensing. This gives 
\be 
(\bbox{\beta} - \bbox{\theta})D_{\rm s} =-\bbox{\alpha} D_{\rm ls} 
\end{equation} 
or 
\be \bbox{\beta} =
\bbox{\theta} - \frac{D_{\rm {ls}}}{D_{\rm s}} \bbox{\alpha}\,\,.
\label{eq:lenseq}
\end{equation}
With the deflection angle (Eq.~\ref{eq:deflection-angle}), we can write the
lens equation as \be \beta=\theta +\frac{4G|M|}{c^2
\xi}\frac{D_{ls}}{D_s}=\theta + \frac{4G|M|}{c^2}\frac{D_{ls}}{D_s
D_l} \frac{1}{\theta}\,\,. \end{equation}

\subsection{Einstein radius and the formation of images}

A natural angular scale in this problem is given by the quantity
\be \theta^2_{\rm E}=\frac{4 G|M|}{c^2} \frac{D_{\rm ls}}{D_{\rm
s} D_{\rm l}}\,\,, \end{equation} which is called the Einstein angle. In the
case of a positive point mass lens, this corresponds to the angle at which
the Einstein ring is formed when the source, lens and observer are perfectly
aligned. As we will see later in this section, this does not happen if the
mass of the lens is negative. There are other differences as well. A typical
angular separation of images is of order $2\,\theta_{\rm E}$ for a positive
mass lens. Sources which are closer than about $\theta_{\rm E}$ to the
optical axis are significantly magnified, whereas sources which are located
well outside the Einstein ring are magnified very little. All this is
different with a negative mass lens, but nonetheless, the Einstein angle
remains a useful scale for the description of the various regimes in the
present case and, therefore, we shall use the same nomenclature for its
definition.

The Einstein angle corresponds to the Einstein radius in the linear scale (in
the lens plane): 
\be 
R_{\rm E} = \theta_{\rm E} D_{\rm l} = \sqrt{\frac{4
GM}{c^2} \frac{D_{\rm ls}D_{\rm l}}{D_{\rm s}}}\,\,. 
\end{equation} 
In terms of Einstein angle the lens equation takes the form 
\be 
\beta=\theta+ \frac{\theta_{\rm E}^2}{\theta}\,\,, 
\label{eq:negleq}
\end{equation} 
which can
be solved to obtain two solutions for the image position $\theta$: \be
\theta_{1,2} = \frac{1}{2} \left( \beta\pm \sqrt{\beta^2 - 4 \theta_{\rm
E}^2}\right). \end{equation} Unlike in the lensing due to positive masses, we
find that there are three distinct regimes here and, thus, can classify the
lensing phenomenon as follows:

\begin{enumerate}

\item[I:] {$\beta < 2 \theta_{\rm E}$}~~There is no real solution for the
lens equation. It means that there are no images when the source
is inside twice the Einstein angle.

\item[II:] { $\beta > 2 \theta_{\rm E}$}~~There are two solutions,
corresponding to two images both on the same side of the lens and
between the source and the lens. One is always inside the Einstein
angle, the other is always outside it.

\item[III:] { $\beta=2\,\theta_{\rm E}$}~~This is a degenerate case,
$\theta_{1,2}=\theta_{\rm E}$; two images merge at the Einstein
angular radius, forming the {\it radial} arc (see Section~\ref{sec:mag}).
\end{enumerate}
\begin{figure}[ht!]
\centerline{\epsfig{figure=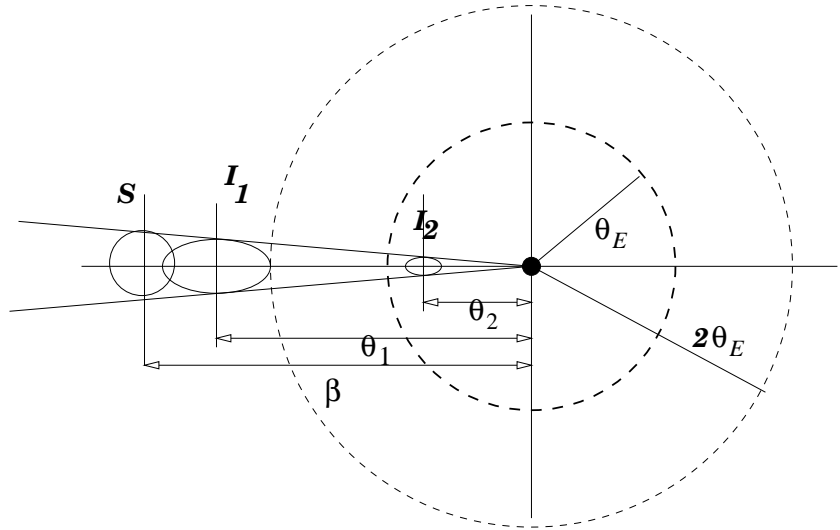,width=0.7\textwidth}}
\vspace*{.25cm} \caption[Formation of images by a point negative mass
lens]{Formation of images by a point negative mass lens. {\bf S} is the
source, $\bf I_1$ is the outer image and $\bf I_2$ is the inner image. Lens
is in the centre, $\te$ is the angular scale of the lensing. This Figure
shows schamaticall the radial distortion of images in the case of a negative
mass lens.}  \label{fig:shape}   \end{figure} 
We also obtain two important scales, one is the Einstein angle ($\theta_{\rm
E}$)---the angular radius of the radial critical curve, the other is twice the
Einstein angle ($2\,\theta_{\rm E}$)---the angular radius of the caustic.
Thus, we have two images, one is always inside the $\theta_{\rm E}$, one is
always outside; and as a source approaches the caustic ($2\,\theta_{\rm
E}$) from the positive side, the two images come closer and closer
together, and nearer to critical curve, thereby brightening. When the source
crosses the caustic, the two images merge on the critical curve ($\theta_{\rm
E}$) and disappear. In Fig.~\ref{fig:shape} we show schematically the
formation and positions of images for the negative mass lensing. 

\subsection{Time gain and time-offset function}
\begin{figure}
\centerline{
%
\includegraphics[width=0.7\textwidth]{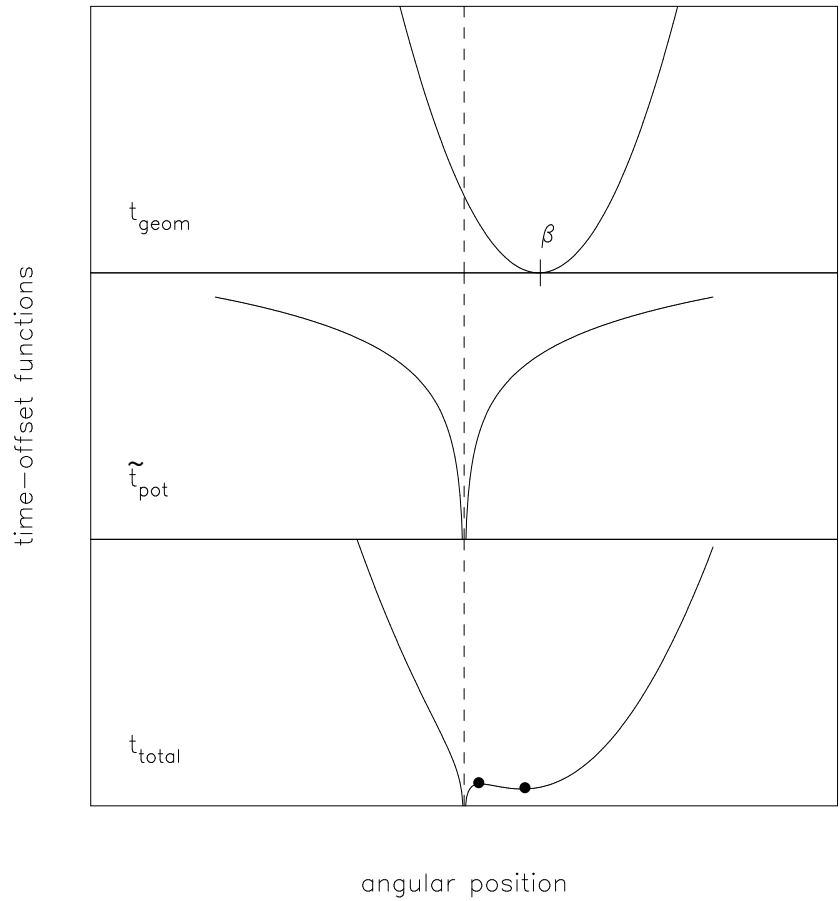}}
\vskip 0.6in
\caption[Geometric time delay, gravitational time gain and total time offset
produced by a point negative mass lens]{Geometric time delay, gravitational
time gain and total time offset produced by a point negative mass lens for a
source that is slightly off the optical axis.} \label{fig:time} 
\end{figure}

Following \cite{Narayan}, we define a scalar potential,
$\psi(\bbox{\theta})$, which is the appropriately scaled projected Newtonian
potential of the lens, \be \psi(\bbox{\theta}) = \frac{D_{\rm ls}}{D_{\rm l}
D_{\rm s}}\frac {2}{c^2} \int{\Phi(D_{\rm l}\bbox{\theta},z) dz}\,\,.
\end{equation} For a negative point mass lens it is \be \psi(\bbox{\theta}) =
\frac{D_{\rm ls}}{D_{\rm l} D_{\rm s}} \frac{4 G |M|}{c^2} \ln{|\theta|}\,\,.
\end{equation} The derivative of $\psi$ with respect to $\bbox{\theta}$ is
the deflection angle \be \bbox{\nabla}_{\theta}\psi = D_l
\bbox{\nabla}_b \psi = \frac{2}{c^2}\frac{D_{ls}}{D_s}
\int{\bbox{\nabla}_{\bot}\Phi dz} = \bbox{\alpha}\,\,. \end{equation}
Thus, the deflection angle is the gradient of $\psi$---the deflection
potential, 
\be
\bbox{\alpha}(\bbox{\theta})= \bbox{\nabla}_\theta
\psi\;.
\end{equation}
From this fact and from the lens equation (\ref{eq:lenseq}) we obtain
\be (\bbox{\theta} -\bbox{\beta}) + \bbox{\nabla}_{\theta}
\psi(\bbox{\theta}) = 0 \,\,. \end{equation} This equation can be written as
a gradient, \be
\bbox{\nabla}_{\theta}\left[\frac{1}{2}(\bbox{\theta} -
\bbox{\beta})^2 + \psi(\bbox{\theta}) \right] = 0\,\,. \end{equation} If we
compare this equation with that for the Fermat's principle
\cite{Narayan} \be \bbox{\nabla}_{\theta}\,t(\bbox{\theta}) =
0\,\,, \end{equation} we see that we can define the time-offset function
(opposite to time-delay function in the case of positive mass
lens) as \be t(\bbox{\theta}) = \frac{(1+z_{\rm l})}{c}
\frac{D_{\rm l} D_{\rm s}}{D_{\rm ls}} \left[\frac{1}{2}
(\bbox{\theta}-\bbox{\beta})^2 + \psi(\bbox{\theta}) \right] =
t_{\rm geom} + \tilde{t}_{\rm pot}\,\,. \end{equation} Here $t_{\rm geom}$ is
the geometrical time delay due to the extra path length of the
deflected light ray relative to the unperturbed one. It remains
the same as in the positive case---increase of light-travel-time
relative to an unbent ray. The coefficient in front of the square
brackets ensures that the quantity corresponds to the time offset
as measured by the observer. The second term $\tilde{t}_{\rm pot}$
is the time gain a ray experiences as it traverses the deflection
potential $\psi(\bbox{\theta})$, with an extra factor $(1+z_l)$
for the cosmological `redshifting'. Thus, cosmological geometrical
time delay is \be t_{\rm geom} = \frac{(1+z_{\rm l})}{c}
\frac{D_{\rm l} D_{\rm s}}{D_{\rm ls}} \frac{1}{2}(\bbox{\theta} -
\bbox{\beta})^2\,\,, \end{equation} and cosmological potential time gain is
\be \tilde{t}_{\rm pot}=\frac{(1+z_{\rm l})}{c} \frac{D_{\rm l}
D_{\rm s}}{D_{\rm ls}} \psi(\bbox{\theta})\,\,. \end{equation}
In Fig.~\ref{fig:time} we show the time delay and time gain functions. The
top panel shows $\rm t_{\rm geom}$ for a slightly offset source. The curve is
a parabola centered on the position of the source, $\beta$. The central panel
displays $\rm \tilde t_{\rm pot}$ for a point negative mass lens. This curve
is centered on the lens. The bottom panel shows the total time-offset. Images
are located at the stationary points of $\rm t_{\rm total}$. Here we
see two extrema---maximum and minimum---on the same side (right)
from the optical axis (marked by dots).
We can find the time difference between the two images, $\theta_1$
and $\theta_2$. If a source has intrinsic variability, it will appear in the
two images at an interval  \be 
\D t_{12} =
\frac{r_{\rm s}}{c} (1+z_{\rm l}) \left( \nu^{1/2} - \nu^{-1/2} -
\ln{\nu} \right)\,\,, 
\end{equation} 
where by $\nu$ we denoted the ratio of absolute values of magnifications of
images, 
\be 
\frac{\mu_{1}}{\mu_{2}} = \left[\frac{\sqrt{u^2-4}
+u}{\sqrt{u^2-4} - u} \right]^2\,\,, \label{eq:ratio} 
\end{equation} 
and $u$ is the scaled angular position of the source $u=\beta/\theta_{\rm
E}$ and $r_{\rm s}$ is the Schwarzschild radius of the lens.

\subsection{Magnifications}\label{sec:mag}

\begin{figure}
\centerline{
\includegraphics[width=1.\textwidth]{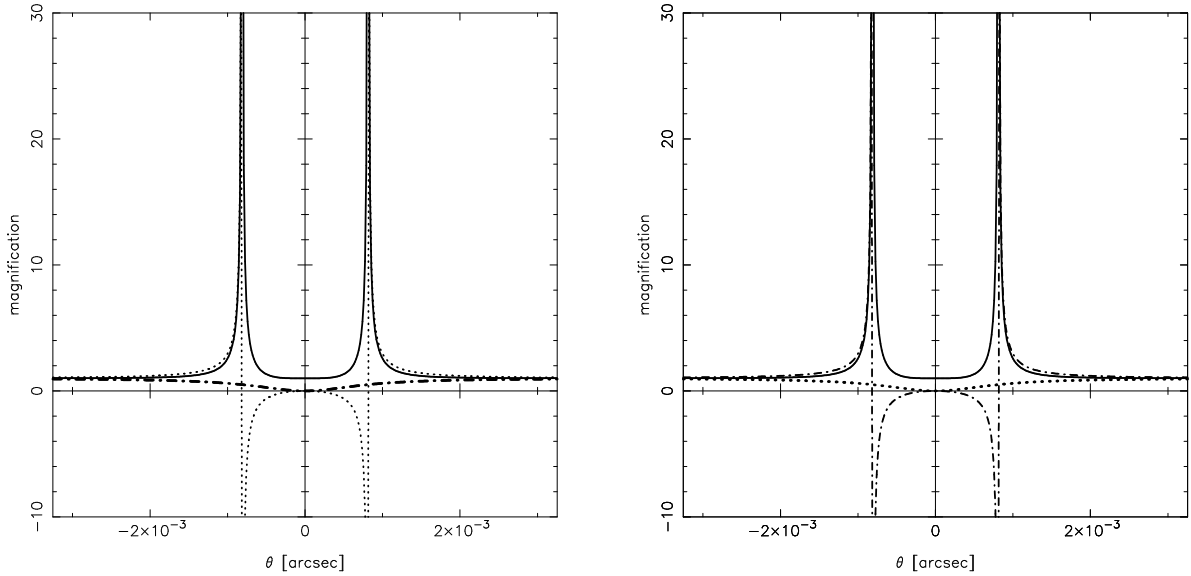}}
\vspace{.25cm}\caption[The magnifications for the positive point mass lens and
negative point mass lens]{The magnifications: tangential $\mu_{\rm tan}$
(dotted lines), radial $\mu_{\rm rad}$ (dash-dotted lines), and total $\mu$
(continuous curves), are plotted as functions of the image position $\theta$
for two cases; in the left panel for the positive mass, in the right panel for
the negative mass. The singularities of $\mu_{\rm tan}$ and
$\mu_{\rm rad}$ give the positions of the tangential and radial
critical curves, respectively. In the left panel the singularity
is in the tangential critical curve.  In the right panel, instead,
in the radial critical curve. Here $|M|=1\, M_{\odot}$, $D_{\rm
s}=0.05$ Mpc and $D_{\rm l}=0.01$ Mpc. Angles are in arcseconds.}
\label{fig:magnifications}
\end{figure}

For a point mass lens the magnification is given by (see
Section~\ref{sec:magnifications}) \be \mu^{-1} =
\left|\frac{\beta}{\theta}\frac{d\beta}{d\theta}\right|\,\,. \end{equation}
From the lens equation (\ref{eq:negleq}), we find \be \frac{\beta}{\theta} =
\frac{\theta^2 + \theta_{\rm E}^2}{\theta^2}\,\,\,\,\,;\,\,\,
\frac{d\beta}{d\theta} = \frac{\theta^2 - \theta_{\rm
E}^2}{\theta^2}\,\,. \end{equation} Thus, \be \mu^{-1}_{1,2} = \left| 1 -
\frac{\theta_{\rm E}^4}{\theta_{1,2}^4}\right|\,\,. \end{equation} Using
definition for $u$ in (\ref{eq:ratio}), we find the total magnification
(Fig.~\ref{fig:magnifications}, right panel, continuous curve) as \be
\mu_{\rm tot} = |\mu_1| +|\mu_2| = \frac{u^2-2}{u \sqrt{u^2-4}}\,\,.
\end{equation}

The tangential and radial critical curves follow from the singularities in
tangential \be \mu_{\rm tan}=\left|\frac{\beta}{\theta}\right|^{-1}
=\frac{\theta^2}{\theta^2+\theta_{\rm E}^2} \end{equation} 
and radial magnifications \be \mu_{\rm
rad} = \left|\frac{d\beta}{d\theta}\right|^{-1} =
\frac{\theta^2}{\theta^2-\theta_{\rm E}^2}\,\,. \end{equation} $\mu_{\rm
rad}$ diverges when $\theta=\theta_{\rm E}$---angular radius of
the radial critical curve. $\mu_{\rm tan}$ always remains finite,
which means that there are no tangential critical curves---no
tangential arcs can be formed by the negative point mass lens. In
Fig.~\ref{fig:magnifications} we show the magnification curves (radial,
tangential and total) for both positive (left panel) and negative mass lenses
(right panel). The difference can be seen as follows---in the left panel
there is no singularity in the radial curve (no radial arcs are formed by the
positive mass lens), whereas in the right panel we see that the curve for the
radial magnification experiences a singularity.

\section{Microlensing}

\subsection{Light curves of the point source}\label{sec:microlensing} 
\begin{figure}
\centerline{
\includegraphics[width=0.5\textwidth]{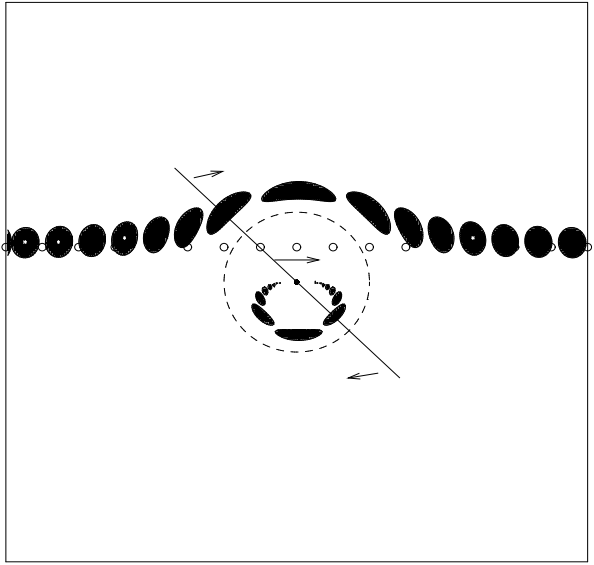}}
\vspace{.25cm} \caption[Schematic representation of the geometry
of the positive mass lensing]{Schematic representation of the geometry
of the positive mass lensing due to the motion of the source, lens and the
observer (in this case we can consider only the motion of the source in the
plane perpendicular to the optical axis). The lens is indicated with a dot at
the center of the Einstein ring, which is marked with a dashed line. The
positions of the source center are shown with a series of small open circles.
The locations and the shapes of the two images are shown with a series
of dark ellipses. At any instant, the two images, the source and the lens are
all on a single line, as shown in the figure for one particular instant.
\label{fig:positive}} \end{figure}

When the angular separation between the images
\be 
\D\theta= \sqrt{\beta^2 - 4\theta_{\rm E}^2} \,\, 
\end{equation} 
is of the order of milliarcseconds, we cannot resolve the two images with
existing telescopes and we can only observe the lensing effect through their
combined light intensity. This effect is called {\it microlensing}. Both the
lens and the source are moving with respect to each other (as well as the
observer). Thus, images change their position and brightness. Of particular
interest are sudden changes in luminosity, which occur when a compact source
crosses a critical curve. For a positive mass the situation is quite simple
(Fig.~\ref{fig:positive}). Positive mass microlensing is extensively reviewed
in \cite{sazhin-review}).

\begin{figure}
\centerline{
\includegraphics[width=0.7\textwidth]{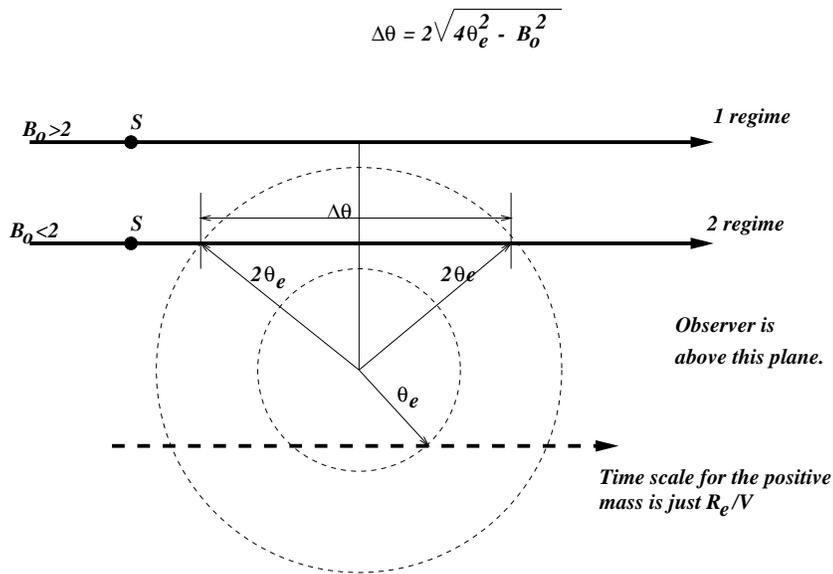}}
\caption[Schematic representation of the geometry of the
microlensing]{Schematic representation of the geometry of the microlensing.
All teh quantities are defined in the text.} \label{fig:definitions}
\end{figure}

For a negative mass lens the situation is different. We define a
dimensionless minimum impact parameter $B_0$ in terms f the Einstein radius
as the shortest distance between the path line of the source and the lens (all
necessary definitions are illustrated in the Fig~\ref{fig:definitions}).

\begin{figure}
\centerline{
\includegraphics[width=0.5\textwidth]{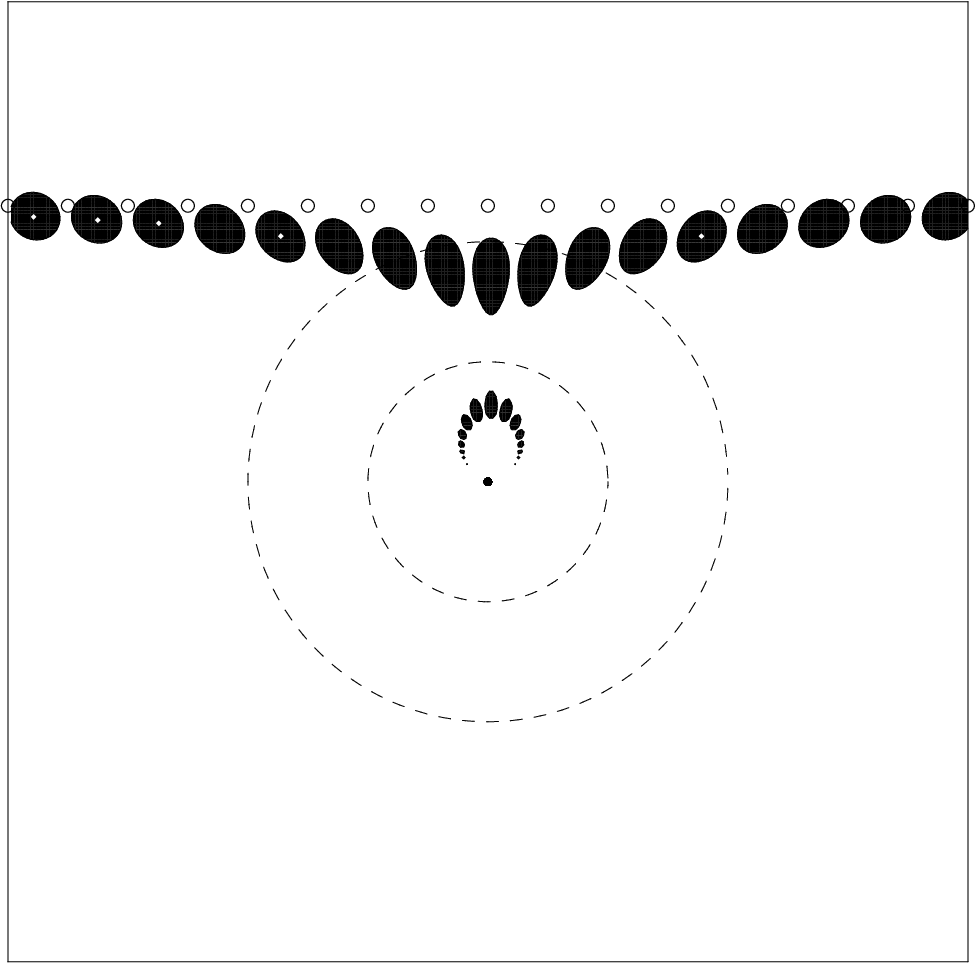}}
\vspace{.25cm} \caption[True motion of the source and apparent
motion of the images for minimum impact parameter $B_0 > 2$]{True motion of
the source and apparent motion of the images for $B_0 > 2$. The inner dashed
circle is the Einstein ring, the outer dashed circle is twice the Einstein
ring. The rest is as in Fig.~\ref{fig:positive}.}
\label{fig:bmin_gt2}
\end{figure}

\begin{figure}
\centerline{
\includegraphics[width=0.5\textwidth]{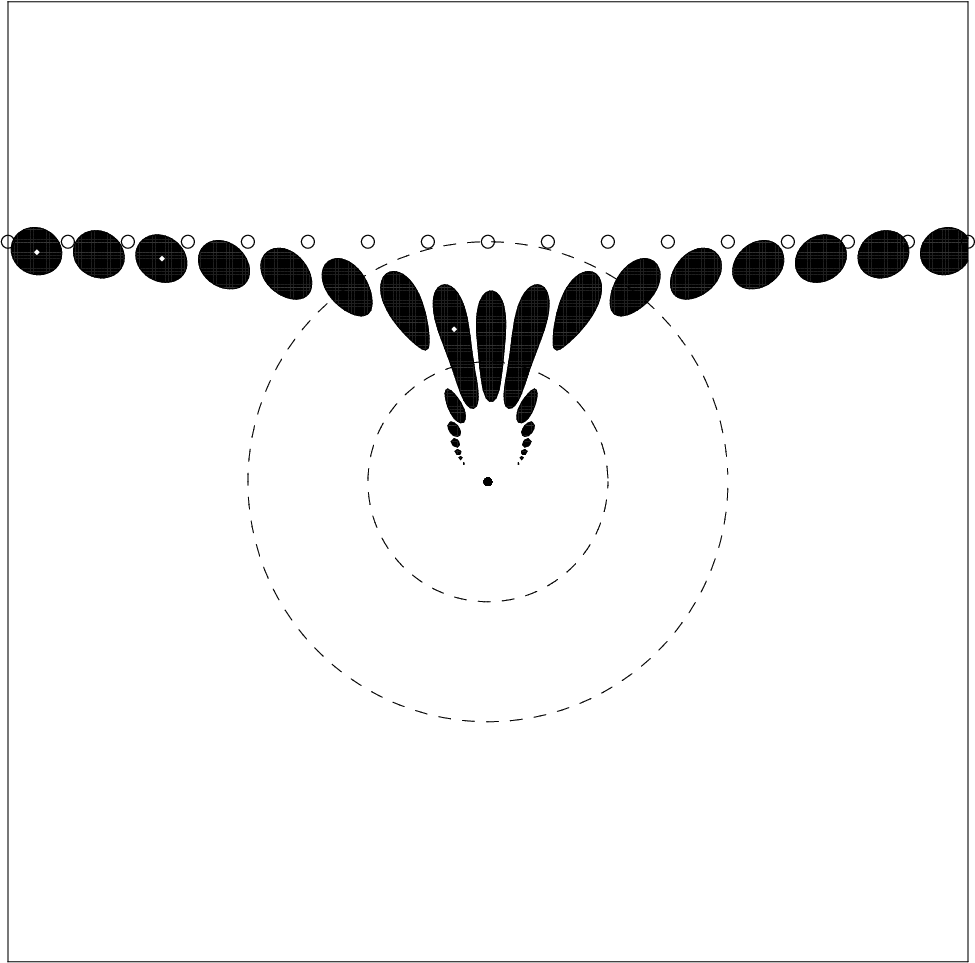}}
\vspace{.25cm} \caption[True motion of the source and apparent
motion of the images for $B_0= 2$]{True motion of the source and apparent
motion of the images for $B_0= 2$. The inner dashed circle is the
Einstein ring, the outer dashed circle is twice the Einstein ring.
The rest is as in Fig.~\ref{fig:positive}.}
\label{fig:bmin_eq2}
\end{figure}

\begin{figure}
\centerline{
\includegraphics[width=0.5\textwidth]{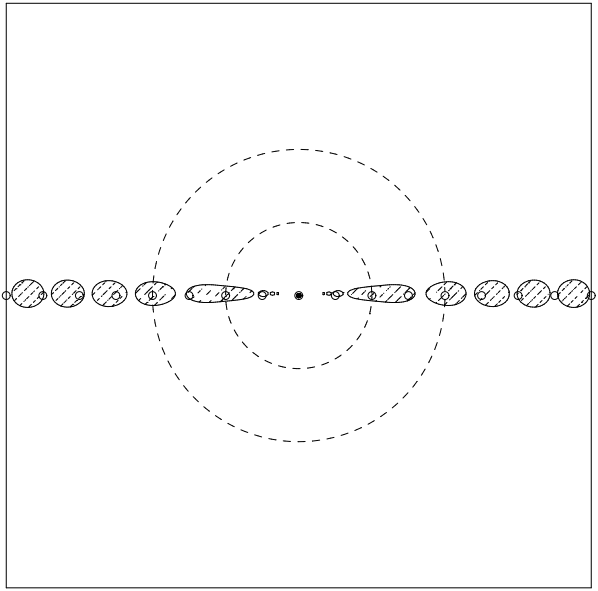}}
\vspace{.25cm} \caption[True motion of the source and apparent
motion of the images for $B_0=0$]{True motion of the source and apparent
motion of the images for $B_0=0$. The inner dashed circle is the
Einstein ring, the outer dashed circle is twice the Einstein ring.
Images here are shown with the shaded ellipses. The rest is as in
Fig.~\ref{fig:positive}.}
\label{fig:bmin_0}
\end{figure}

For three different values of $B_0$ we have three different lensing
configurations shown in Figs.~\ref{fig:bmin_gt2},~\ref{fig:bmin_eq2} and
~\ref{fig:bmin_lt2}. Note the large difference in the shapes of the images for
these three regimes. Fig.~\ref{fig:bmin_0} shows the case of a minimum
impact parameter equal to zero, $B_0=0$ (the path of the source
goes through the lens).

\begin{figure}
\centerline{
\includegraphics[width=0.5\textwidth]{chapter3/bmin_lt2.ps}}
\vspace{.25cm} \caption[True motion of the source and apparent
motion of the images for $B_0 < 2$]{True motion of the source and apparent
motion of the images for $B_0 < 2$. The inner dashed circle is the
Einstein ring, the outer dashed circle is twice the Einstein ring.
The rest is as in Fig.~\ref{fig:positive}.}
\label{fig:bmin_lt2}
\end{figure}

\begin{figure}
\begin{center} \epsfxsize=7cm \epsfysize=7cm \epsffile{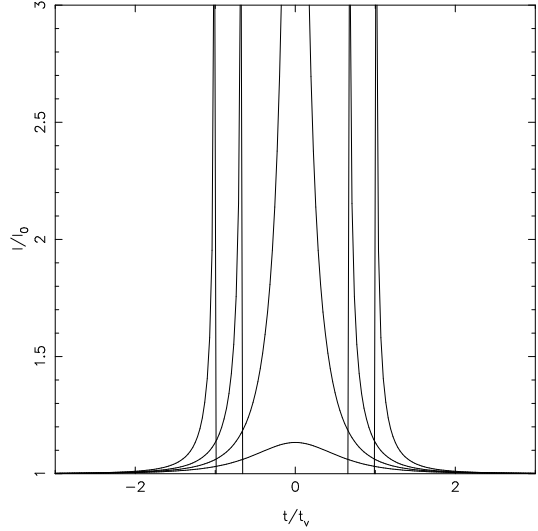}
\vspace{.25cm}\end{center} \caption[Light curves for the negative
mass lensing of a point source]{Light curves for the negative
mass lensing of a point source. From the center of the graph towards
the corners the curves correspond to $B_0=2.5$, $2.0$, $1.5$, $0.0$.
The time scale here is $\xi_0$ divided by the effective transverse
velocity of the source.
\label{fig:light}}
\end{figure}

It can be assumed without any loss of generality that the observer and the
lens are motionless and the source moves in the plane perpendicular to the
line of sight (thereby, changing its position in the source plane). We adopt
the treatment given in \cite{Kayser1986} for the velocity $V$, and consider
effective transverse velocity of the source relative to the critical curve.
We define the time scale of the microlensing event as the time it takes the
source to move across the Einstein radius, projected onto the source plane,
$\xi_0=\theta_{\rm E} D_{\rm s}$, \be t_{\rm v}=\frac{\xi_0}{V}\,\,.
\label{eq:tscale} \end{equation} Angle $\beta$ changes with time according to
\be 
\beta(t) = \sqrt{\left(\frac{Vt}{D_{\rm s}}\right)^2 + \beta_0^2}\,\,.
\label{eq:beta(t)}
\end{equation} 
Here the moment $t=0$ corresponds to the smallest angular
distance $\beta_0$ between the lens and the source. Normalizing
(\ref{eq:beta(t)}) to $\theta_E$, \be u(t) = \sqrt{\left(\frac{Vt}{\theta_{\rm
E} D_{\rm s}}\right)^2 + \left(\frac{\beta_0}{\theta_{\rm E}}\right)^2}\,\,,
\end{equation} where the dimensionless impact parameter $u$ is defined in
(\ref{eq:ratio}). Including the time scale $t_v$ and defining \be B_0=
\frac{\beta_0}{\theta_{\rm E}}\,\,, \end{equation} we obtain \be u(t)=
\sqrt{B_0^2 + \left(\frac{t}{t_v}\right)^2}\,\,. \label{eq:u(t)}
\end{equation} Finally, the total amplification as a function of time is
given by \be A(t)=\frac{u(t)^2-2}{u(t)\sqrt{u(t)^2-4}}\,\,. 
\end{equation}
Comparing this analysis with that of Cramer et al. \cite{cramer},
we note that they wrote the equation for the time dependent
dimensionless impact parameter as (cf. our Eq.~\ref{eq:u(t)})
$$
B(t) = B_0 \sqrt{1 + \left( \frac{t}{T_0} \right)^2}\,\,,
$$
and defined the time scale for the microlensing event as the time
it takes to cross the minimum impact parameter (cf. our Eq.~\ref{eq:tscale})
$$
T_0=\frac{b_0}{V}\,\,,
$$
where $b_0$ is the minimum impact parameter and other variables carry the
same meaning as in our case. While there is no mistake in using such
definitions, there is a definite disadvantage in doing so. Using Eq.~10 of
\cite{cramer} for $B(t)$ we cannot build the light curve for the case of the
minimum impact parameter $B_0=0$. In this case their Eq.~8 diverges, although
there is nothing wrong with this value of $B_0$ (see our Figs.~\ref{fig:bmin_0}
and~\ref{fig:light}). In the same way, their definition of a time scale does
not give much information on the light curves. With our definition
(Eq.~\ref{eq:tscale}) we can see in Fig.~\ref{fig:light} that in the extreme
case of $B_0=0$ the separation between the half-events is exactly
$2\theta_{\rm E}$; it is always less than that with any other
value of $B_0$.\\
In Fig.~\ref{fig:light} we show the light curves for the point source for
four source trajectories with different minimum impact parameters $B_0$. As
can be seen from the light curves, when the distance from the point mass to
the source trajectory is larger than $2\theta_{\rm E}$, the light curve is
identical to that of a positive mass lens light curve. However, when the
distance is less than $2\theta_{\rm E}$ (or in other terms, $B_0 \le 2.0$), the
light curve shows significant differences. Such events are characterized by
the {\it asymmetrical} light curves, which occur when a compact source crosses
a critical curve. A very interesting, eclipse-like, phenomenon occurs here. We
have a zero intensity region (disappearance of images) with an angular radius
$\theta_0$ \be \theta_0 = \sqrt{4\theta^2_{\rm E}- \beta_0^2}\,\,,
\end{equation} or in terms of normalized unit $\theta_{\rm E}$, \be \D =
\sqrt{4 - B_0^2}\,\,. \label{eq:Delta} \end{equation}

In the next subsection we shall see how these features get affected by the
presence of an extended source.

\subsection{Extended sources}

In the previous subsections we considered magnifications and light curves for
point sources. However, sources are extended, and although their size may be
small compared to the relevant length scales of a lensing event, this
extension definitely has an impact on the light curves as will be
demonstrated below. From variability arguments, the optical and X-ray
continuum emitting regions of quasars are assumed to be much less than 1
pc~\cite{Schneider2}, whereas the broad-line emission probably has a radius as
small as $0.1$ pc~\cite{Osterbrock1976}. The high energy gamma-spheres have a
typical radius of 10$^{15}$ cm \cite{blan}. Hence, one has to consider a
fairly broad range of source sizes.

We define the dimensionless source radius, $\tilde R$, as \be
\tilde R = \frac{\rho}{\theta_{\rm E}} = \frac{R}{\xi_0}\,\,, \end{equation}
where $\rho$ and $R$ are the angular and the linear physical size
of the source, respectively, and $\xi_0$ is the length unit in
the source plane (see Eq.~\ref{eq:tscale}).

\subsection{Comments on numerical method and simulations}\label{sec:algorithm}

It is convenient to write the lens equation in the scaled scalar
form \be y = x + \frac{1}{x}\,\,, \label{eq:lens} \end{equation} where we
normalize the coordinates to the Einstein angle:\footnote{Note,
that for the case in which $x$ and $y$ are expressed in length
units, we obtain a different normalization in each plane, which is
not always convenient.} \be x=\frac{\theta}{\theta_{\rm E}}
\,\,\,;\,\,\,\,\,\,\,y= \frac{\beta}{\theta_{\rm E}}\,\,.
\label{eq:eqscaled} \end{equation} The lens equation can be solved
analytically for any source position. The amplification factor,
and thus the total amplification, can be readily calculated for
point sources. However, as we are interested in extended sources,
this amplification has to be integrated over the source
(Eq.~\ref{eq:total}). Furthermore, as we want to build the
light curves, the total amplification for an extended source has
to be calculated for many source positions. The amplification
${\cal A}$ of an extended source with surface brightness profile
$I({\bf y})$ is given by \be {\cal A} = \frac{\int d^2y I({\bf y})
{\cal A}_0({\bf y})} {\int d^2y I({\bf y})}\,\,, \label{eq:total}
\end{equation} where ${\cal A}_0({\bf y})$ is the amplification of a point
source at position ${\bf y}$. We have used the numerical method
first described in \cite{Schramm1987}. We cover the lens plane
with a uniform grid. Each pixel on this grid is mapped, using
Eq.~\ref{eq:lens}, into the source plane. The step width ($5000
\times 5000$) is chosen according to the desired accuracy (i.e. the
observable brightness). For a given source position
($y_{10},y_{20}$) we calculate the squared deviation function
(SDF) \be S^2=(y_{10} - y_1(x_1,x_2))^2 +
(y_{20}-y_2(x_1,x_2))^2\,\,. \label{eq:sdf} \end{equation} The solutions of
the lens equation (Eq.~\ref{eq:eqscaled}) are given by the zeroes of the SDF.
Besides, Eq.~\ref{eq:sdf} describes circles with radii $S$ around
($y_{10},y_{20}$) in the source plane. Thus, the lines $S=$ constant are just
the image shapes of a source with radius $S$, which we plot using standard
plotting software. Therefore, image points where SDF has value $S^2$
correspond to those points of the circular source which are at a distance $S$
from the center. The surface brightness is preserved along the ray and  if
$I(R_0)=I_0$ for the source, then the same intensity is given to those pixels
where SDF$\,=R_0^2$. In this way an intensity profile is created in the image
plane and integrating over it gives the total intensity of an image.
Thus, we obtain the approximate value of the total magnification by estimating
the total intensity of all the images and dividing it by that of the unlensed
source, according to the corresponding brightness profile of the source. For
a source with constant surface brightness the luminosity of the images is
proportional to the area enclosed by the line $S=$ constant. The total
magnification is obtained by estimating the total area of all images and
dividing it by that of the unlensed source. For calculations of light curves
we used a circular source which is displaced along a straight line in the
source plane with steps equal to 0.01 of the Einstein angular radius.

\subsection{Shapes of images and light curves for the uniform brightness
source}

For a circular source of radius $R$ and uniform brightness,
equation (49) transforms into 
\be 
{\cal A} = \frac{\int{d^2 y I({\bf y}) {\cal A}_0({\bf y})}} {\pi R^2} \,\,.
\end{equation}

\begin{figure}
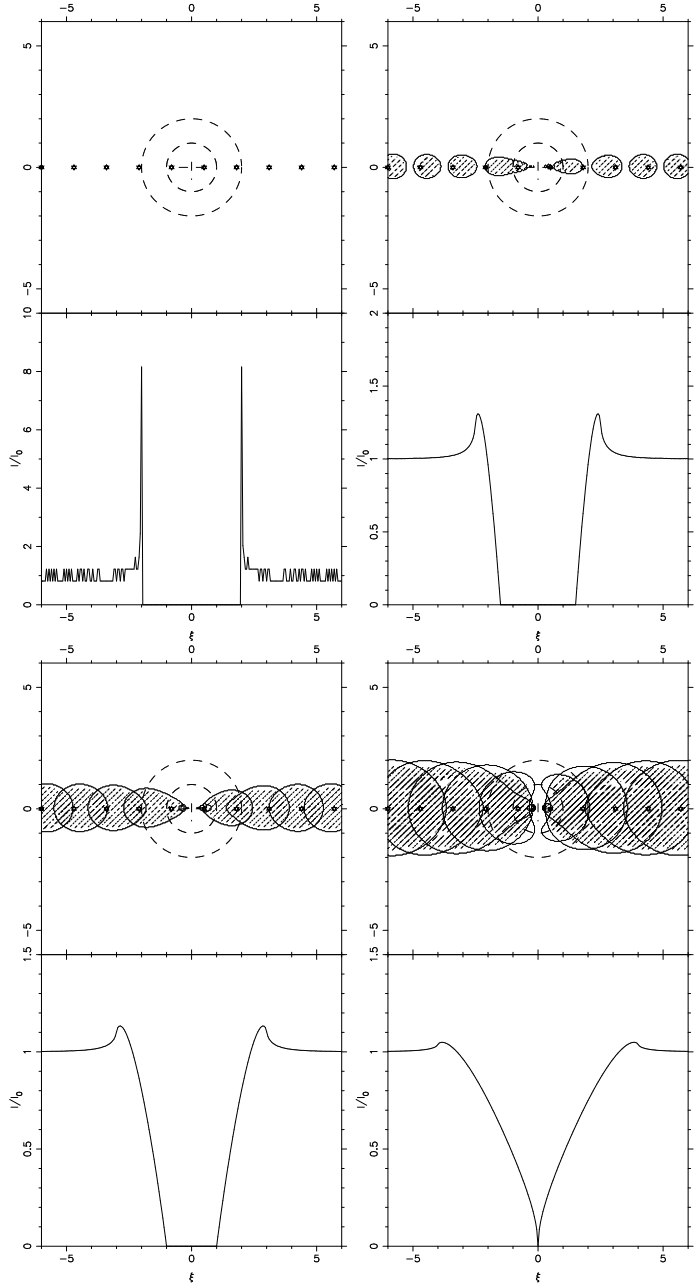

\begin{center}
\epsfxsize=4.5cm \epsfysize=8.5cm
\epsffile{chapter3/ext0_1.ps}  \epsfxsize=4.5cm \epsfysize=8.5cm
\epsffile{chapter3/ext0_2.ps} \epsfxsize=4.5cm \epsfysize=8.5cm
\hspace{1in}
\epsffile{chapter3/ext0_3.ps} \epsfxsize=4.5cm \epsfysize=8.5cm
\epsffile{chapter3/ext0_4.ps}
\end{center}
\vspace{.2cm}
\caption[Four sets of lens-source configurations for four
different values of the dimensionless source radius]{Four sets of
lens-source configurations (\emph{upper panels}) and corresponding
amplification as a function of source's center position (\emph{bottom panels})
are shown for four different values of the dimensionless source radius $\tilde
R $ (0.01, 0.5, 1.0, 2.0, in normalized units, $\theta_{\rm E}$). Each
of the four upper panels display the time dependent position of
the source's center, the shapes of images (shaded ellipses) and
critical curves (dashed circles). The series of open small circles
show the path of the source center. The lens is marked by the
central cross. Minimum impact parameter $B_0=0$. By replacing
$\xi$ with $\xi \theta_{\rm E} D_{\rm s} V^{-1} = \xi t_{\rm v}$
we get corresponding time depending light curve.}\label{fig:ext0}
\end{figure}

\begin{figure}
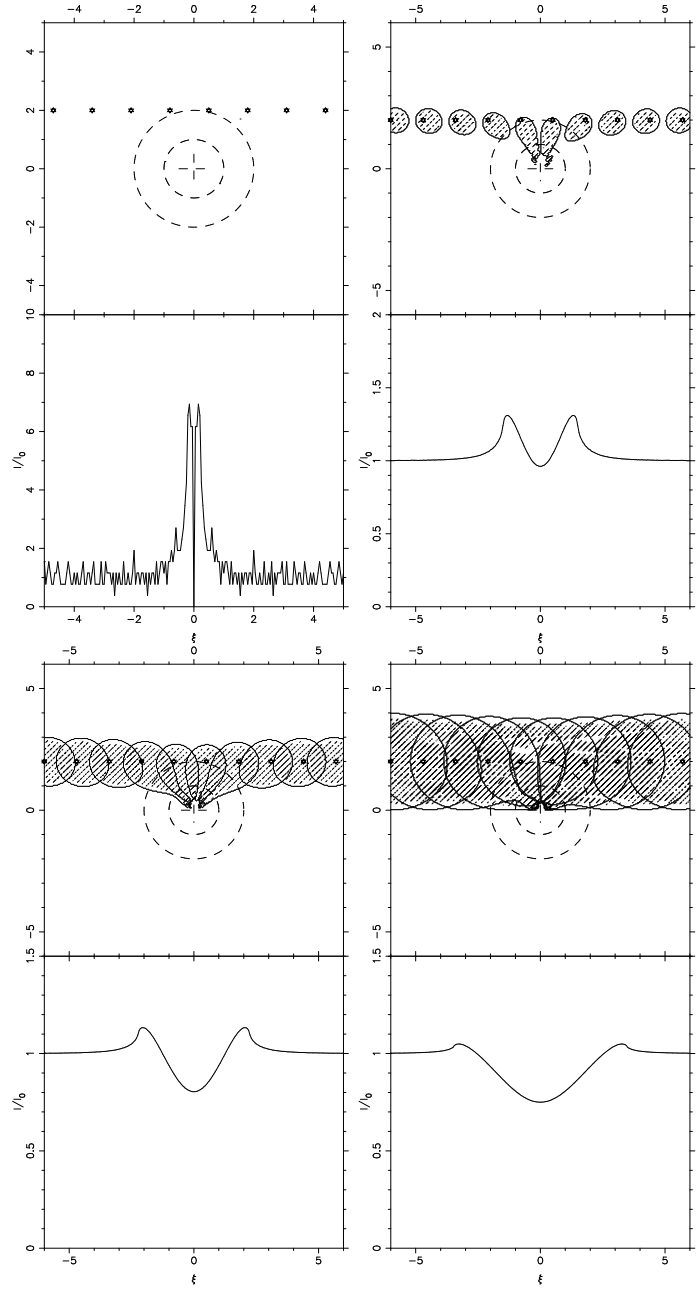

\begin{center}
\epsfxsize=4.5cm \epsfysize=8.5cm
\epsffile{chapter3/ext2_1.ps}  \epsfxsize=4.5cm \epsfysize=8.5cm
\epsffile{chapter3/ext2_2.ps} \epsfxsize=4.5cm \epsfysize=8.5cm
\hspace{1in}
\epsffile{chapter3/ext2_3.ps} \epsfxsize=4.5cm \epsfysize=8.5cm
\epsffile{chapter3/ext2_4.ps}
\end{center}
\vspace{.25cm}
\caption{Same as in Fig.~\ref{fig:ext0}, but for minimum impact parameter
$B_0=2.0$.}\label{fig:ext2}
\end{figure}

Figs.~\ref{fig:ext0} and~\ref{fig:ext2} show four projected source and
image positions, critical curves/caustics in the lens/source plane and
representative light curves for different normalized source sizes. The
sources are taken to be circular disks with constant surface brightness. In
order to get absolute source radii and real light curves we need the value of
$\theta_{\rm E}$, the normalized angular unit, the distance to the source, as
well as the velocity $V$ of the source relative to the critical curves in the
source plane. We have used $M=M_{\odot}$, $H_0=100$ km
s$^{-1}$ Mpc$^{-1}$ and a standard flat cosmological model with zero
cosmological constant. Here and in all subsequent simulations the
redshift of the source is $z_{\rm s}=0.5$ and the redshift of the
lens is $z_{\rm l}=0.1$.

We display two cases for two different impact parameters. It must be noted
here that the minimum impact parameter $B_0$ defines now the shortest distance
between the line of path of the center of the source and the lens. For each
$B_0$, the dimensionless radius of the source $\tilde R$ increases from 0.01
to 2.0 in normalized units of $\theta_{\rm E}$. The shape of produced images
changes notably with the increase of the source size, as can be seen in the
bottom right panel of Figs.~\ref{fig:ext0} and~\ref{fig:ext2}. At the same
time the smaller the source the greater the magnification, since when the
source radius is greater than the Einstein radius of the lens, the exterior
parts, which are amplified, compete with the interior ones, which are
demagnified.

It can be noted that despite the noise in some of the simulated light curves
the sharp peaks which occur when the source is crossing the critical line are
well defined even for the smallest source. Note that all infinities are
replaced now by finite amplifications, and that the curves are softened; all
these effects being generated by the finite size of the source. Indeed, while
the impressive drop to zero in the light curve is maintained, the divergence
to infinity, that happens for a point source, is very much reduced. Note, that
in cases of a large size of the source,  the magnification is very small. If
we would like to see bigger enhancement than that, we should consider sources
of smaller sizes, approaching the point source situation (cf.
Fig.~\ref{fig:ext0}, upper left plot). It is also interesting to note here
that for the impact parameter $B_0=2.0$ the light curve of a small extended
source, though approaching the point source pattern (Fig.~\ref{fig:light}),
still differs considerably from it (Fig.~\ref{fig:ext2}, upper left plot).

\subsection{Shapes of images and light curves for the sources with
non-uniform brightness profiles}

In order to compare a constant surface brightness source with more
realisitc distributions, we simulate image configurations and
calculate light curves for two different assumed profiles with
radial symmetry. 

\subsubsection{Source with Gaussian brightness distribution}

For a Gaussian source we have $ I(r)=I_0 e^{-r^2/r_0^2}$, where we normalized
the profile such that the maximum value of $I$ equals unity. We define the
radius containing $90\%$ of all the luminosity as the effective radius of the
source, $R_{\rm S}$. To find the relation between $R_{\rm S}$ and $r_0$, we
write the total luminosity as \be L(\infty) = \int_{0}^{\infty}
e^{-{r^2}/{r_0^2}} 2 \pi r\,dr= \pi r_0^2 \,\,, \end{equation} \be L(R) =
\int_{0}^{R} e^{- {r^2}/{r_0^2}} 2 \pi r\,dr= \pi r_0^2 \left[1 -
e^{- {R^2}/{r_0^2}} \right]\,\,, \end{equation} then \be \frac{L(R_{\rm
S})}{L(\infty)} = 0.9 = \left[ 1- e^{-{R_{\rm S}^2}/{r_0^2}}
\right]\,\,. \end{equation} Thus, effective radius relates to the parameter
$r_0$ as 
\be 
\frac{R_{\rm S}}{r_0} = \sqrt{\ln{10}}\,\,. 
\end{equation}

\begin{figure}
\centerline{\epsfig{figure=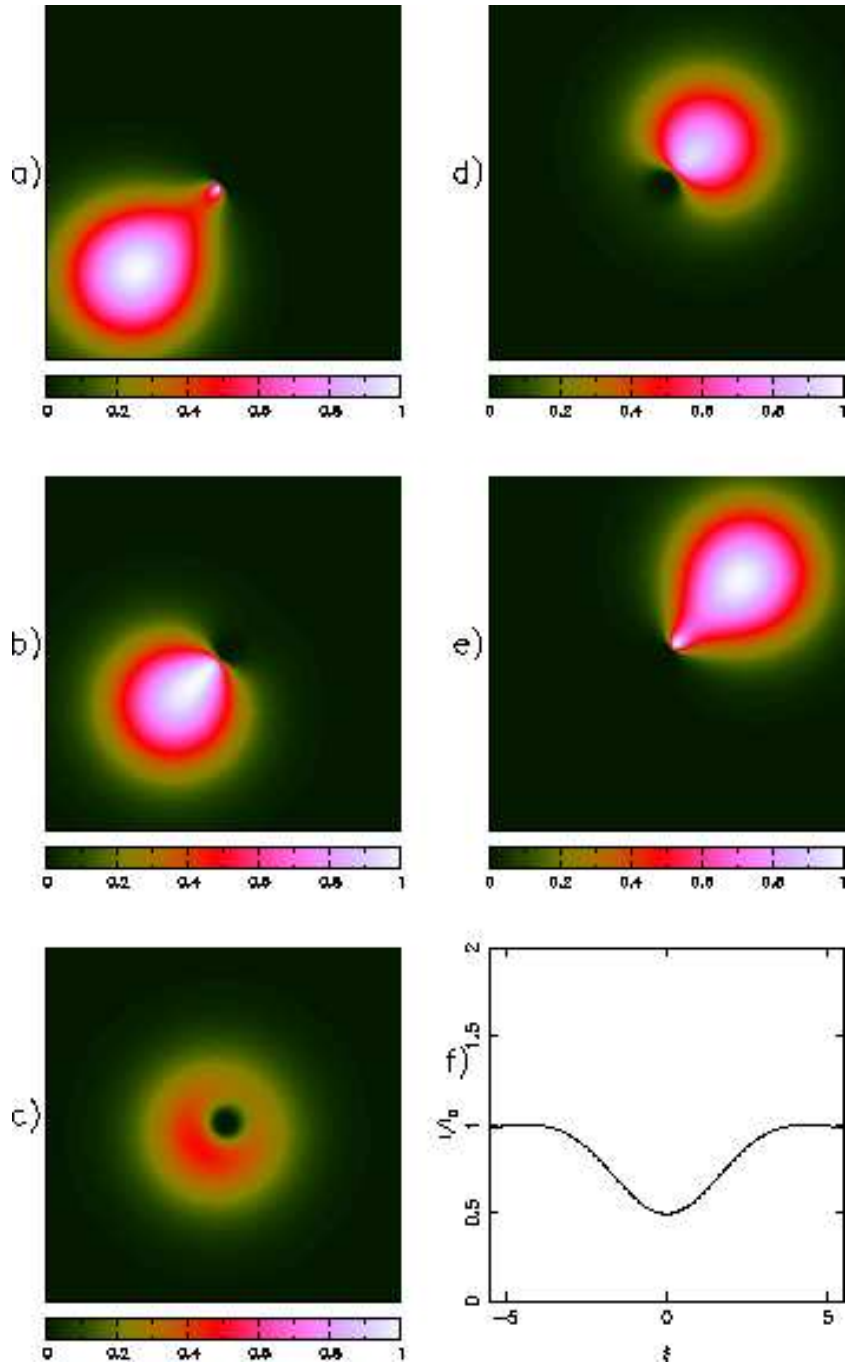,width=0.7\textwidth}}
\vspace{.25cm} 
\caption[Image configurations and a corresponding light curve for a Gaussian
source]{Image configurations (frames a to e) and a corresponding light curve
(f) for a Gaussian source with effective radius $\widetilde R_{\rm S}=3.0$, in
units of the Einstein angle. The source is moving from the lower left corner
(frame a) to the right upper corner (frame e), passing through the lens
($B_0=0$). The lens is in the center of each frame. Size of each frame is $5
\times 5$, in the normalized units. Wedges to each frame provide
the brightness scale for the images. Note the eclipse-like
phenomenon, consistent with the incomplete demagnification showed
in the light curve (frame f). \label{fig:gauss3}}
\end{figure}

\begin{figure}
\centerline{\epsfig{figure=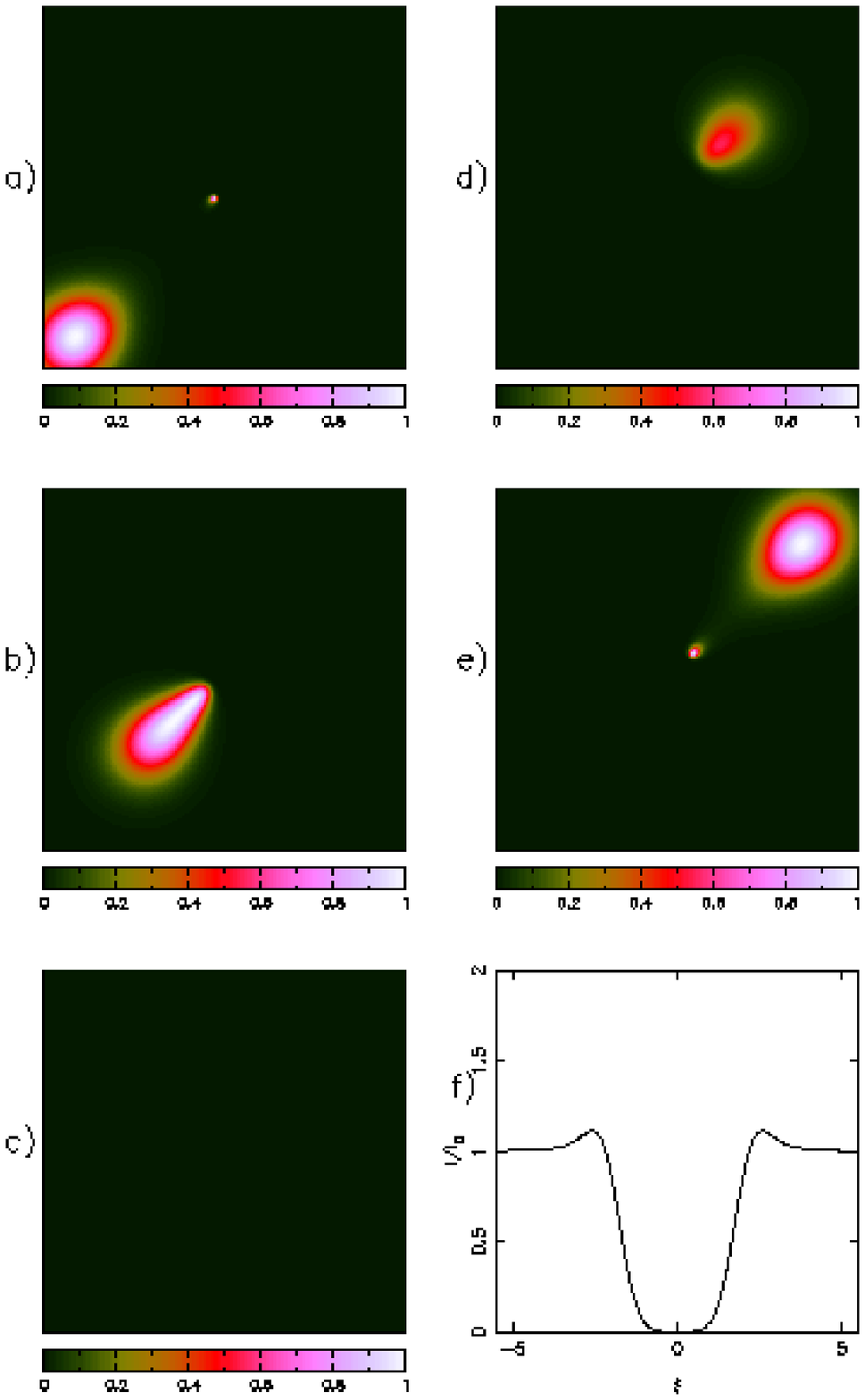,width=0.7\textwidth}}
\vspace{.25cm} \caption[Same as in the previous Fig. , but with the
different effective source radius]{Image configurations (frames a to e) and a
corresponding light curve (f) for a Gaussian source with effective radius
$\widetilde R_{\rm S}=1.0$, in units of the Einstein angle. The source is
moving from the lower left corner (frame a) to the right upper corner (frame
e), passing through the lens ($B_0=0$). The lens is in the center of each
frame. Size of each frame is $3 \times 3$, in the normalized units. Wedges to
each frame provide the brightness scale for the images. Note the complete
disappearence of the source when it is inside the double Einstein
radius of the lens (frame c), corresponding to the drop of
magnification to zero in the light curve (frame f).
\label{fig:gauss1}}
\end{figure}

In Figs.~\ref{fig:gauss3} and~\ref{fig:gauss1} we show the images of an
extended source with a Gaussian brightness distribution for two effective
dimensionless source radii $\tilde R_{\rm S}$, 3.0 and 1.0, in units of the
Einstein angle (Fig.~\ref{fig:gauss3}, frames a--e and Fig.~\ref{fig:gauss1},
frames a--e, respectively), together with the corresponding light curves
(Fig.~\ref{fig:gauss3}, frame f and Fig.~\ref{fig:gauss1}, frame f,
respectively). Here the source path passes through the lens ($B_0=0$), which
lies exactly in the center of each frame. In Fig.~\ref{fig:gauss3} the source's
extent in the lens plane is greater than the Einstein radius of the lens. We
notice there that there is an eclipse-like phenomenon, occurring most notably
when most of the source is near or exactly behind the lens. This is
consistent with the light curve (frame f), where there is a
de-magnification. For the source with radius smaller than the
double Einstein radius of the lens (Fig.~\ref{fig:gauss1}), the low
intensity region is replaced by the zero intensity region; the source
completely disappears from the view (frame c).

\begin{figure}
\centerline{\epsfig{figure=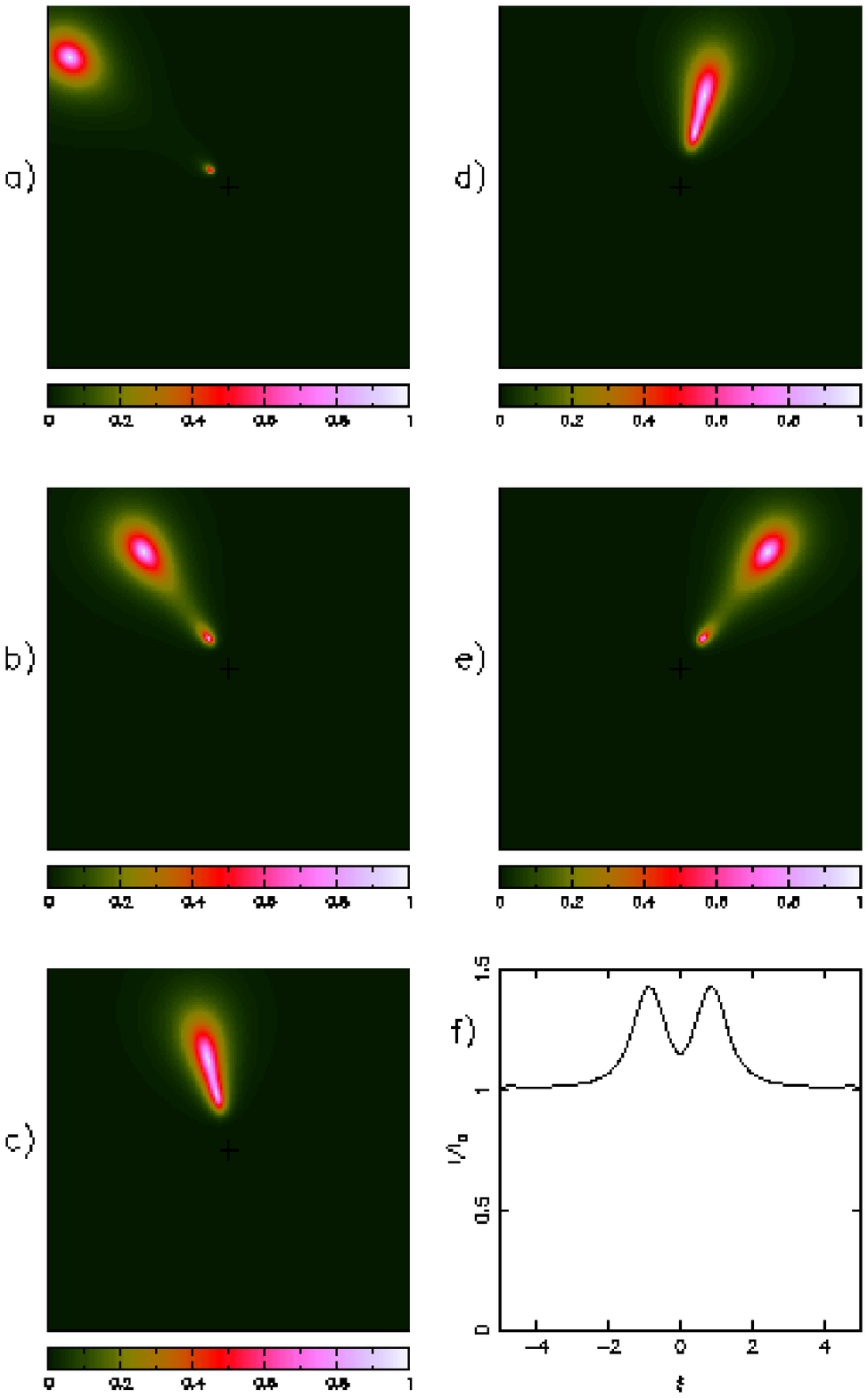,width=0.7\textwidth}}
\vspace{.25cm}
\caption[Image configuration for a source with exponential
brightness distribution]{Image configurations (frames a to e) and a
corresponding light curve (f) for a source with exponential
brightness distribution and effective radius $\widetilde R_{\rm
S}=1.0$, in units of the Einstein angle. The source is moving from
the upper left corner (frame a) to the upper right corner (frame
e) with the impact parameter $B_0=2.0$. The lens is in the center
of each frame. Size of each frame is $2.5 \times 2.5$, in the
normalized units. Wedges to each frame provide the brightness
scale for the images. \label{fig:exponent}}
\end{figure}

\subsubsection{Source with exponential brightness distribution}

We have for this brightness distribution $I(r) = I_0 e^{-r/r_0}$. In the same
manner as above, $R_{\rm S}$ is defined as radius, containing $90\%$ of
total luminosity. In the same way as above, total luminosity 
\be
L(\infty) = \int_{0}^{\infty} e^{- {r}/{r_0}} 2 \pi r\,dr= 2 \pi
r_0^2 \,\,, 
\end{equation} 
then 
\be 
L(R) = \int_{0}^{R} e^{- {r}/{r_0}} 2 \pi
r\,dr = 2 \pi \left[r_0^2 - \left(R \,r_0 + r_0^2\right)e^{-
{R}/{r_0}} \right]\,\,, 
\end{equation} 
and 
\be 
\frac{L(R_{\rm S})}{L(\infty)}= 0.9 \,\,. 
\end{equation}
From where we find that effective radius relates to the parameter $r_0$ as
\be 
e^{- {R_{\rm S}}/{r_0}}\left(1+
\frac{R_{\rm S}}{r_0}\right) = 0.1 \,\,. 
\end{equation} 
Solution to this equation gives $R_{\rm S}/r_0 \approx~ 3.89$. This profile
is also normalized such that the maximum value of $I$ equals unity.

In Fig.~\ref{fig:exponent} we display the images of the source with the
exponential brightness profile and the corresponding light curve (frame f).
The effective radius of the source is 1.5. The impact parameter here is
$B_0=2.0$; the lensing regime corresponds to the one schematically depicted in
Fig.~\ref{fig:bmin_eq2}. We see how shapes of the images change, becoming
elongated and forming the radial arc (frames c and d).

\subsubsection{Comparing light curves for sources with different brightness
profiles}
\begin{figure}[ht!]
\centerline{\epsfig{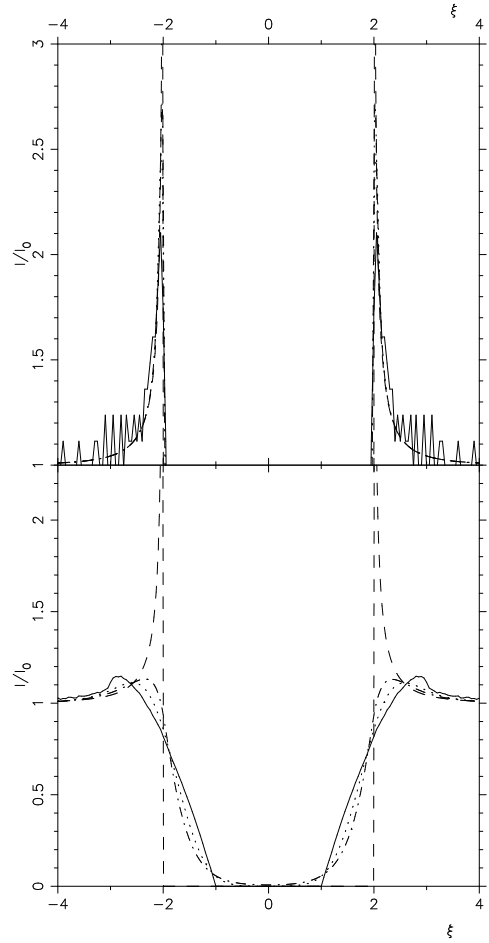}}
\caption[Light curves for sources with different brightness profiles]{Light
curves for the point mass source (dashed line), source with constant surface
brightness (solid), source with Gaussian brightness distribution (dash-dotted)
and exponential brightness distribution (dotted) for two different effective
dimensionless source radii, $0.1$ (\emph{upper panel}) and $1.0$
(\emph{bottom panel}). \label{fig:lcurves}}
\end{figure}

In Fig.~\ref{fig:lcurves} we compare light curves for three
different radially symmetric source profiles: uniform, Gaussian
and exponential, for two dimensionless source radii $\widetilde R =
\widetilde R_{\rm S}^{\rm gauss} = \widetilde R_{\rm S}^{\rm expon} = 0.1$
and $\widetilde R=\widetilde R_{\rm S}^{\rm gauss} = \widetilde R_{\rm S}^{\rm
expon} = 1.0$. As a reference curve we show the light curve of the
point source. All curves are made for the impact parameter $B_0=0$. We can
see larger noise in the uniform source curve, since the source with uniform
brightness has an extremely sharp edge, whereas Gaussian and exponential
sources are extremely smooth. Though we considered the sources with the same
effective radius, we can see from the plot that for a small source size, the
maximum magnification is reached by the  source with exponential profile (upper
panel). This is explained by the fact that this profile has a narrower central
peak than the Gaussian.

For the larger source, this behaviour smoothens, though we still can see
large differences in the light curves (bottom panel). Here the uniform source
experiences darkening, while sources with other profiles only undergo
demagnification.

\subsection{Time scales of microlensing}

Let the source have a transverse velocity ${\bf v}_{\rm s}$
measured in the source plane, the lens a transverse velocity ${\bf
v}_{\rm l}$ measured in the lens plane, and the observer a
transverse velocity ${\bf v}_{\rm obs}$ measured in the observer
plane. The effective transverse velocity of the source relative to
the critical curves with time measured by the observer is
\be
{\bf V} = \frac{1}{1+z_{\rm s}}{\bf v}_{\rm s} - \frac{1}{1+z_{\rm l}}
\frac{D_{\rm s}}{D_{\rm l}}{\bf v}_{\rm l} + \frac{1}{1+z_{\rm l}}
\frac{D_{\rm ls}}{D_{\rm l}}{\bf v}_{\rm obs}\,\,.
\label{eq:velocity}
\end{equation}
This effective velocity is such that for a stationary observer and lens, the
position of the source will change in time according to $\delta \bbox{\xi}={\bf
V} \Delta t$.

We basically have two time scales of interest here. The first one is the
typical rise time to a peak in the amplification. We can estimate that it
corresponds to a displacement of $\Delta y \sim \widetilde R$ of the source
across a critical line. The corresponding time scale is $ \tau_1 = t_{\rm v}
\widetilde R$, with $t_{\rm v}$ given by (\ref{eq:tscale}). In terms of the
physical source size $R=\xi_0 \widetilde R$, \be \tau_1=\frac{R}{V}\,\,,
\end{equation} where effective transverse velocity of the source $V$ is given
by Eq.~\ref{eq:velocity}. The second time scale of interest is the time between
two peaks $\tau_2$. For a point source we can estimate it as $ \tau_2=t_{\rm v}
\Delta $,  where $\Delta$ is given in Eq.~(\ref{eq:Delta}). For a source with
radius $\widetilde R$ and impact parameter $B_0=0.0$ it can shown to be \be
\tau_2=t_{\rm v} \left(\Delta + \widetilde R\right)  \,\,. \end{equation} In
Table 1 we list the time scales $\tau_1$ and $\tau_2$ for different values of
a source radius, and the time delay between two images $\Delta t_{12}$ for
the point source. The value of $V$ is estimated to be $V=5000$ km s$^{-1}$.

\begin{table}
\caption[Microlensing time scales for several source radii]
{Time scales for several source radii. $\tilde R$ is the
dimensionless source radius, in units of Einstein angle,
$|M|=1.0\,M_{\odot}$, redshift of the lens is $z_{\rm l}=0.1$,
redshift of the source is $z_{\rm s}=0.5$,
$\xi_0=5.42\times10^{11}$ km is the normalized length unit in the
source plane. The time scales correspond to apparent source
velocity (see Eq.~\ref{eq:velocity}) $V=5000$ km s$^{-1}$.}
\vskip 0.1in
\tabcolsep 0.3in
\begin{center}
\begin{tabular}{|l|c|c|c|r|}
\hline \hline
\emph{$\widetilde R$} & \emph{$R$} (pc) & \emph{$\Delta
t_{12}$}(sec)$^a$ & \emph{$\tau_1$} (yr) & \emph{$\tau_2$}(yr)$^b$ \\ \hline
        &                     &                               &          &           \\
$0.0$   & point source        &      $2.0\times 10^{-4}$      &    --    & $6.78$    \\
$0.01$  & $1.07\times10^{-4}$ &             --                & $0.03$   & $6.81$    \\
$0.1$   & $1.07\times10^{-3}$ &              --               & $0.34$   &  $7.11$    \\
$1.0$   & $1.07\times10^{-2}$ &              --               &  $3.38$   & $10.16$   \\
$2.0$   & $0.3\times10^{-1}$  & --   & $6.75$  &$13.6$ \\
\hline \hline
\end{tabular}
\end{center}
$^a$  u=4.0 (definition in Eq.~\ref{eq:ratio})\\
$^b$  $B_0=0.0$
\end{table}  
 
\subsection{Concluding remarks}
In these two sections we have explored the consequences of gravitational
microlensing following from the existence of matter violating the energy
conditions. We have also quantitatively analyzed, using numerical simulations,
the influence of a finite size of the source on the gravitational lensing
negative-mass event. We have thus enhanced and completed previous works where
the focus was put on the point source light curves and no discussion was given
concerning the shapes of images, actual simulations of microlensing events,
time gain function, and other features presented here.
Figs.~
\ref{fig:magnifications},~\ref{fig:bmin_gt2}--\ref{fig:bmin_0},~\ref{fig:ext0}--\ref{fig:lcurves}
and Table 1 comprise our new results: a useful comparison arena where possible
existence of wormholes or any other kind of negative mass compact objects can
be observationally tested.

The next step would be to test these predictions using archival, current, and
forthcoming observational microlensing experiments. The only search done up to
now included the BATSE database of $\g$-ray bursts. There is still much
unexplored territory in the gravitational microlensing archives. We suggest
adaption of alert systems of these experiments in order to include the possible
effects of negative masses as well. This would lead to a whole new world of
discoveries.

\section{Macrolensing}

In this section we present a set of simulations showing macrolensing
effects we could observe if a large amount of negative energy density exists
in our universe. The physical systems could have an energy density equivalent
to a total negative mass of the size of a galaxy or even a cluster of
galaxies. In what follows, we used a background cosmology described by a FRW
flat universe with $\O_{\rm m}=1$ and a zero cosmological constant. In all
numerical computations a dimensionless Hubble parameter $h$ is put equal to 1
($H_0 = 100\,h$ km sec$^{-1}$ Mpc$^{-1}$). The relationship $D_{\rm
eff}=\dl \dls \ds^{-1}$ is a measure of the lensing efficiency of a given
mass distribution. $D_{\rm eff}$ peaks, quite independently of the
cosmological model assumed, at a lens redshift of $\sim 0.2-0.4$ for sources
at a typical redshift $z_{\rm s} \sim 1-1.3$ \cite{3}. To be more realistic,
we  placed the lens at a redshift of $z_l = 0.3$ and generated a random sample
of galaxies in the redshift range $ 0.3 < z_{\rm source} < 2.0$. The redshift
distribution conserves their comoving number density. This number density and
the projected sizes for these background sources were taken to be close to the
Tyson population of faint blue galaxies \cite{5}. The luminous area of each
galaxy was taken to be a circular disk of radius $R$ with a uniform brightness
profile and orientations of disk galaxies randomly placed in space. This was
done by defining in the code the ellipticity $e$ ($e = (1-r)/(1+r)$, $r$ being
the ratio of the minor axis to the major axis) and the position angle
$\varphi$, and randomly chosing the values of $e$ from the range $0<e<0.7$ and
the values of $\varphi$ from the range $0<\varphi<2\pi$.\footnote{The random
number generator needed in the code was taken from the book by Press et al.
\cite{1}, and we use the algorithm described in Ref. \cite{schneider}, p.298.
PGPLOT routine PGGRAY was employed in the code.}

We write the gravitational lens equation, which governs the mapping from the
lens to the source plane, in dimensionless form
\be 
\mbox{\boldmath $\beta$} = \mbox{\boldmath $\theta$} 
\left [1 + \frac{\theta^2_{\rm E}}{\theta^2} \right]\,\,. \label{eq:leq}
\end{equation} 
The lens equation (\ref{eq:leq}) describes a mapping ${\bf \t}
\mapsto {\bf \beta}$, from the lens to the source plane. For
convenience, we redefine the lens plane as ${\bf x}$ and the
source plane as ${\bf y}$. Then, Eq. (\ref{eq:leq}) can be written
as
\be
{\bf y} = {\bf x} \left(1+ \fr{\t^2_{\rm E}}{x^2}\right),
\end{equation}
where $x=|{\bf x}|=\sqrt{x_1^2+x_2^2}$.

We now consider a source, whose shape---either circular or elliptical---can
be described by a function $\chi({\bf y})$. Curves of constant $\chi$ are the
contours of the source. One can as well consider $\chi$ as a function of ${\bf
x}$, $\chi({\bf y}({\bf x}))$, where ${\bf y}({\bf x})$ is found using the
lens equation. Thus, all points ${\bf x}$ of constant $\chi$ are mapped onto
points ${\bf y}$, which have a distance $\sqrt{\chi}$ from the centre of the
source. If the latter contour can be considered an isophote of a source, one
has thus found the corresponding isophotes of the images (see
Section~\ref{sec:algorithm} for details on the algorithm).  

\subsection{Simulations results}
\begin{figure}[ht!]
\centerline{
\vspace{-1.5 cm}
\hspace{-1.0 cm}
\makebox{
\epsfig{figure=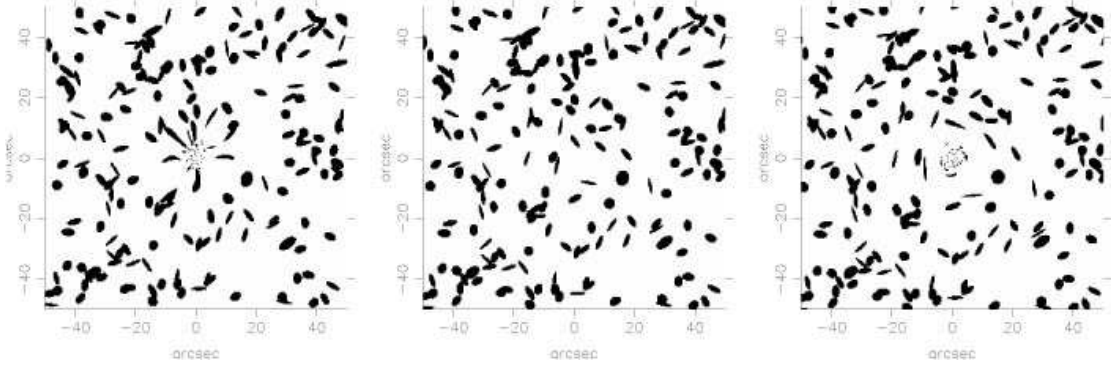,width=0.93\textwidth}}}
\vskip 0.6in
\caption[Appearance of a background field of sources in a
range of redshifts for lens mass equal to $1\times 10^{13}\,M_{\odot}$]
{{\it Left}: Appearance of a background field of sources in a range of
redshifts (see text) (200 galaxies, intrinsic radius 7 Kpc), when lensed by a
negative mass of $|M|_{\rm lens} = 1\times 10^{13}\,M_{\odot}$. {\it
Center}: Unlensed background field. {\it Right}: Appearance of the same
background field of galaxies when it is lensed by an equal amount of positive
mass; redhsifts are the same as for the negative mass case.} \label{fig:13}
\end{figure}

\begin{figure}[ht]
\centerline{
\vspace{-1.5 cm}
\hspace{-1.0 cm}
\makebox{
\epsfig{figure=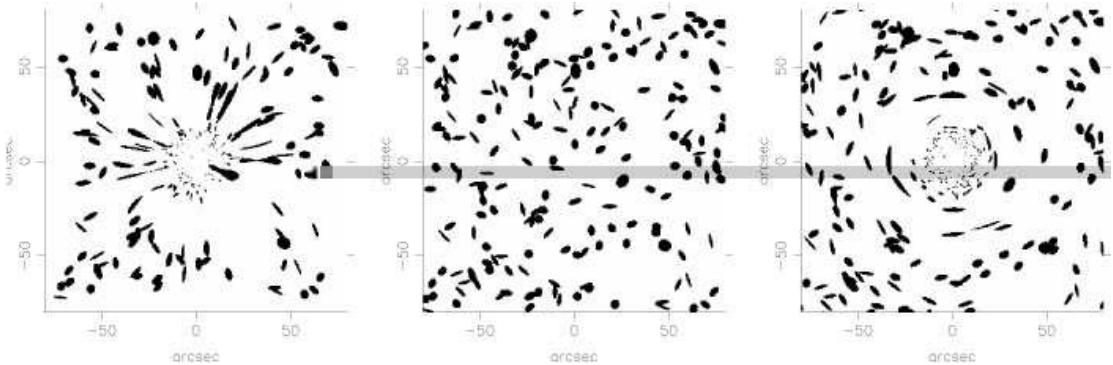,width=0.93\textwidth}}}
\vskip 0.6in
\caption[Appearance of a background field of sources in a
range of redshifts for lens mass equal to $1\times 10^{14}\,M_{\odot}$]{{\it
Left}: Appearance of a background field of sources (200 galaxies, intrinsic
radius 10 Kpc), when it is lensed by a negative mass of $|M|_{\rm lens} =
1\times 10^{14}\,M_{\odot}$. {\it Center}: As in Figure~\ref{fig:13}. {\it
Right}: As in Figure~\ref{fig:13}.} \label{fig:14}

\end{figure}

\begin{figure}[htb!]
\centerline{
\vspace{-1.5 cm}
\hspace{-1.0 cm}
\makebox{
\epsfig{figure=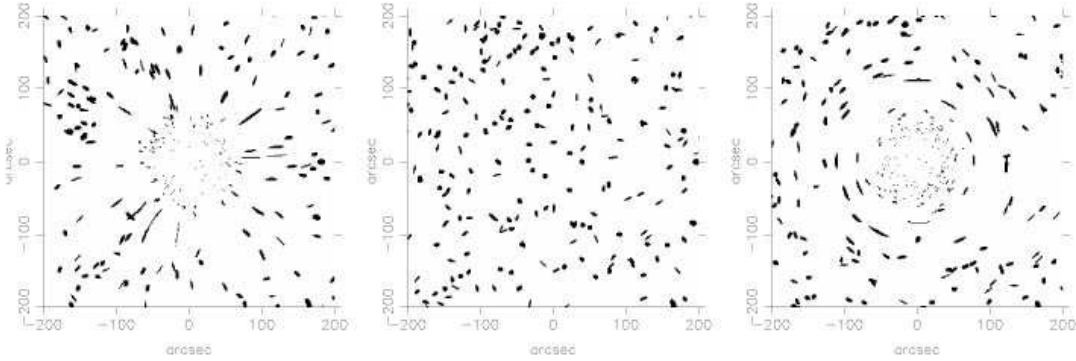,width=0.9\textwidth}}}
\vskip 0.6in
\caption[Appearance of a background field of sources in a
range of redshifts for lens mass equal to $1\times 10^{15}\,M_{\odot}$]{{\it
Left}: Appearance of a background field of sources (300 galaxies, intrinsic
radius 15 Kpc), when it is lensed by a negative mass of $|M|_{\rm lens} =
1\times 10^{15}\,M_{\odot}$. The simulation was made taking into account
sources located within 1.2 of the size of the shown window. {\it Center}: As in
Fig.~\ref{fig:13}. {\it Right}: As in Fig~\ref{fig:13}.} \label{fig:15}
\end{figure}

\begin{figure}
\centerline{
\vspace{-0.5 cm}
\hspace{-0.5 cm}
\makebox{
\epsfig{figure=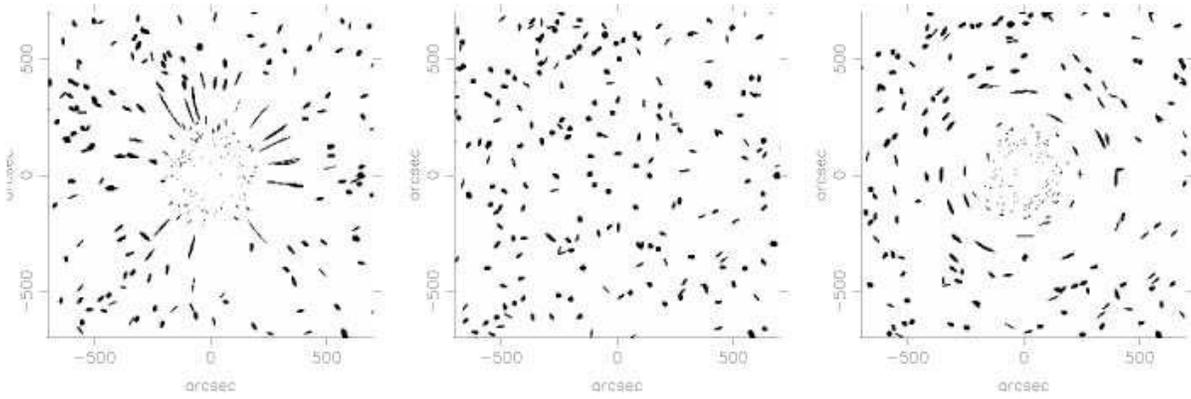,width=1.\textwidth}}}
\vskip 0.5in
\caption[Appearance of a background field of sources in a
range of redshifts for lens mass equal to $1\times 10^{16}\,M_{\odot}$]{{\it
Left}: Appearance of a background field of sources (300 galaxies, intrinsic
radius 50 Kpc), when it is lensed by a negative mass of $|M|_{\rm lens} =
1\times 10^{16}\,M_{\odot}$. The simulation was made taking into account
sources located within 1.2 of the size of the shown window. {\it Center}: As in
Fig.~\ref{fig:13}. {\it Right}: As in Fig.~\ref{fig:13}.} \label{fig:16}
\end{figure}

\begin{figure}
\centerline{
\vspace{-1.5 cm}
\hspace{-0.5 cm}
\makebox{
\epsfig{figure=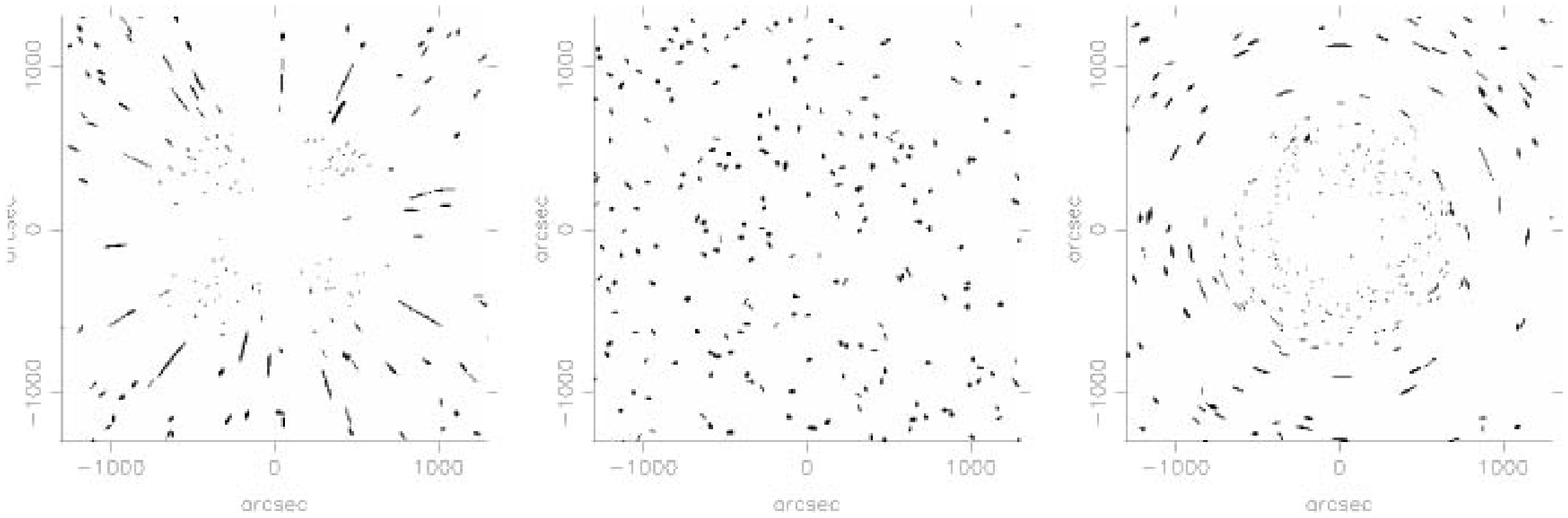,width=1.\textwidth}}}
\vskip 0.6in
\caption[Appearance of a background field of sources in a
range of redshifts for lens mass equal to $1\times 10^{17}\,M_{\odot}$]{{\it
Left}: Appearance of a background field of sources (500 galaxies), intrinsic
radius 60 Kpc, when it is lensed by a negative mass of $|M|_{\rm lens} =
1\times 10^{17}\,M_{\odot}$. The simulation was actually made taking into
account sources located within 1.5 of the size of the shown window. {\it
Center}: As in Fig.~\ref{fig:13}. {\it Right}: As in Fig.~\ref{fig:13}.}
\label{fig:17} 
\end{figure}

\begin{figure}
\centerline{
\vspace{-1.5 cm}
\hspace{-1.0 cm}
\makebox{
\epsfig{figure=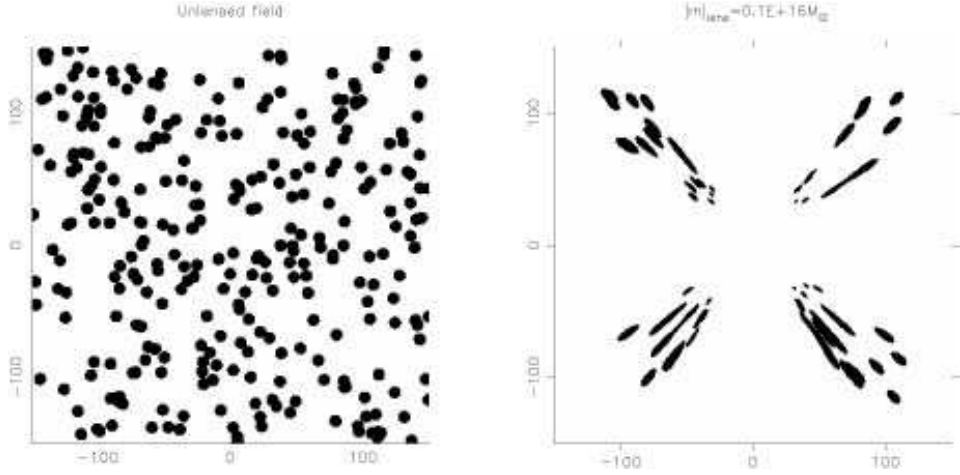,width=0.8\textwidth}}}
\vskip 0.6in
\caption[Problems with simulations of a large mass lens]{The need for the
increase of the background number of sources is shown by the appearance of a
four-fold symmetric pattern, which occurs because only the galaxies at the
corners of the left window are being affected by lensing effects. Axis are
marked in arcseconds. See next figure.}\label{fig:four-fold}
\label{pr}
\end{figure}

\begin{figure}
\centerline{
\vspace{-1.5 cm}
\hspace{-0.5 cm}
\makebox{
\epsfig{figure=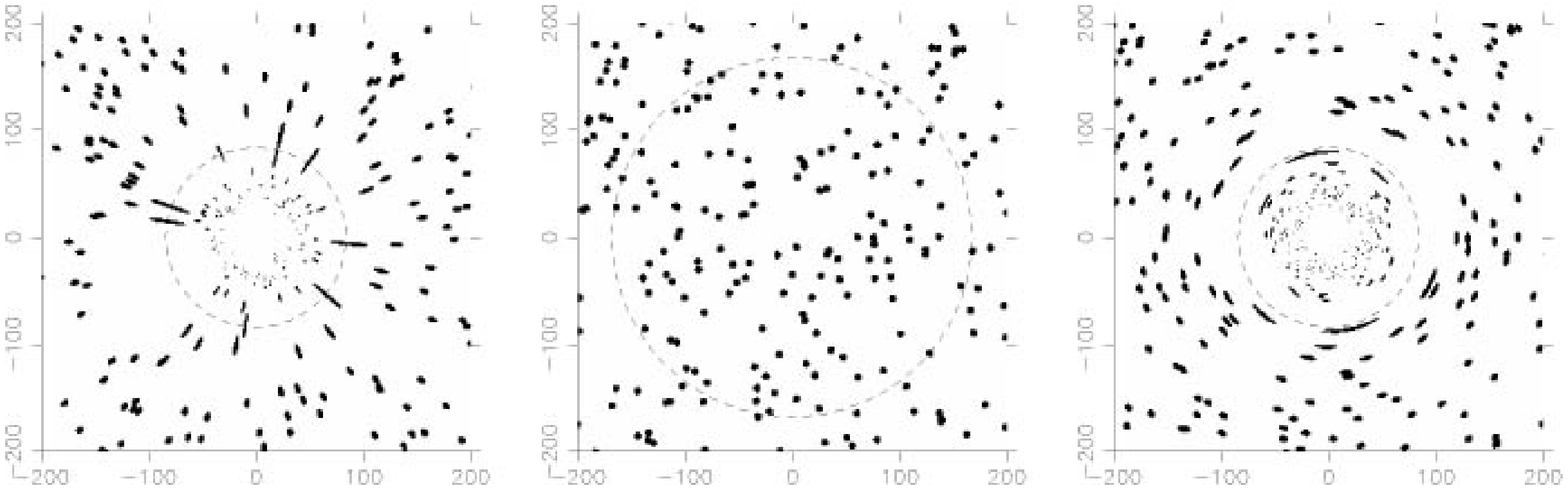,width=1.\textwidth}}}
\vskip 0.6in
\caption[Problems with simulations of a large mass lens; resolution]{{\it
Left}: Appearance of a background field of circular sources (300 galaxies),
each of them of 15 Kpc radius, when it is lensed by a negative mass of
$|M|_{\rm lens} = 1\times 10^{15}\,M_{\odot}$ with an Einstein angular radius
equal to $\theta_{\rm E}=84$ arcsec. The dashed circle is the Einstein
ring, the radial arcs are centered on it while the images inside
it are demagnified. Here $z_{\rm lens} = 0.4$, $z_{\rm
source}=1.4$. The simulation was made taking into account sources
located within 1.2 of the size of the shown window. {\it Center}:
Background field in the absence of the lens, dashed circle is the
double Einstein radius, all sources inside this radius are
shadowed. {\it Right}: Macrolensing effects produced by a positive mass
lens of $10^{15}\,M_{\odot}$, a dashed circle is the Einstein
radius shown here for comparison, sources inside it are strongly
lensed.} \label{fig:resolution} 
\end{figure}

In the first set of figures (Figs.~\ref{fig:13}-\ref{fig:17}), we show
the results of our simulations. Some special precautions must be taken for the
largest masses. The problem is that for a very massive lens the Einstein ring
becomes very large. Since for the negative mass lensing all sources inside
the double Einstein radius are shadowed (i.e. we can see the images of only
those sources which are outside the double Einstein radius), if we were to use
for lensing only the sources shown on the window, a fold-four symmetry pattern
would appear (Fig.~\ref{fig:four-fold}). Only the sources at the corners of the
current window (and outside the double Einstein radius) are lensed. In order to
solve this problem we have to consider also the sources from outside the
current window; then the lensing picture is restored and the scale
of the simulation is consistently increased. For this reason we increase the
number of background galaxies in Figs.~\ref{fig:15}-\ref{fig:17}. We show this
in detail in Figs.~\ref{fig:four-fold} and~\ref{fig:resolution}.


As a general feature of our simulations we can remark that, contrary to the
standard positive mass case, where ring-like structures appear, the negative
mass lensing produces finger-like, apparently ``runaway" structures, which seem
to escape from a central void. This is in agreement with the appearance of a
central umbra in the case of a point-like negative mass lensing (see previous
sections of this chapter). In the case of macrolensing, the umbra (central
void) is maintained on a larger scale which, depending on the negative mass of
the lens, can reach hundreds of acrsec in linear size. This umbra is always
larger than the corresponding one generated in positive macrolensing (see
Figures~\ref{fig:13}-\ref{fig:17}) and totally different in nature
\cite{cramer}. At least qualitatively, the existence of a macroscopic amount
of negative mass lens mimics the appearance of galaxy voids.

\begin{figure}
\centerline{
\vspace{-1.5 cm}
\hspace{-0.5cm}
\makebox{
\epsfig{figure=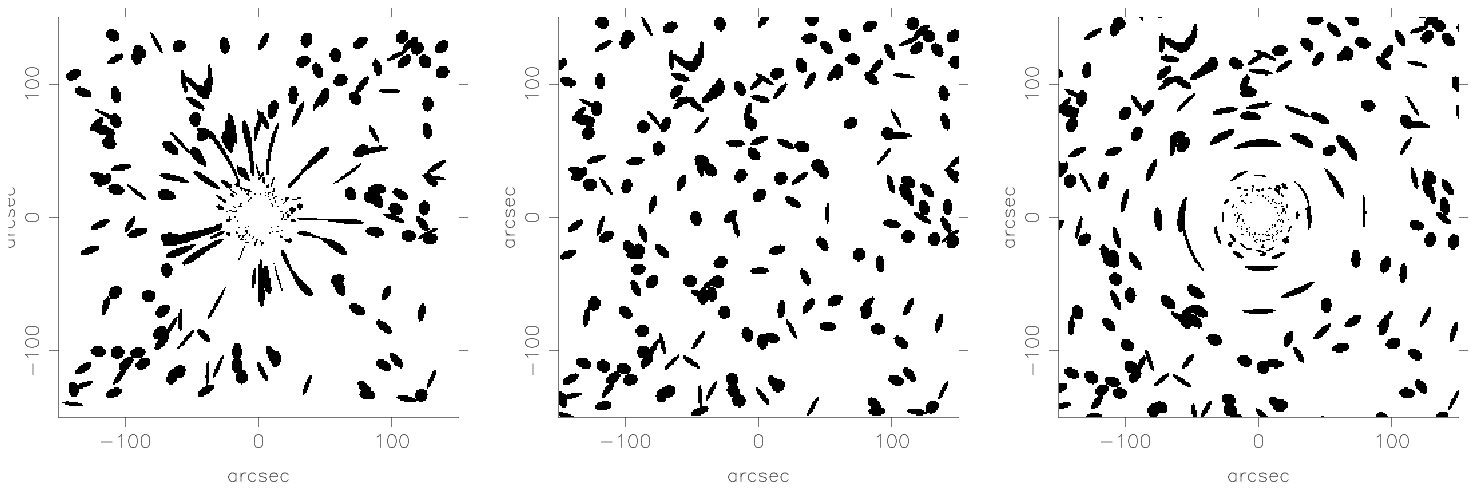, width=1.\textwidth}}}
\vskip 0.6in
\caption[Appearance of a background field of sources at a fixed redshift,
lensed by a $1 \times 10^{14}\,M_{\odot}$ lens]{{\it Left}:
Appearance of a background field of sources (200 galaxies), each of them 5 Kpc
radius, when it is lensed by a negative mass of $|M|_{\rm lens} = 1 \times
10^{14}\,M_{\odot}$ with an Einstein angular radius equal to
$\theta_{\rm E}=47$ arcsec. {\it Center}: Unlensed background field.
{\it Right}: Appearance of the same background field of galaxies when
lensed by an equal amount of positive mass, located at the same
redshift (see text).} \label{fig:void1}
\end{figure}

\begin{figure}
\centerline{
\vspace{-1.5 cm}
\hspace{-0.5 cm}
\makebox{
\epsfig{figure=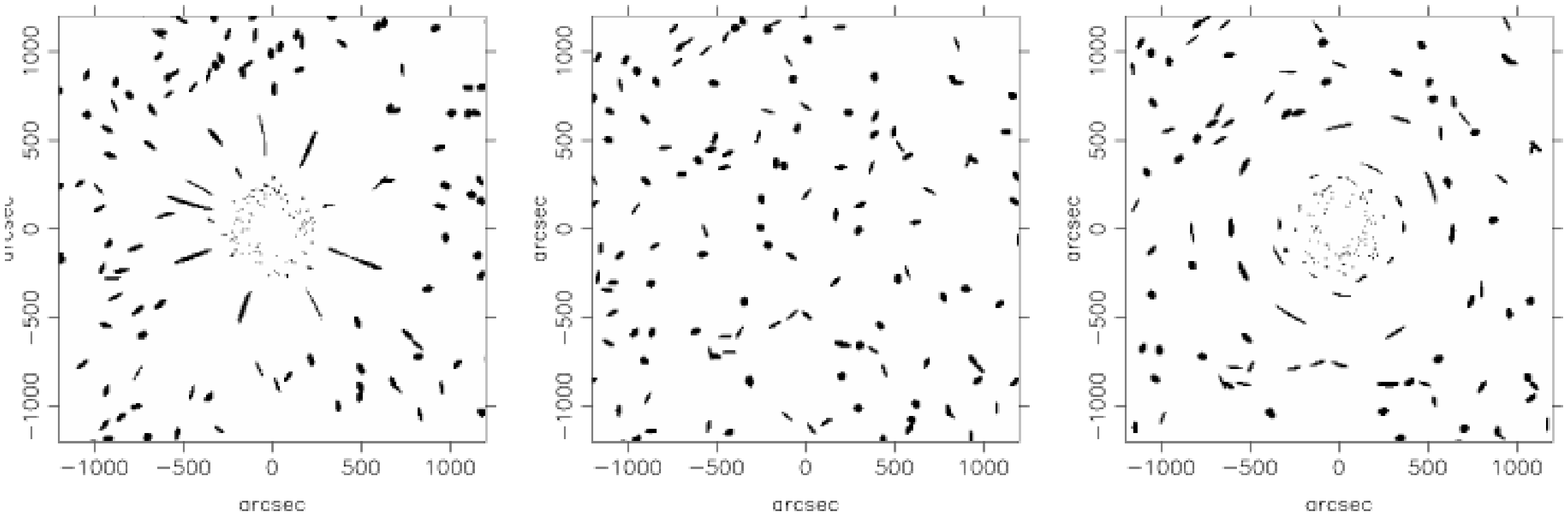, width=1.\textwidth}}}
\vskip 0.6in
\caption[Appearance of a background field of sources at a fixed redshift,
lensed by a $1 \times 10^{16}\,M_{\odot}$ lens]{{\it Left}: Appearance of a
background field of sources (300 galaxies), each of them of 25 Kpc radius,
when it is lensed by a negative mass of $|M|_{\rm lens} = 1\times
10^{16}\,M_{\odot}$ with an Einstein angular radius equal to
$\theta_{\rm E}=467$ arcsec. {\it Center}: As in Figure~\ref{fig:void1}. {\it
Right}: As in Figure~\ref{fig:void2}.} \label{fig:void2}
\end{figure}

In order to explore the influence of the adopted redshift values, we turn now
to the case where $z_{\rm sources} = 0.08$ and $z_{\rm lens} = 0.05$. The
Bootes void is the closest void to us, and lies between the supercluster
Corona Borealis ($z \approx 0.08$) and Hercules ($z \approx 0.03$) \cite{4}.
This serves as motivation for the selection of these redshift values. As an
example of the results for different lens masses, we show in
Figs.~\ref{fig:void1} and~\ref{fig:void2} the cases with $|M|_{\rm lens} =
1\times 10^{14}\,M_{\odot}$ and $|M|_{\rm lens} = 1\times 10^{16}\,M_{\odot}$.

It is interesting to note that since the background population of galaxies
is very dense, one would expect a lot of lensing in the standard
model of cosmology. However, there is still a surprising dearth of
candidates for (positive mass) lensed sources \cite{Cooray99}. Some of the
richest clusters do not display arcs in the deepest images. CL0016+16
($z=0.56$), for instance, is one of the richest and strongest X-ray emitting
clusters. It is rather extended on the sky and the light from many background
sources should cross this cluster. Neither arcs nor arclets have been found,
though weak lensing has been reported. This may be pointing towards a
cautionary note: if the kind of finger-like structures displayed in our
figures is not directly seen in its full pattern, that does not necessarily
mean that they are absent. Even the presence of one radial arc (without
tangential counter arc and/or tangential arcs) may be significant.

\subsection{Concluding remarks}

The null EC (NEC) is the weakest of the EC. Usually, it was considered that
all reasonable forms of matter should at least satisfy the NEC. However, even
the NEC and its averaged version (ANEC) are violated by quantum effects and
semi-classical quantum gravity (quantized matter fields in a classical
gravitational background). Moreover, it has recently been shown that there are
also large classical violations of the energy conditions \cite{ULT}. Here we
have shown that disregarding the fundamental mechanism by which the EC are
violated (e.g. fundamental scalar fields, modified gravitational theories,
etc.), if large localized violations of NEC exist in our universe, we shall be
able to detect them through cosmological macrolensing. Contrary to the usual
case, where arc-like structures are expected, finger-like ``runaway"
filaments and a central void appear. In Figures~\ref{fig:13}-\ref{fig:17}
and~ \ref{fig:void1}-\ref{fig:void2} we compare the effects of negative
masses with the case of macrolensing effects on background fields produced by
equal amounts of positive mass located at the same redshift. The differences
are obvious. These results make the cosmological macrolensing produced by
matter violating the null energy condition observationally distinguishable
from the standard situation. Whether large-scale violations of the EC,
resulting in space-time regions with average negative energy density indeed
exist in the universe can therefore be decided through observations.

\chapter{Gravitational Lensing as a Tool to Test Alternative
Cosmologies} 
\section{Introduction}

Standard cold dark matter FRW cosmology presents serious theoretical and 
observational difficulties as an acceptable description
of the Universe. For example, a study conducted on a sample of 256
ultra-compact sources \cite{jackson96} showed that the canonical CDM
model ($\O_0=1$, $\O_{\L}=0$) is ruled out at the 98.5\% confidence level. An
overview in the literature shows the existence of a growing body of work
discussing alternative cosmologies \cite{alternative}. The first motivation
comes from the conflict between the age of the Universe and the age of the
oldest stars in Galactic globular clusters. The ages of the globular clusters
typically fall in the interval $12\,-\,14$ Gyr (\cite{Alcaniz}, and
references within), while measurements of the Hubble parameter were recently
updated to $h=0.7 \pm 0.1$ \cite{Freedman00} with the value for
the age in the interval 8.1 Gyr $\le t_0\le$ 10.8 Gyr. ``The age problem'' is
even more acute if we consider its variant based on the age constraints from
old galaxies at high redshifts \cite{Krauss97}. 

Another important motivation is the cosmological constant problem.
Historically,  it was Einstein who introduced the $\L$ term in his field
equations in order to make them compatible with a static universe. The $\L$
terms was introduced several times in cosmology but was discarded when
improved data became available. Recently, observations by Perlmutter et al
\cite{perl99} and Riess et al \cite{riess} of more than 50 type Ia supernovae
suggested Friedmann models with negative pressure matter, such as a
cosmological constant, domain walls or cosmic strings
\cite{Vilenkin,Garnavich98}. The main conclusion of these works was that the
expansion of the universe is accelerating. Though a cosmic acceleration can
also be accounted for by invoking inhomogeneity (though at a cost of the
cosmological principle \cite{pascual,Dabrowski99}), a $\L$-dominated model was
revived again. The cosmological constant solves a lot of problems at once (for
ex., \cite{vishwakarma}). It supplies the `missing matter' required to make
$\O_{\rm tot}=1$, it modifies CDM by putting more power on large scales, as is
compatible with the CMBR anisotropy limits. It also removes the inconsistency
between the age of the universe and that of the globular clusters for larger
values of $H_0$. However, we face
the problem that the upper limit of $\L$ from observations ($\L \lapprox
10^{-56}$ cm$^{-2}$) is120 orders of magnitude below the value for the vacuum
energy density predicted by quantum field theory \cite{Weinberg89,Caroll92}.
(It is customary to associate a positive cosmological constant $\L$ with a
vacuum density $\r_{\rm v} \equiv \L/8\pi G$.) According to gravitational
lensing statistics, a universe with the a large cosmological constant should
have more multiple-image systems  than are actually observed. It was shown,
for example, that HDF data may be incompatible with large $\L$
\cite{Cooray99}. 

These (and other) problems generated a lot of interest in an open FRW model
with linear evolution of the scale factor, $a(t) \propto t$. This universe
expands with a constant speed, hence the term coasting cosmology. Notable
among such models is a recent idea of Allen \cite{allen}, in which  such a
scaling results in an $SU(2)$ cosmological instanton dominated  universe. Yet
another possibility derives from the Weyl gravity theory of Manheim and Kazanas
\cite{mann}. Here again the FRW scale factor appoaches a linear evolution at
late times.   

There are several motivations for investigating such models. Many of the
problems of the standard model are naturally resolved in such a cosmology. Such
a cosmology does not suffer from the horizon problem. Horizons occur in models
with $a(t) \approx t^{\a}$ for $\a <1$. Linear evolution of a scale factor is
supported in alternative gravity theories (eg. non-minimally coupled
scalar-tensor theories), where it turns out to be independent of the matter
equation of state. The scale factor in such theories does not constrain the
matter density parameter, thereby curing the flatness problem. The age
estimate in a coasting universe, deduced from the measurement of the Hubble
parameter, is given by $t_0=1/H_0$. This is 50\% greater than the age in
standard CDM cosmology, thus making it comfortably concordant with the ages of
globular clusters. Finally, a linear coasting cosmology independent of the
equation of state of matter, is a generic feature in a class of models that
attempt to dynamically solve the $\L$ problem \cite{models}. Such models have
a scalar field non-minimally coupled to the large scale curvature of the
universe. With the evolution of time, the non-minimal coupling diverges, the
scale factor quickly approaches linearity and the non-minimally coupled field
acquires a stress energy that cancels the vacuum energy in the theory. This
model is consistent with nucleosynthesis in the early universe \cite{annu} and
presents a good fit to the current SNE Ia data \cite{abha}.

After the discovery of the first multipy imaged quasars, gravitational arcs
and arclets, gravitational lensing has rapidly become one of the  most
promising tools for cosmology. It is now well known that gravitational lensing
is a useful probe of the geometry of the universe. With the rapid growth of
the number of lensed systems, proposals were made to apply the statistical
analysis to the samples of lenses in order to extract the cosmological
parameters. The use of gravitational lensing statistics as a cosmological tool
was first considered in detail by Turner et al. \cite{TOG} and Fukugita et al
\cite{fukugita}. More recently, Kochanek \cite{K96} and Falco et al
\cite{Falco98} have laid the groundwork for using gravitational lensing
statistics for a detailed analysis of the extragalactic surveys. It has been
pointed out \cite{fukugita,turner90} that the expected frequency of multiple
imaging lensing events for high-redshift sources is sensitive to cosmology.
In view of the successfull results of the above mentioned works
\cite{annu,abha}, it is tempting to use this test to constrain the power
index of the scale factor of a general power law cosmology, $a(t) \propto
t^{\alpha}$. In the next section we consider constraints on this index. The 
expected number of lens systems depends upon the index $\alpha$ through the
angular diameter distances. By varying $\alpha$, the number of lenses changes
and on comparison with the observations we obtain constraints on
$\alpha$.  
  
One of the remarkable results of field theory is the existence of stable
classical states of a field with non-vanishing energy. Such fields include
topological solitons, domain walls, strings and monopoles. In addition to
these field configurations, which are stabilized by their topological
properties, non-topological solitons (NTS) produced by scalar fields have
appeared in the literature and their relevance to cosmology has been assessed
(for example, \cite{gelmini}). Variations on this theme include cosmic
neutrino balls \cite{holdom,dolgov}, Q-balls \cite{coleman} and soliton stars
\cite{lee-pang}. NTS are rendered stable by the existence of a conserved
Noether charge carried by the fields confined to a finite region of space. The
theory essentially contains an additive quantum number $N$, carried either by
a spin-1/2 field $\psi$ (for example, fermion number), or a spin-0 complex
field $\varphi$. In addition there is a scalar field $\sigma$ whose coupling
gives  $\varphi$ or $\psi$ a mass. The soliton contains the interior in which
$\sigma \approx \sigma_0$, surrounded by a shell of width $\approx \mu^{-1}$,
over which $\sigma$ changes from $\sigma_0$ to $0$. $\mu$ is the mass
associated with $\sigma$.  $\sigma=0$ and $\sigma=\sigma_0$ are two minima of
the effective potential of the scalar field. The $N$-carrying field $\psi$, or
$\varphi$, is confined to the interior where it is effectively massless at the
local minimum of the potential, $\sigma=\sigma_0$. At the global minumum,
$\sigma=0$, the field ($\psi$ or $\varphi$) has a non-vanishing mass, $m$.
This leads to a stable configuration of massless particles trapped inside a
region with $\sigma=\sigma_0$ separated from the true vacuum $\sigma=0$ by a
wall of thickness $\approx \mu^{-1}$. Confinement occurs for all particle
states that are not on-shell in the exterior region. The NTS is stable as
long as the kinetic energy $E_k$ of the particles inside the bag is less
than $E_{\rm free}$---the minimum on-shell energy in the exterior region. This
was explicitly demonstrated by Lee and Wick \cite{lee-wick,friedberg}.

The role of scalar fields in effective gravity models stems from the
classical work of Brans and Dicke. In these approaches the gravitational
action is induced by a coupling of the scalar curvature with a function of a
scalar field. Lee and Wick's results can be carried over to a curved spacetime
with a scalar curvature $R$ non-minimally coupled to $\sigma$ through
arbitrary function $U(\s)$  in the class of theories described by the
effective action 
\be  
S = \int \sqrt{-g} \,d^4x \left [ U( \s)R + \fr{1}{2}
(\del_{ \mu} \s)^2 -V( \s) +  L_{\rm M} \right ]\,\,.
\end{equation}
Here $V( \s)$ is effective potential of a scalar field, $L_{\rm M}$ is the
matter field(s) Lagrangian which includes a Higgs coupling of $\s$ to a
fermion. Establishing $\sigma_{\rm in}$ and $\sigma_{\rm out}$ as the
interior and exterior values of $\sigma$, the effective gravitational constant
would be given by $$
G_{\rm eff}^{\rm in} = [U(\sigma_{\rm in}]^{-1}
$$
and
$$
G_{\rm eff}^{\rm out} = [U(\sigma_{\rm out}]^{-1}
$$
If $U$ goes to infinity at some point, defined as $\sigma = 0$ without
loss of generality, the theory gives rise to a solution with a spatial
variation of the effective gravitational constant, $G_{\rm eff}^{\rm out}=0$
and $G_{\rm eff}^{\rm in}=$constant. This would be a generic feature of a
Lee-Wick solution in which the scalar field is non-minimally coupled to the
scalar curvature. It would give rise to a stable ``ball of gravity''. 

A gravity ball is characterised by an effective refractive index in its
interior. This would cause bending of light incident on it. The G-ball thus
acts as a spherical lens. In the last section we investigate gravitational
lensing of an empty gravity ball situated at a cosmological distance. The
lens---gravity ball---has interesting features which are not shared by other
known lenses. In the last section we examine gravitational lesning properties
of this special kind of NTS solutions referred to as G-balls \cite{dakshmeetu}.

\section{Gravitational lensing statistics as a test for coasting cosmology}
\subsection{Linear coasting cosmology}

We consider a general power law cosmology with the scale factor
given in terms of two parameters
\be
a(t)=B\fr{c}{H_0}\left(\fr{t}{t_0}\right)^{\a}\,\,,
\label{eq:ansatz}
\end{equation}
for an open FRW metric. However, the light propagates through the
inhomogeneous rather than the averaged smooth spacetime. The light ray feels
the local metric which deviates from the smooth FRW metric---FRW metric
excludes gravitational lensing in principle. We may assume, however, that the
relation between the affine distance of the null geodesic and the redshift of
the source object in a clumpy universe is the same as in the FRW cosmology.
This may be regarded as a mathematical expression of the assumption that the
universe is described by FRW geometry on large scales (for the metric
see Eq.~\ref{eq:frw}).

\subsubsection{Dimensionless Hubble parameter}

The expansion rate of the universe is described by a Hubble parameter, $H =
\dot{a}/a$. Using (\ref{eq:ansatz}), \be
H(t)=\fr{\a}{t}\,\,.
\label{eq:exact}
\end{equation}
The {\it present} expansion rate of the universe is defined by a Hubble {\it
constant}, equal in our model to $H_0=\a/t_0$ (here and subsequently the
subscript 0 on a parameter refers to its present value). With the usual
definition of the redshift $z$
\be
\frac{a_0(z)}{a(z)} = 1+z
\label{eq:redshift}
\end{equation}
we obtain $t_0/t = (1+z)^{1/\a}$, and thus
\be 
H(z) = H_0 (1+z)^{1/\a}\,\,.
\end{equation}
The dimensionless Hubble parameter is
\be
h(z)=\frac{H(z)}{H_0}=(1+z)^{1/\a}.
\label{eq:dimensionless_hubble}
\end{equation}

\subsubsection{Present-day scale factor}

We define the present `radius' of the universe (\ref{eq:ansatz}) as
\be
a_0 =B\fr{c}{H_0}\,\,.
\label{eq:a_0}
\end{equation}

\subsubsection{Angular diameter distance }
There are many ways to define a distance in cosmology---parallax distance, 
angular diameter distance, luminosity distance, proper motion distance, etc
(see~\ref{sec:distance}). It is the angular diameter distance that is relevant
to the angular separation of images. 
 
Consider a source at $r=r_1$ which emitted light at $t=t_1$. The
observed angular diameter of the source, $\delta$, is related to
the proper diameter of the source, D, by
\be
\delta = \fr{D}{a(t_1)\,r_1}\,\,.
\end{equation}
The angular diameter distance $\rm D_A$ is defined to be 
\be
{\rm D_A}(z) = \fr{D}{\d}= a(t_1)\,r_1 = \frac{a_0 r_1}{1+z},
\label{eq:D_A_definition}
\end{equation}
where $r_1(z)$ is the coordinate distance to a redshift $z$. Angular
diameter distance of a source at $z_2$ measured by an (fictitious) observer at
$z_1$ is given by \be
D_{\rm A}(z_1,z_2) = \fr{a_0 r_{12}}{1+z_2}\,\,.
\label{eq:z_12}
\end{equation}
Light travels on
null geodesics, $ds^2=0$. If a comoving observer is at coordinates 
$(r_0,\theta_0,\phi_0)$ at time $t_0$, the geodesics intersecting 
$r_0=0$ are lines of constant $\t$ and $\phi$. Geodesic equation becomes
\be
0=cdt^2-a^2(t) \frac{dr^2}{(1+r^2)}
\end{equation}
and the light signal emitted from coordinates $(r_1,\t,\phi)$ at time
$t$ will reach the observer at time $t_0$ determined by
\be
\int_{t}^{t_0} \frac{c\,dt'}{a(t')} = \int^{r_1}_0
\frac{dr}{\sqrt{1+r^2}}
\label{eq:over_r}
\end{equation}
We can convert the integrals over $t$ into integrals over $z$ by
differentiating (\ref{eq:redshift}),  
$$
\frac{dz}{dt} = -\frac{a_0}{a^2} 
\dot{a}=-\frac{a_0}{a}\frac{\dot{a}}{a} = -(1+z)H(z)\,\,,
$$
to obtain
\be
dt=-\frac{dz}{(1+z) H(z)}\,\,.
\label{eq:dt}
\end{equation}
Thus, equation (\ref{eq:over_r}) becomes:
\be
\int_{0}^{z}\frac{c\,dz}{H(z)(1+z) a} = \int_{0}^{r_1}
\frac{dr}{\sqrt{1+r^2}}
\end{equation}
Since $a_0/a=1+z$, we use it to obtain
\be
\frac{c}{a_0}\int_{0}^{z_1}\frac{a_0\,dz}{H(z)(1+z) a} =
\frac{c}{a_0}
\int_{0}^{z_1}\frac{dz(1+z)}{H(z)(1+z)}=\frac{c}{a_0}\int_{0}^{z_1}
\frac{dz}{H(z)}\,\,,
\end{equation}
or, in terms of a dimensionless Hubble parameter $h(z)$,
Eq.~\ref{eq:over_r}
becomes
$$
\frac{c}{H_0 a_0}\int_{0}^{z_1}\frac{dz}{h(z)} =
\int_{0}^{r_1}\frac{dr}{\sqrt{1+r^2}}\,\,.
$$

\SS{Comoving Coordinate Distance $r$}

For an open universe, $k=-1$, we obtain
$$
\frac{c}{H_0 a_0}\int_{0}^{z_1}\frac{dz}{h(z)} = \sinh^{-1}{r_1}\,\,.
$$
Thus, 
$$
r_1=\sinh{\left[\fr{c}{H_0a_0}\int_0^{z_1}\fr{dz}{h(z)}\right]}\,\,.
$$
 
\SS{$D_{\rm A}$ formulae}
If we take the value of $a_0$ from Eq.~\ref{eq:a_0}, 
we obtain
\be
r_1=\sinh{\left[\fr{1}{B}\int_0^{z_1}\fr{dz}{h(z)}\right]}\,.
\end{equation}
Then,
\be
D_{\rm A}(z,\a)= \fr{Bc}{(1+z)H_0}\sinh{\left[ 
\fr{1}{B} \int_0^{z_1} \fr{dz}{h(z)}\right]}\,,
\end{equation} 
with the integral 
$$
\int_0^{z_1}\fr{dz}{h(z)} = \fr{\a}{\a -1} \left\{(1+z)^{\fr{\a -1}{\a}}-
1^{\fr{\a -1}{\a}} \right\}\,\,.
$$
Finally, the distance formula is
\be
D_{\rm A}(z,\a) = \fr{Bc}{(1+z)H_0}
\sinh{\left[ \fr{1}{B}\frac{\a}{\a-1} 
\left\{ (1+z)^{\frac{\a-1}{\a}} - 1^{\frac{\a-1}{\a}} \right\}
\right]}\,\,.
\label{eq:D_A}
\end{equation}
The distance between two objects at different redshifts is:
\be
D_{\rm A}(z_1,z_2 ,\a) = \fr{Bc}{(1+z_2) H_0}
\sinh{\left[ \fr{1}{B}\frac{\a}{\a-1}
\left\{ (1+z_2)^{\frac{\a-1}{\a}} - (1+z_1)^{\frac{\a-1}{\a}} 
\right\} \right]}\,\,.
\end{equation} 

\SS{Limiting ($\a \rightarrow 1$) formulae}

We know that  $h(z)=(1+z)^{1/\a}$, with $\a \rightarrow 1$, thus,
$h(z)=(1+z)$,
then the integral 
\be
\int_{0}^{z}\fr{dz'}{1+z'} = \ln{(1+z)}\,\,,
\end{equation}
and the Eq.~\ref{eq:D_A} reduces to
\be
{\rm D_A}(z) = \fr{Bc}{H_0} \fr{1}{1+z}
\sinh{\left[\fr{1}{B}\ln{(1+z)}\right]}=
\fr{Bc\left[(1+z)^{2/B} - 1\right]}{2\,H_0(1+z)^{\fr{B+1}{B}}}\,\,.
\end{equation}
The distance between two objects at different redshifts is, accordingly,
\be
{\rm D_A}(z_1,z_2) = \fr{Bc}{2 H_0}\fr{\left[(1+z_2)^{2/B} -
(1+z_1)^{2/B} \right]}{(1+z_1)^{1/B}(1+z_2)^{\fr{B+1}{B}}}\,\,.
\end{equation}

\subsubsection{Look-back time}

The look-back time, $\Delta t=|t-t_0|$, is the difference between the age of
the universe when a particluar light ray was emitted and the age of the
universe now: \be
\Delta t(a;a_0)=\int_{a}^{a_0}\fr{da}{\dot{a}(a)}\,\,\,.
\end{equation}
We can obtain the expression for the look-back time  from Eq.~\ref{eq:dt} and
Eq.~\ref{eq:dimensionless_hubble}. Since we define $t_0$ as zero and time is
increasing  as we look back, we drop the negative sign in that expression.
Thus,
\be
\fr{dt}{dz}=\fr{1}{H_0 (1+z)^{\fr{\a +1}{\a}}}\,\,,
\end{equation}
and finally the expression we seek is:
\be
\fr{c\,dt}{dz_{\rm L}}=\fr{c}{H_0 (1+z_{\rm L})^{\fr{\a
+1}{\a}}}\,\,. 
\label{eq:backtime}
\end{equation}    

\subsection{Basic equations of gravitational lensing statistics}
Given a lens model, the number counts of galaxies--lenses and their
properties, we compute the probability $p_i$ that quasar $i$ is lensed and
the probability $p_i(\D\t_i)$ that quasar $i$ is lensed and has image
separation $\D\t_i$. We shall consider only early types of galaxies,
elliptical and lenticulars, E/S0, neglecting the contribution of spirals as
lenses. This is because the velocity dispersion $v$ of spirals is small
compared to E/S0 galaxies, outweighing the larger number density. The numbers
of E/S0 and spirals are roughly 30\% and 70\%, respectively, whereas in our
adopted sample the velocity dispersion of ellipticals is about 200 km s$^{-1}$
and that of spirals is about 130 km s$^{-1}$ \cite{Loveday}. The calculations
for the lensing statistics depend strongly on this quantity, since the lensing
optical depth goes as the fourth power of the velocity dispersion (see
Eq.~\ref{eq:F*}).  

It is increasingly clear \cite{K96} that E/S0 galaxies are effectively
singular. The almost uniform absence of central images in the observed lenses
and models of individual lenses imply core radius smaller than 100 $h^{-1}$
pc and essentially suggest that the galaxy lenses are nearly singular
\cite{Wallington93,K95}. We will model the lensing galaxies as singular ($=$
zero core radius) isothermal spheres (SIS). The detailed lensing properties of
singular isothermal spheres are described in Section~\ref{sec:SIS}. Here we
simply state the most important relations for the present calculations. A SIS
lens with a velocity dispersion $v$ at a redshift $z_{\rm L}$ will produce two
images of a quasar at a redshift $z_{\rm S}$ separated by an angle $\D\t$ 
\be
\D\t = 2 \a_{\rm d} \fr{D_{\rm LS}}{D_{\rm S}}\,\,,
\label{eq:separation}
\end{equation}
if angular position of the source is less than the critical angle $\b_{\rm
cr} \equiv \a_{\rm d} D_{\rm LS}/D_{\rm S}$. The deflection angle, $\a_{\rm
d}$, is given for all impact parameters as:
\begin{equation}
\a_{\rm d} = 4\pi \left(\fr{v}{c}\right)^2\,\,.
\label{eq:defangle}
\end{equation}
The angular separation of the two images is independent of the impact parameter
as long as it is small enough to produce two images. The critical impact
parameter is defined by $a_{\rm cr} \,\equiv \, D_{\rm L} \b_{\rm cr}$ or, with
the help of the previous definitions,
\begin{equation}
a_{\rm cr}=4\pi \left( \fr{v}{c} \right)^2 \fr{D_{\rm L}D_{\rm LS}}{D_{\rm S}}\,\,.
\end{equation}
The quasars are treated as point sources of radiation. We assume that
no evolution of galaxies or quasars with cosmic time (quasar evolution is
irrelevant, since the quasar redshift distribution will be drawn from observed
samples). In other words, the comoving number density of lenses is 
conserved
\be
n_{\rm L} = n_0 (1+z_{\rm L})^3\,\,,
\label{eq:n_L}
\end{equation} 
where $n_0$ is an average comoving density measured at the present epoch.
Merging between galaxies and infall of surrounding mass onto galaxies are two
posible processes that can change the comoving density of galaxies and/or
their mass. Under the generic relation between the velocity dispersion and
mass of early-type galaxies, Rix et al \cite{rix} and Mao \& Kochanek
\cite{mao} found that merging and/or evolution does not significantly change
the statistics of lensing. 
 
\subsubsection{Cross-section and optical depth of lensing.}

The cross-section $\s$ for "strong" lensing events is given by$\s = \pi
a_{\rm cr}^2$. Using (\ref{eq:defangle}) and the definitions from the
previous subsection, we write the cross-section as
\be
\s = 16 \pi^3 \left(\fr{v}{c}\right)^4 \left(\fr{D_{\rm L}D_{\rm LS}}{D_{\rm S}}
\right)^2\,\,.
\end{equation} 
The cross-section is largest approximately halfway in the distance between
the source and the observer; and $\s$ vanishes at the two endpoints. To find
the average effect on an image passing within  $a_{\rm cr}$ of a mass
(``scoring a hit"), we average over the cross-section to find the mean image
separation as \be
\overline{\D \t} = 8\pi \left(\fr{v}{c}\right)^2 \fr{D_{\rm LS}}{D_{\rm S}}\,\,.
\end{equation}
The differential probability (or the optical depth) $d\tau$ that a line of sight
intersects a galaxy-lens at $z_{\rm L}$ in traversing the path of $dz_{\rm
L}$ from a population with number density $n_{\rm L}$ is given by a  ratio of
differential light travel distance $cdt$ to its mean free path between
successive encounters with galaxies, $1/n_{\rm L}(z)\s$, 
\be
d\tau=n_{\rm L}(z)\s\fr{cdt}{dz_{\rm L}}dz_{\rm L}\,\,.
\end{equation}
For our case $n_{\rm L}(z)$ is given by (\ref{eq:n_L}) and the quantity 
$ cdt/dz_{\rm L}$ is obtained in the previous section (Eq.~\ref{eq:backtime}). 
Substituting for $\s$ and $n_{\rm L}(z)$, we get
\be
d\tau = \fr{16\pi^3}{c^4}(1+z_{\rm L})^3 \la n_0v^4 \ra 
\left(\fr{D_{\rm L} D_{\rm LS}}{D_{\rm S}}\right)^2\fr{cdt}
{dz_{\rm L}}dz_{\rm L}\,\,.
\label{eq:dtau}
\end{equation}

Assuming that the brightness distribution of galaxies at any given redshift is 
described by a Schechter function, the comoving density of galaxies at
redshift $z$ and with luminosity between $L$ and $L+dL$ is
\be
\P(L,z) dL = n_{*}(z) \left( \fr{L}{L_{*}(z)} \right)^{\hat \a} 
\exp{\left( -\fr{L}{L_{*}(z)} \right)} \fr{dL}{L_{*}(z)}\,\,.
\label{eq:lumfunz}
\end{equation}
The parameter $n_{*}(z)$ is the average comoving density, 
$L_{*}(z)$ is the characteristic luminosity at which the luminosity function
exhibits a rapid change in the slope in the $\log{n},\,\log{L}$)-plane and
$\hat \a$  gives the slope of the luminosity function in the
($\log{n},\,\log{L}$)-plane, when $L \ll L_{*}$. The comoving number density
of galaxies, characteristic luminosity and mass of a galaxy at any redshift
remain constant, therefore, $n_{*}(z) = n_{*}(0) = \mbox{constant}$ and
$L_{*}(z) = L_{*}(0) = \mbox{constant}$. ``0" refers to present-day values.
This is the most commonly used luminosity function for early type galaxies.
Thus, Eq.~\ref{eq:lumfunz} becomes \be
\P(L,z=0)\,dL = n_{*} \left( \fr{L}{L_{*}} \right)^{\hat \a} 
\exp{\left( -\fr{L}{L_{*}} \right)} \,\fr{dL}{L_{*}}\,\,.\,
\label{eq:lumfun}
\end{equation}
where $n_{*}$, $\hat \a$ and $L_{*}$ are the normalization factor, index of the
faint-end-slope and the characteristic luminosity at the present epoch, 
respectively. These values are fixed in order to fit the current luminosities and
densities of galaxies. From Eq.~\ref{eq:dtau} and Eq.~\ref{eq:lumfun} we have
\be
\la n_0 v^4 \ra =  v_{*}^4 n_{*} \int_0^{\infty}
\left( \fr{L}{L_{*}} \right)^{\hat \a} \exp{\left( -\fr{L}{L_{*}} \right)}\, 
\fr{dL}{L_{*}} \left( \fr{v}{v_{*}} \right)^4 \,.
\label{eq:n_0}
\end{equation}  
    We assume the velocity dispersion $v$ is related to the luminosiy $L$ by
the empirical Faber-Jackson relation for E/S0 galaxies:
\be
\left(\fr{L}{L_{*}}\right) = \left( \fr{v}{v_{*}}\right)^{\g}\,\,.
\label{eq:F-J}
\end{equation}
Therefore, Eq.~(\ref{eq:n_0}) becomes 
\be
\la n_0 v^4 \ra =  v_{*}^4 n_{*} \int_0^{\infty}
\left( \fr{L}{L_{*}} \right)^{4\hat \a +\g} \exp{\left( -\fr{L}{L_{*}}
\right)}  \fr{dL}{L_{*}} \,\,.
\end{equation} 
By integrating it over the luminosity function at redshift $z$, we obtain the 
differential probability 
\be 
d\tau=\fr{16 \pi^3}{c^4} (1+z_{\rm L})^3 n_{*}v_{*}^4 \G \left(
\hat \a +\fr{4}{\g} +1 \right) \left(\fr{D_{\rm L} D_{\rm LS}}{D_{\rm
S}}\right)^2 \fr{cdt}{dz_{\rm L}}dz_{\rm L}\,\,
\end{equation}
where $\G$ is the normal gamma function. If we define the `dimensionless' image
splitting as $\phi = \D \t/8\pi \left(v_*/c\right)^2$, we can find the
differential optical depth of lensing in traversing $dz_{\rm L}$ with the
angular separation between $\phi$ and $\phi +d\phi$ as $\fr{d^2\tau}{dz_{\rm
L}d\phi} d\phi dz_{\rm L}$. Using luminosity-velocity relation
(Eq.~\ref{eq:F-J}), we obtain the relation \be \fr{L}{L_{*}}=\left(\fr{D_{\rm
S}}{D_{\rm LS}}\phi \right)^{\g/2}\,\,. \end{equation}
Taking the differential,
\be
\fr{dL}{L_{*}}=\fr{\g}{2}\left(\fr{D_{\rm S}}{D_{\rm LS}}\phi\right)^{\g/2}
\fr{d\phi}{\phi}\,\,.
\label{eq:lum-vel}
\end{equation}
Returning to the Eq.~\ref{eq:dtau}), we write
$d\tau/dz_{\rm L}$ as \be
\fr{d\tau}{dz_{\rm L}}=n_{\rm L}(z)\fr{16\pi^3}{c^4}v^4\left(\fr{D_{\rm L}
D_{\rm LS}}{D_{\rm S}}\right)^2\fr{cdt}{dz_{\rm L}}\,\,.
\end{equation}
The differential optical depth of lensing in traversing $dz_{\rm L}$ with
the angular separation between $\phi$ and $\phi +d\phi$ is
\berr
&&\hskip -0.8in \int_0^{\infty}\fr{d^2\tau}{dz_{\rm L}d\phi}d\phi= \no \\
&& \\
&& n_{*}v_{*}^4 \int_0^{\infty}\left(\fr{L}{L_{*}}\right)^{\hat \a}
\exp{\left(-\fr{L}{L_{*}}\right)}\fr{dL}{L_{*}}\left(\fr{L}{L_{*}}\right)^{4/\g}
\left(1+z_{\rm L}\right)^3\fr{16\pi^3}{c^4} 
\left(\fr{D_{\rm L}D_{\rm LS}}{D_{\rm S}}\right)^2 \fr{cdt}{dz_{\rm L}}\,\,.\no
\err
Using the relations (Eq.~\ref{eq:lum-vel}), we obtain
\berr
\hspace{-1 cm}\lefteqn{\fr{d^2\tau}{dz_{\rm L}d\phi}d\phi dz_{\rm L}=}\no \\
&& \\
&&\hspace{-0.7 cm}F^{*}(1+z_{\rm L})^3\fr{H_0}{c}
\fr{\g/2}{\G \left(\hat \a +1+\fr{4}{\g}\right)}\fr{cdt}{dz_{\rm L}}
\left(\fr{H_0 D_{\rm L}D_{\rm LS}}{c\,\, D_{\rm S}}\right)^2
\left(\fr{D_{\rm S}}{D_{\rm LS}}\phi\right)^{\fr{\g}{2}(\hat \a +1+\fr{4}{\g})}
\exp{\left[\left(-\fr{D_{\rm S}}{D_{\rm LS}}\phi \right)^{\fr{\g}{2}}\right]}
\fr{d\phi}{\phi}dz_{\rm L}\,. \no
\label{eq:d2tau}
\err
Here we have introduced the useful dimensionless quantity $F^*$ which
measures the effectiveness of matter in producing multiple images \cite{TOG}
\begin{equation}
F^{*} \equiv \fr{16\pi^3}{cH_0^3}n_*v_*^4\G \left(\hat \a
+1+\fr{4}{\g}\right)\,\,. \label{eq:F*}
\end{equation}

\subsubsection{Lens and source parameters.}   
The variables relating galaxy numbers counts and the isothermal lens model
are the number density of E/S0 galaxies $n_{*}$, the Schechter function slope 
$\hat \a$, the Faber-Jackson exponent $\g$ and the velocity dispersion of the
dark matter for an $L_{*}$ galaxy $v_{*}$. The Schechter function slope
$\hat \a$ controls the relative number of low- and high-mass galaxies. We use
the following list of lens parameters for our calculations \cite{Loveday}:
\vspace{0.2in}
\begin{center}
\begin{tabular}{|c|ccccc|}\hline\hline
$Survey$ & $ \alpha$ & $ \gamma$ & $ v^{*} (Km /s)$ & $\phi^{*}
(Mpc^{-3})$ & $ F^{*}$
\\ \hline
\hline
$LPEM$ & $+ 0.2$ & $4.0$ &$205$ &$3.2 \times 10^{-3}$& $0.010$   \\
\hline
\end{tabular}
\end{center}
\vspace{0.2in}
We consider a sample of 867 ($z > 1$) high luminosity optical quasars which
include 5 lensed quasars ($1208 +1011$,  H $1413+117$, LBQS $1009+0252$, PG
$1115+080$, $0142+100$). This sample  is taken from optical lens surveys, such
as the HST Snapshot survey, the Crampton survey, the Yee survey, Surdej
survey, the NOT Survey and the FKS survey \cite{surveys}. The lens surveys and
quasar catalogues usually use V magnitudes, so we transform $m_{V}$ to a
B-band magnitude using $B-V=0.2$ \cite{Bahcall}. 

We make two corrections to the optical depth to get the lensing probability:
magnification bias and selection function.

\paragraph{Magnification bias. }
Lensing increases the apparent brightness of a quasar causing 
over-representation of multiply imaged quasars in a flux-limited sample. This
effect is called \emph{magnification bias}. It is an enhancement of the
probability that a quasar is lensed, that is why we have to include this
correction in the probability function. The bias for a quasar at a redshift
$z$ with apparent magnitude $m$ is given as \cite{fukugita,K93,K96}
\be
B(m,A_{\rm lim},A_2,z)=\left(\fr{dN(m.z)}{dm}\right)^{-1}
\int_{A_{\rm lim}}^{A_2}\fr{dN(m_{\rm A},z)}{dm} p(A)dA\,\,.
\end{equation}
Here $A$ is the total magnification, $A_{\rm lim}$ the magnification 
at which the images have the minimum detectable flux ratio (for 
a SIS model, $A_{\rm lim}=A_0=2$), $m_{\rm A}=m+2.5\log{A}$ is the lensing
enhanced magnitude and $p\,(>A)$ is the probability distribution for 
the two images having a total magnification larger than $A$, $p(A)=8/A^3$
for a SIS model. We can allow the upper limit on magnification $A_2$ 
to be infinite, but in practice we set it to be $A_2=10^4$. For the quasar
apparent magnitude number counts we use Kochanek's broken power law
\cite{K96}  
\be
\fr{dN(m,z)}{dm} \propto \left(10^{-a(m-m_0(z))} + 
10^{-b(m-m_0(z))}\right)^{-1}\,\,,
\end{equation}
where the bright-end slope index $a$ and faint-end slope index $b$ are
constants and the break magnitude $m_0$ evolves with redshift as
$$
m_0(z)=
\left\{
\barr{lcl}
m_0+(z+1)  & \mbox{if} & z<1 \\
m_0        & \mbox{if} & 1<z<3\\
m_0-0.7(z-3) & \mbox{if} & z>3\,\,.
\earr
\right.
$$
Fitting this model to the quasar luminosity function data in \cite{Hartwich}
for $z>1$, Kochanek finds that the `best model' has $a=1.07$, $b=0.27$ and
$m_0=18.92$ at B magnitude. Thus, magnitude corrected probability $p_i$ for
the quasar $i$ with  apparent magnitude $m_i$ and redshift $z_i$ to get lensed
is $p_i=\tau(z_i) B(m_i,A_{\rm lim},A_2,z_i)$.

\vspace{-0.1in}
\paragraph{Selection function. }
Selection effects are caused by limitations on dynamic range, limitations on
resolution and presence of confusing sources such as stars. In the SIS model
the selection function is modeled by a maximum magnitude difference $\D m(\t)$
that can be detected for two images separated by $\D \t$. This is equivalent
to a limit on the flux ratio between two images $f=10^{0.4 \D m(\t)}$. The
total magnification becomes $A_f=A_0(f+1)/(f-1)$. So, the survey can only
detect lenses with magnifications greter than $A_f$. This sets the lower
limit on magnifications and in the bias function $A_{\rm lim}$ gets replaced
by $A_f$. To get selection function corrected probabilities, we divide our
sample into two parts: the ground based surveys and the HST Snapshot survey and
use the selection function for each survey as suggested by Kochanek
\cite{K93}. 

The corrected lensing probability and image separation distribution function
for a single source at redshift $z_{\rm S}$ are given as \cite{K96}
\be
p^{'}_{i}(m,z) = p_{i}\int \frac{ d(\Delta\theta)\,
p_{c}(\Delta\theta)\emph{B}(m,z,M_{f}(\Delta\theta),M_{2})}
{\emph{B}(m,z,M_{0},M_{2})},
\label{prob2}
\end{equation}
and
\be
p^{'}_{ci} =p_{ci}(\Delta\theta)\,\frac{p_{i}}{p_{i}^{'}}\,
\frac{\emph{B}(m,z,M_{f}(\Delta\theta),M_{2})}
{\emph{B}(m,z,M_{0},M_{2})}\,\,,
\label{confi}
\end{equation}
where
\be
p_{c}(\Delta\theta) =
\frac{1}{\tau(z_{S})}\,\int_{0}^{z_{S}}\frac{d^{2}\tau}{dz_{L}d(\Delta\theta)}
\,dz_{L}\,\,.
\label{pcphi}
\end{equation}
Equation (\ref{confi}) defines the configuration probability. It is the
probability that the lensed quasar $i$ is lensed with the observed image
separation.
 
\subsection{Testing the model against observations}
 
The above basic equations were used to perform the following test:
\begin{trivlist}
\item The sum of the lensing probabilities $p_{i}^{'}$ for the optical 
QSOs gives the expected number of lensed quasars, $n_{\rm L} = \sum\,
p_{i}^{'}$. The summation is over the given quasar sample. We look for those
values of the parameter for which the adopted optical sample has exactly five
lensed quasars (that is, those values of the parameters for which $n_{\rm
L}=5$). 

We started with a two parameter fit. We allowed $\a$ to vary in the range ($0.0
\leq \a \leq 2.0$) and $B$ to vary in the range ($0.5\leq B \leq 10.0$). We
observe that for $B\ge 1$, $D_A$ becomes independent of it (as is also
obvious from equation (\ref{eq:D_A})). In the previous works constraining
power law cosmology \cite{annu,abha}, the value of $B = 1$ was found to be
compatible with the observations. Incidentally, we can estimate the present
scale factor of the universe as $a_0 \approx c/H_0$.  Therefore, we used $B =
1$ in further analysis.     

Fig.~\ref{fig:NlA} shows the predicted number of lensed quasars for the above 
specified range of $\alpha$. We obtained $n_{L} = 5$ for $\a = 1.06$. We
further generated $10^{4}$ data sets using bootstrap method (see description in
\cite{abhathesis}) and found the best fit for $\a$ for each data set in order
to obtain error bars on $\a$. We finally obtained $\a = 1.09 \pm 0.3$.

\begin{figure}[ht]
\vspace{-1in}
\centerline{
\epsfig{figure=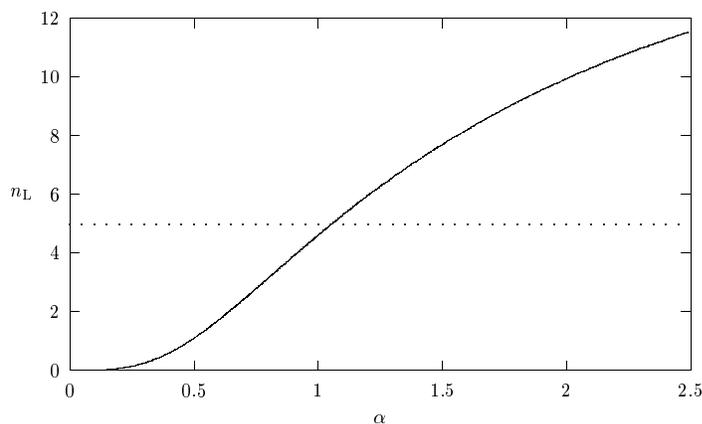,width=1.\textwidth}}
\vspace{-5.2in} 
\caption[Predicted number of lensed quasars $n_{\rm L}$ vs. power index $\a$]
{Predicted number of lensed quasars $n_{\rm L}$ in the adopted optical
quasar sample, with  $\Delta\theta \leq 4''$, vs. power index $\a$.}
\label{fig:NlA} 
\end{figure}
\end{trivlist}

The maximum likelihood analysis was also performed \cite{abhathesis} to
determine the value of $\alpha$, for which the observed sample becomes the
most probable observation. It was found that $0.85 \leq \a \leq 1.56$ at
$1\sigma$  ($68\%$ confidence level), and $0.65 \leq \a \leq 2.33$ at $2\sigma$
($95.4\%$ confidence level). It is also interesting to notice that for a
general power law cosmology the simplest constraint on $\a$ comes from the
relation $\a=H_0 t_0$ (Eq.~\ref{eq:exact}). With updated value of $H_0=70
\pm7$ km/sec/Mpc \cite{Freedman00} and $t_0=14\pm2$ Gyr \cite{pont}, this
constraint gives $\a=0.98 \pm 0.25$. 

\subsection{Concluding remarks}

In this section we discussed the general power law cosmology and the ways
to  constraint its power index. We have derived angular diametere distances for
this cosmology. We described the general formulation of the gravitational
lensing statistics and used it to constraint the parameters for that
cosmology. These results together with other tests were used in \cite{rita4}
in the combined form and the conclusion made was that open linear coasting
cosmology, $a(t)=t$, is consistent with the present observations.

\section{Gravity balls as gravitational lenses}
\subsection{Gravity ball as a NTS solution}
Here we will discuss the general formalism of the problem for a spherically
symmetric system consisting of fermion field $\psi$, scalar field $\sigma$ and
gravitational field $g_{\mu \nu}$. We follow the theory presented by Sethi and
Lohiya \cite{dakshmeetu}. Gravity balls are the NTS solutions of the field
equations arising from (1)  
\be
U(\sigma) [R^{\mu \nu} - \frac{1}{2} g^{\mu \nu} R] = -\frac{1}{2} 
\left [T^{\mu \nu}_{\omega} +
T^{\mu \nu}_{\sigma} + T^{\mu \nu}_{\sigma,\psi} + T^{\mu \nu}_{\psi} + 
2U(\sigma)^{;\mu;\nu} - 2 g^{\mu \nu} U(\sigma)^{;\lambda}_{;\lambda} 
\right ]  
\end{equation}
\begin{equation}
g^{\mu \nu} \sigma_{;\mu;\nu} + \pder{V}{\s} - R
\pder{U}{\s} = 0
\end{equation}
Here $T^{\mu \nu}_{\sigma}$, $T^{\mu \nu}_{\psi}$,
$T^{\mu\nu}_{\sigma,\psi}$ and $T^{\mu \nu}_{\omega}$ are energy
momentum tensors constructed from action for the scalar field, the fermion
field, together with its Higgs coupling to $\sigma$, and the rest of the
matter fields, respectively. We consider an NTS with the scalar field held to
a value $\sigma_0$ in the interior and making a fast transition to $\sigma=0$
outside a thin shell. Thus, we have essentially three regions: (i) the interior
of a soliton ($r<R_0$); (ii) a shell of thickness $\sim \mu^{-1}$ and surface
energy density $s \approx \mu \sigma_0^2/6$; and (iii) the exterior ($r>R_0$).
The total energy of a NTS has contributions from: (1) the surface tension
energy $E_s \approx s R_0^2\,$; (2) the energy of the fermions $E_f \approx
N^{4/3}/R_0$ and; (3) the volume energy $E_V \approx V(\sigma_{\rm in})R_0^3$.
For the degenerate case $V(\sigma_{\rm in})=0$, a NTS has total mass
constrained by the stability against gravitational collapse to a value
determined by the surface tension $s$. The soliton mass, obtained by
minimizing the total energy, is $M=12 \pi s R_0^2$. For $s \sim
(\mbox{MeV})^3$ and $N \approx 10^{75}$, the size of the NTS is of the order
of tens of kiloparces, while it is still away from the Schwarzschild bound.
For configurations with $R$ much greater than the Schwarzschild radius, the
effects of gravity can be treated as a small perturbation. Thus, the form of
the metric for the NTS satisfies a weak-field approximation. 

\vspace{-1in} 
\paragraph{A. Interior: $r < R_0 + {\cal O}\left(\mu^{-1}\right)$.}
The metric inside is described as
\begin{equation}
ds^2 = e^{2u(r)}\,dt^2 - e^ {2v(r)}\,dr^2 - r^2 [d\theta^2 +
\sin^2\theta\,d\varphi^2]
\end{equation}
The interior metric has one specific solution: $v = -\hat{C} r^2/6$ and
\begin{equation} 
u = u_0 + \frac{r^2}{2} \left[ \frac{\tilde{C}}{2} + \frac{\hat{C}}{3}
\right]\;, \end{equation}
with $\tilde{C}$ and $\hat{C}$---constants depending on the fermionic
energy inside the soliton. In the weak field approximation used here we
consider the interior with
\begin{equation}
ds^2 \approx e^{2u_0}\,dt^2 - dr^2 - r^2\left[d\theta^2 + \sin^2{\theta\,
d\varphi^2}\right]
\end{equation}

\paragraph{B. Exterior: $r > R_0 + {\cal O}\left(\mu^{-1}\right)$.}
In the exterior region we have essentially a Minkowskian metric
\begin{equation}
ds^2 = dt^2 - dr^2 - r^2 \left[d\theta^2 +\sin^2{\theta\,d\varphi}^2
\right]
\end{equation}
To be consistent with the observations we will see that the $u_0$ has
to be a small negative constant. The propagation of light inside the G-ball is
equivalently described by using the Fermat principle with the effective
refraction index $n_{\rm eff}$ inside the ball given by $n_{\rm eff} = 1 -
u_0$. This gives straight trajectories of light rays inside the ball. But since
outside the ball $n_{\rm eff}=1$, we see that G-ball behaves as a spherical
lens (deflection of light occurs only at the boundaries).
 
\subsection{Lens model for a G-ball and lensing properties}

We restrict our discussion to a cosmological model which is a variant of a
Milne universe \cite{dakshmeetu} with a scale factor $a(t)=t$. In this
cosmology angular diameter distances are given by
\begin{align}
&\dl= \frac{c z_{\rm l}}{2 H_0}\frac{(2+z_{\rm l})}{(1+z_{\rm l})^2}\,,\no\\
&\ds = \frac{c z_{\rm s}}{2 H_0}\frac{(2+z_{\rm s})}{(1+z_{\rm s})^2}\,,\no\\  
&\dls = \frac{c z_{\rm ls}}{2H_0(1+z_{\rm l})}\frac{(2+z_{\rm ls})}{(1+z_{\rm
ls})^2}\,\,.  
\end{align}
with 
$$
z_{\rm ls} = \frac{z_{\rm s}-z_{\rm l}}{1+z_{\rm l}}\,\,,
$$
and $H_0=100\,h$ km s$^{-1}$ Mpc$^{-1}$. Subscripts
`l', `s' and `ls' stand for lens, source and lens-source, respectively.

We consider the gravity ball to be of the size of a typical cluster of
galaxies at the cosmological distance and embedded in empty space region
with no matter concentrations close to it. The deflector is transparent. We
also assume a thin-lens approximation, since all the deflection occurs within
$\Delta z \le \pm R_{\rm ball}$; the extent of the deflector is thus taken to
be small compared with its distance from both the observer and the source. 
Typical distances from us to the cluster at $z=0.3$ and to the source at $z=1$
are $\sim 1\,\,Gpc$ and $\sim 2\,\, Gpc$, respectively, while $R_{\rm ball}
\approx 0.5$ to $1$ Mpc. We consider deflection angles to be small. We also
assume that the source, lens, and observer are stationary with respect to
comoving coordinates.

The magnification of images in the axisymmetric case is given by
\begin{equation}
\mu = \left( \frac{\beta \,d\beta}{\theta\,d \theta} \right)^{-1}\,\,.
\label{eq:sec4mag}
\end{equation}
The tangential and radial critical curves folow from the singularities of the
tangential and radial magnification
\begin{equation}
\mu_{\rm t} \equiv \left (\frac{\beta}{\theta} \right)^{-1}\,;\qq\qq\mu_{\rm r}
\equiv \left( \frac{d\beta}{d\theta} \right)^{-1}\,\,. \end{equation}

\paragraph {Lens equation}

\begin{figure}[ht]
\centerline{
\epsfig{figure=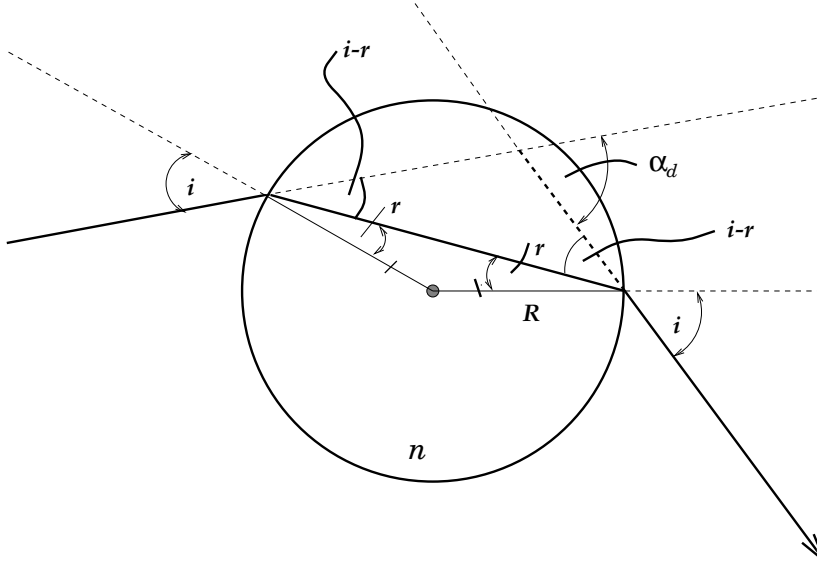,width=0.8\textwidth}}
\caption[G-ball as a gravitational lens]{Geometry of lensing by a gravity ball.
$R_{\rm ball}$---radius of the  ball, $i$---angle of incidence, $r$---angle of
refraction, $\a_{\rm d}$--- deflection angle, $n$---ratio of the refractive
index inside the ball to the  refractive index outside; $n>1$.}
\label{fig:ball} 
\end{figure}

In the Fig.~\ref{fig:ball} we present the geometry of G-ball as a lens. 
From this figure and Snell's law the deflection angle $\a_{\rm d}$ is
\begin{equation}
\a_{\rm d} = 2 \left[i - \sin^{-1} \left( \frac{1}{n}\sin i \right)
\right]\,\,. \end{equation}
Since we consider all angles to be small, the following approximations 
are valid. Defining the quantity $\t_{\rm C}= R_{\rm ball}/\dl$,  which is the
radius of the region inside which the refraction occurs, we obtain the the
expression for the deflection angle $\a_{\rm d}$  \begin{equation}
\a_{\rm d} = 2 \left[ \sin^{-1} \left( \frac{\theta}{\t_{\rm C}} \right) -
\sin^{-1} \left( \frac{\theta}{n \t_{\rm C}} \right) \right]\,\,.
\label{eq:ch4deflection}
\end{equation}
Thus, we obtain the lens equation for the gravity ball  
\begin{equation}
\left \{
\begin{array}{lll}
\bs{\beta} & = & \bs{\theta} - \frac{2 \dls}{\ds} \frac{\bs{\theta}}
{\theta} \left \{ \sin^{-1} \left( \frac{\theta}{\theta_{\rm C}} \right) -
\sin^{-1} \left( \frac{\theta}{n \theta_{\rm C}} \right) \right \} \\
\bs{\beta} & = & \bs{\theta}
\end{array}
\right.
\begin{array}{l}
\mbox{for $ 0 \le \theta \le \theta_{\rm C}$}\,; \\
\mbox{for $ \theta > \theta_{\rm C}$}\,\,. 
\end{array} \label{eq:ch4leq}
\end{equation}
where $\theta \equiv |\bs{\theta}| = \sqrt{\theta_1^2+\theta_2^2}$ is the 
radial position of the image in the lens plane.

\paragraph{Multiple image diagram and conditions for multiple imaging.}

\begin{figure}[ht]
\centerline{
\epsfig{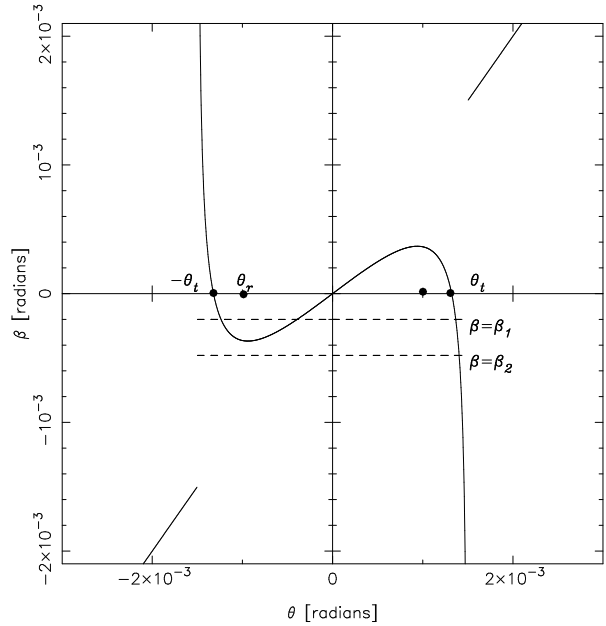}}
\caption[Solution to the lensing equation by a gravity ball]{Solution to the
lensing equation by a gravity ball. The solid curve represents the lensing
equation curve with $n=1.0005$, $\t_{\rm C}=5'$, $z_{\rm source}=1$,
$z_{\rm lens}=0.3$, together with lines $\beta =\beta_i$ (dashed lines) for
various source positions $\beta_i$. The intersections of the lines $\beta =
const$ with the lensing equation curve give the number and positions of the
lensed images. The source at $\beta_2$ lies on a caustic point and
$\theta_{\rm r}$ is the radius of the radial critical curve; source at
$\beta_1$ has three images and point $\beta=0$ produces a ring $\t_{\rm
C}=\theta_{\rm t}$; in addition, it has an image at $\theta=0$. The function
$\beta(\theta)$ is continuos except at $\t_{\rm C}$ due to the edge of the
ball.}\label{fig:diagram} 
\end{figure}

To illustrate the lensing properties described by equation (\ref{eq:ch4leq}) we
show in Figure~\ref{fig:diagram} the lensing curve for the gravity ball with
parameters $n=1.0005$, $\t_{\rm C}=5'$ and for $z_{\rm source}=1$, $z_{\rm 
lens}=0.3$. The intersections of the lines $\beta = const$ with the
curve given by equation (\ref{eq:ch4leq}) give the solutions to the lensing
equation. The source at $\beta_2$ lies on a caustic point and
$\t_{\rm r}$ is the radius of the radial critical curve. The source at
$\beta_1$ has three images, a point $\beta=0$ produces a
ring---`Einstein ring'--- with the critical radius $\t=\t_{\rm t}$
and an image at $\theta=0$. The function $\beta(\theta)$ is continuous
except at $\t_{\rm C}$ due to the edge of the ball.

Putting $\beta=0$ in (\ref{eq:ch4leq}) gives
\begin{equation}
\theta -   \frac{2\dls}{\ds} \left( \sin^{-1}{\frac{\theta}{\t_{\rm C}}} -
\sin^{-1}{ \frac{\theta}{n \t_{\rm C}}} \right) = 0 \,\,.
\end{equation}
Denoting $a = 2\dls/\ds$, we obtain
\begin{equation}
\theta = a \left( \sin^{-1}\frac{\theta}{\t_{\rm C}} -\sin^{-1}{ 
\frac{\theta}{n \t_{\rm C}}} \right)\,\,.
\end{equation}
When the alignment of the source, lens and the observer is not perfect,
we see features called arcs, which are the result of very strong
distortion of a background sources. Arcs roughly trace the Einstein ring, so
$\theta_{\rm arc} \approx \te$. Since the radii of most known arcs do not
exceed $30''$ small angle approximation is valid and we obtain
\be
\theta \left\{1 -  \frac{a}{\t_{\rm C}} \sqrt{1-\left(\frac{\theta}{n \t_{\rm C}} 
\right)^2} + \frac{a}{n \t_{\rm C}} \sqrt{1-\left(\frac{\theta}{\t_{\rm C}}\right)^2} 
\right \} = 0\,\,.
\label{eq:ch4theta}
\end{equation}
One solution to this equation is trivial:
\begin{equation}
\theta = 0\,\,,
\end{equation}
which corresponds to the image located at the centre of the lens. To find
other solutions we write the expression  in the curly brackets in
(\ref{eq:ch4theta}) as  
\begin{equation}
\theta = \t_{\rm C}\sqrt{1 - \frac{a^2 n^2}{4 \t_{\rm C}^2} \left 
(1 - \frac{\t_{\rm C}^2}{a^2} - \frac{1}{n^2} \right)^2}\,\,,
\label{eq:ch4solution}
\end{equation}
and in order to find the conditions for the Einstein ring we rewrite it in
the form 
\begin{equation}
\te= \t_{\rm C}\sqrt{1 - \frac{a^2 n^2}{4 \t_{\rm C}^2} \left 
(1 - \frac{\t_{\rm C}^2}{a^2} - \frac{1}{n^2} \right)^2}\,\,.
\label{eq:ch4einstein}
\end{equation}
Assuming for simplicity $a = 1$ (the ball (lens) half-way between the 
observer and the source) and using $n$ from other estimates \cite{dakshmeetu} 
to be from $1.001$ to $1.0001$  we obtain\newline 
for $n = 1.0001$
$$
\te\cong 0.98\,\t_{\rm C}\;,
$$
for $n = 1.001$
$$
\te\cong 0.73\,\t_{\rm C}\,.
$$
Here $\te$ is non-trivial solution of the equation (\ref{eq:ch4theta}). This
leads to the appearence of the image in the form of a ring with the radial
size of $\sim 0.9$ to $0.7$ of the total size of the ball, depending on $n$.
To find the condition for the appearence of multiple images we analyze 
(\ref{eq:ch4solution}). To obtain a physical solution we take 
\begin{equation} 
1 - \frac{a^2 n^2}{4 \t_{\rm C}^2} \left (1 - \frac{\t_{\rm C}^2}{a} - 
\frac{1}{n^2} \right)^2 > 0\,\,.
\end{equation}
The conditions for multiple imaging are
\begin{equation}
R_{\rm ball} > 2 D_{\rm eff} \left(1- \frac{1}{n} \right)\,\,, 
\end{equation}
or expressed through $n$
\begin{equation}
1 < n < \frac{2 D_{\rm eff}}{2 D_{\rm eff} - R_{\rm ball}}\,\,, 
\end{equation}
where $D_{\rm eff} = \dls \dl/\ds$ is the effective distance.

\begin{figure}[ht]
\centerline{
\epsfig{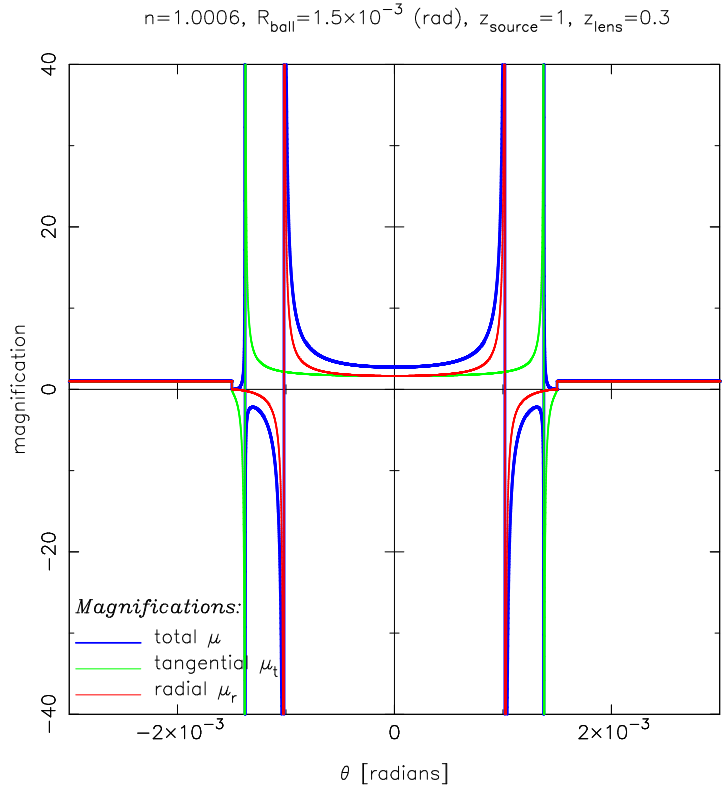}}
\caption[The magnifications for the gravity ball lensing]{The magnifications:
tangential $\mu_{\rm t}$ denoted by green, radial $\mu_{\rm r}$ denoted by red,
and total $\mu$ is shown by blue line. Curves are plotted as a function of the
image positions $\theta$ for the parameters of the lens: $n = 1.0006$,
$\t_{\rm C}=5'$, $z_{\rm l} = 0.3$, $z_{\rm s} =1$. The singularities of
$\mu_{\rm r}$ and $\mu_{\rm t}$ give positions of the tangential and radial
critical curves, respectively.} \label{fig:ballmag} 
\end{figure}

\paragraph{Magnification and critical curves.}
From lens equation for a G-ball (\ref{eq:ch4leq}) and equation
(\ref{eq:sec4mag}) we obtain expression for the total magnification of
the images    
\be
\mu^{-1} = \left \{1 - \frac{a}{\theta} \left [ \sin^{-1}
\left( \frac{\theta}{\t_{\rm C}} \right) - \sin^{-1}\left (\frac{\theta}{n\,\t_{\rm C}} 
\right ) \right ] \right \} \times  \left \{ 1 - a \left ( \frac{1}
{\sqrt{\t_{\rm C}^2-\theta^2}} - \frac{1}{\sqrt{n^2 \t_{\rm C}^2-
\theta^2}} \right ) \right \}\,\,,
\end{equation}
where the first term represents $\mu_{\rm t}$ and second term---$\mu_{\rm r}$.
In Figure~\ref{fig:ballmag} we have plotted tangential magnification $\mu_{\rm
t}$ and radial magnification $\mu_{\rm r}$ against $\theta$ for a G-ball with
refraction index $n=1.0006$. Singularities in these give the angular positions
of tangential critical curves and radial critical curves, respectively. In the
same Figure we have also plotted total magnification $\mu$ vs. $\theta$.
    
\begin{figure}[ht]
\centerline{
\epsfig{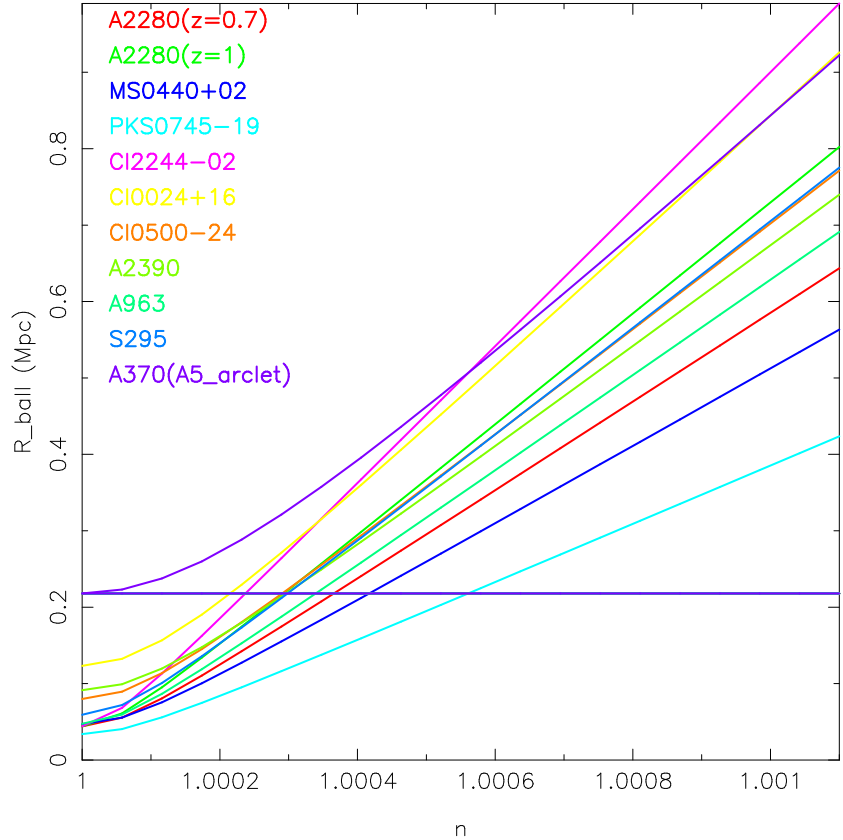}}
\caption[$R_{\rm ball} $ vs. $n$ for different clusters]{$R_{\rm ball} $ vs.
$n$ for different clusters. Clusters are marked by different colours.
Corresponding curves have the same colour as the name of the cluster. Details
are presented in Table~B.1 (App.~B). Horizontal line is the value of
radius of the arc A5 in the cluster A370.}  \label{fig:clusters}
\end{figure}

\paragraph{Limits on $n$ from observations.}
We can rewrite equation (\ref{eq:ch4einstein}) in terms of variables
$R_{\rm ball}$ and $R_{\rm E}$ to get an expression for $R_{\rm ball}$ in
terms of $n$ and a given $R_{\rm E}$.
\begin{equation}
R_{ball}^2 = 4 D_{\rm eff} \left [ 1+ \frac{1}{n^2} -
\frac{\sqrt{4 D_{\rm eff}^2 -R_{\rm E}^2}}{n D_{\rm eff}} \right]\,\,.  
\label{eq:ch4radius}
\end{equation}
Since $R_{\rm ball} \ge R_{\rm E} $, we get a lower bound on the value of
$R_{\rm ball}$ if we know the radius of the Einstein ring. It is clear from
the above equation that if the size of the G-ball is big enough then a large
enough $n$ can give us any desired radius for the Einstein ring. If we assume
that all G-balls to be of the same size, then from observations we can infer a
lower bound on the radius and the refractive index inside a ball by the
following argument. The radius of any gravity ball has to be larger then the
radius of the largest observable Einstein ring. Given this size of the ball
the refractive index should be large enough to give a real Einstein ring for
every other case. This situation is illustrated in the
Figure~\ref{fig:clusters}, where observational data for several clusters with
giant arcs (details are presented in the Table~B.1, App.~B) are used
with the assumption that the raduis of the arc $\theta_{\rm arc} \approx \te$.
Together with the curve (\ref{eq:ch4radius}) for the cluster A370, we plotted
the value of the radius $R_{\rm arc}$ of the A5 arc in that cluster
(horizontal line), which is the largest amongst the clusters considered. It is
clear that the refractive index should be greater than the value where this
line intersects the $R_{\rm ball}$--$n$ curve for this cluster. Assuming all
radii to be the same, we can see that in order to have an arc, a gravity ball
must have $n$ more than $n \approx 1.00056$. Thus, we obtained a lower limit
on $n$. Of course, our present calculations are for the case of an empty
gravity ball. Matter concentrations in the centre of the ball will increase the
deflection angle.  

\subsection{Simulations results} 

\begin{figure}
\centerline{
\vspace{-1.5 cm}
\hspace{-1. cm}
\makebox
{
\epsfig{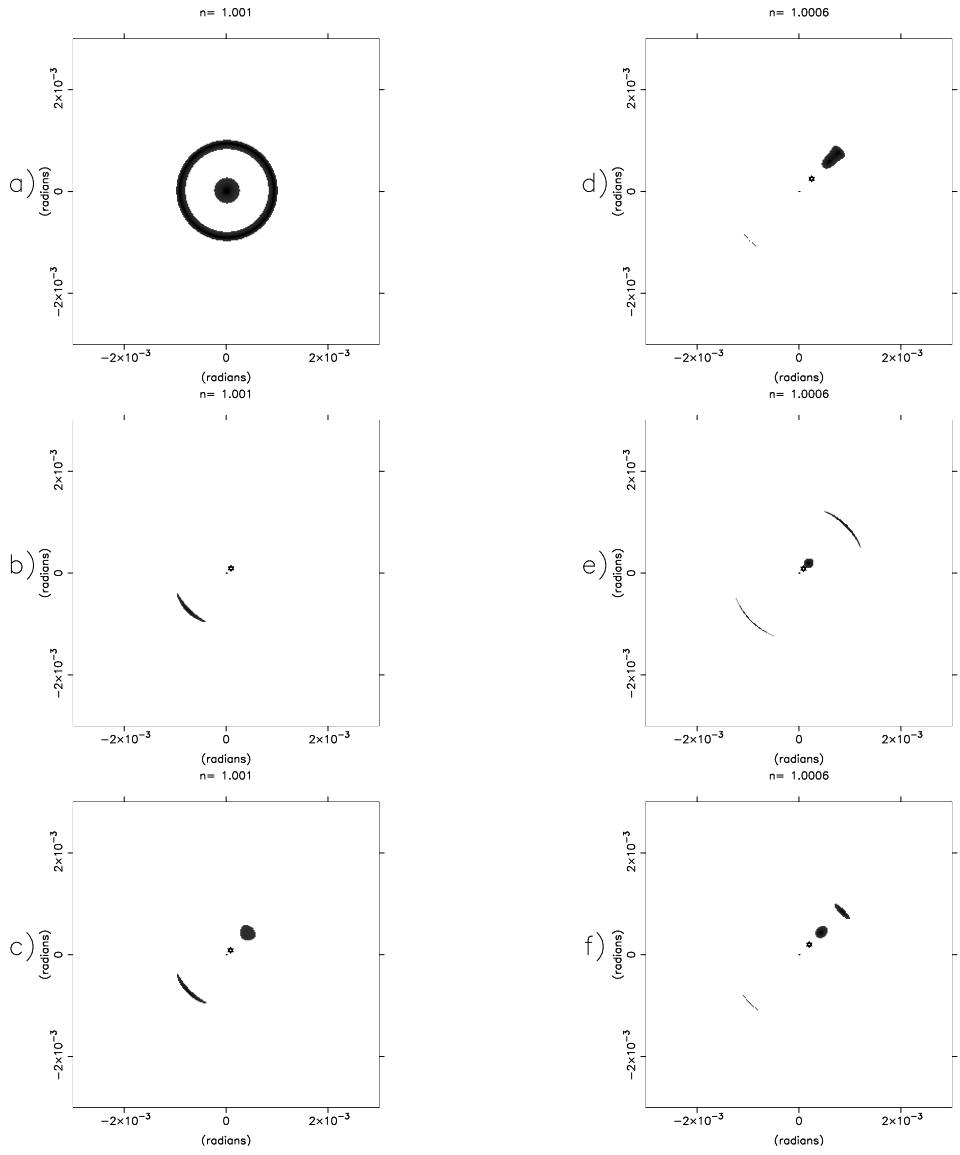}
}}\vskip 0.6in
\caption[Lensing configurations by an empty gravity ball]{Illustration of six
different images connnfigurations for an empty gravity ball. $z_{\rm
l}=0.3$, $z_{\rm s} =1$, $\t_{\rm C}=5'$. Depending on the source position and
$n$ the lens produces images of (a) ring, (b) single arc, (c) arc and opposite
image, (d) radial arc, (e) three images (with two opposite arcs), (f) two
images on one side (with a straight arc). Small dot in the centers of the
panels marks the center of the G-ball. Small star marks the position of the
source.}\label{fig:6} 
\end{figure}

\begin{figure}
\centerline{
\vspace{-1.5 cm}
\hspace{-0.5 cm}
\makebox
{
\epsfig{figure=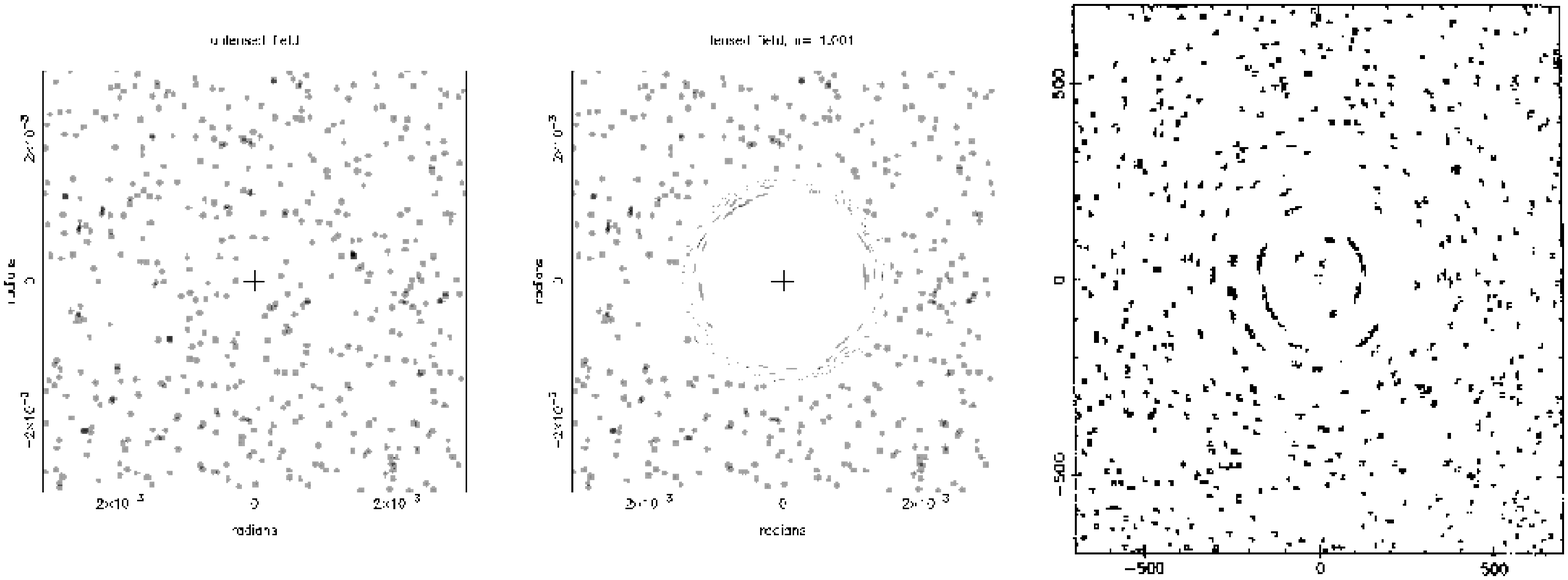,width=1.\textwidth}}}
\vskip 0.6in 
\caption[Distortion field generated by a simulated gravity ball]{Distortion
field generated by a simulated gravity ball. The left panel shows the grid
of randomly distributed background sources as it would be seen in the absence
of the lens. The middle panel shows the same population once they are
distorted by a foreground gravity ball with the parameters: $z_{\rm s}=1$,
$z_{\rm l}=0.3$, $\t_{\rm C}=5'$ rad, $n=1.001$. The right panel shows the same
population distorted by a foreground (invisible) circular cluster ($z_{\rm s}=1.3$,
$z_{\rm l}=0.4$, $\sigma =1000$ $kms^{-1}$); the units for both axis are in
arcsecs.}\label{fig:7}  
\end{figure} 
To demonstrate how a G-ball at intermediate redshift gravitationally
distorts background sources we performed computer simulations and
present the results in Figures~\ref{fig:6} and~\ref{fig:7}. We have simulated
an empty gravity ball the size of a typical galaxy cluster at the reshift of
$z_{\rm l}=0.3$, assuming for simplicity that all source galaxies are at the
same redshift $z_{\rm s}=1$. The luminous area of each background galaxy is
taken to be a circular disk of radius $R$ with constant brightness. In all our
numerical computations we used dimensionless Hubble parameter $h=1$. For the
simulations of G-ball lensing we use the algorithm described in
\cite{schneider}. The image configurations by a simulated gravity ball are
illustrated in Fig.~\ref{fig:6} by changing the relative positions between the
extended circular source and the lens. When the alignment of the source on the
optical axis is perfect, the ring image---Einstein ring---appears
(Fig.~\ref{fig:6}(a)). With a small displacement we see two opposite arcs
approximately located at the Einstein radius (Fig.~\ref{fig:6}(e)). When the
displacement becomes larger different features may appear: single tangential
arc (Fig.~\ref{fig:6}(b)), radial arc (Fig.~\ref{fig:6}(d)), two opposite
images (Fig.~\ref{fig:6}(c)) and straight arc (Fig.~\ref{fig:6}(f)). Figure
\ref{fig:7} displays the distortion the gravity ball produces on the field of
randomly distributed background sources. Figure~\ref{fig:7}(a) shows an
unperturbed galaxy backgound projected randomly in a field of $\sim 10' \times
10'$ at a redshift of $z=1$. Fig.~\ref{fig:7}(b) illustrates the same field
with an empty G-ball at the centre. To compare with the lensing effect a
cluster modelled as a singular isothermal sphere produces on the background
galaxies, we presented in the Fig.~\ref{fig:7}(c) the simulation taken from
\cite{fort-mellier}. We can see from the figures that an empty gravity ball
can indeed act as a strong lens and produce the images of arcs and arclets.
 
\subsection{Concluding remarks}

In this section we have discussed a special kind of nontopological soliton,
known as gravity ball, and examined the possibility for it to be a
gravitational lens. We investigated its lensing properties, calculated the
deflection angle of a point source, derived the lens equation and plotted the
lens curve for the gravity ball with specific parameters. We have shown that
depending on the parameters of the model there can be one, two or three images
(two inside the Einstein radius and one outside). However, the G-ball
gravitational lens has interesting features which are not shared by other known
gravitational lenses. In the case of a source, lens and observer located on the
optical axis together with the Einstein ring there is always an image in the
centre. Besides, the lensing geometry is radically different---there is no
effect outside the lens and thus, the Einstein radius is always less than
radius of the ball. The refractive index of the gravity ball contributes to the
decrease in the radius of the Einstein ring and increase of its thickness. For
the extended sources we showed how a large gravity ball can induce surprisingly
large distortions of images of distant galaxies. It is assumed that in order to
produce large arcs the cluster must have its surface-mass density approximately
supercritical, $\Sigma \ge \Sigma_{crit}$. We found that the empty balls alone
are able to produce arc-like features, including straight arcs, arclets and
radial arcs. We have modelled the gravitational lensing effect of our gravity
ball on a background source field and found that the ball produces distortion
(`shear') of that field, consistent to some extent with the observations. We
also obtained constraints on the size of the large gravity ball, which can be
inferred from the existing observations of clusters with arcs. Empty G-balls
can show in observations as arcs without evident mass distribution---a lens
and/or through the distortion of the background random galaxy field.  

Observationally empty G-balls have not yet been found, but we conclude from
our analysis that the existence of large G-balls cannot be ruled out by
gravitational lensing effects.

\newpage

\chapter{Propagation of Light in Strong Gravitational Fields}
\section{Introduction and main motivations}
\label{sec:5_1} 
As described in the Introduction to the thesis (\ref{sec:introduction}),
gravitational lensing in strong gravitational fields leads to many
astrophysically interesting effects. In this section we discuss yet another
one.

The idea that centres of the globular clusters host massive (up to 
$10^4 \, M_{\odot}$) black holes (BH) goes back to 1975, when scientists
were trying to detect them through X-ray emission. Since then 
other methods were proposed. In particular, the observations of the
millisecond (ms) pulsars in globular clusters can help, since the pulsar
motion would be governed by the presence of massive black hole in the centre
of the globular cluster (\cite{Postnov91},\cite{DePaolis} and references
within). A large number of ms pulsars have been discovered in globular
clusters---more than 35 ms and 4 long-period ones in 13 Galactic globular
clusters have been reported. This is not surprising, since the dense stellar
field in the globular clusters provides a favourable environment for frequent
stellar encounters to form low-mass-X-ray-binaries (LMXB). These are believed
to be the progenitors of the ms pulsars. By some estimates (for
ex.~\cite{Fruchter}), the total number of pulsars in Galactic globular
clusters, beaming at us, most likely lies between 300 and 2000. This provides
an excellent opportunity to use them for revealing the existence of central
black holes. A possible mechanism for establishing the presence of
massive black hole in the centre, which has not so far attracted any
attention, is through the gravitational lensing of the pulsar's radiation by
the central mass. It is only logical to suppose that the black hole in the
centre of the globular cluster would be a rotating black hole. There exist a
number of papers investigating the lensing properties of Kerr black holes.
Bray \cite{bray} was the first to investigate the properties of the isolated
Kerr black hole as a gravitational lens. He demonstrated that effect of spin
of the BH comes in the second order terms of the expansion of null geodesics
in the Kerr metric and considered the conditions for the multiple imaging.
Simulteneously, Dymnikova \cite{dymnikova} studied the time dilation, in the
sense of Shapiro effect, and showed how the asymmetry in the time delay
depends on the mutual orientation of a photon propagation direction and of the
rotation axis. The photon is accelerated, if its motion is along the direction
of the rotation, and it is delayed otherwise \cite{Laguna}. This effect can be
traced back to the dragging of inertial frames. The asymmetry's is very enhanced
in the regions corresponding to lensing and increases sharply for the photons
traveling very close to the central body. If, for example, the motion is
retrograde, the effect of the rotation is to impose an additional time delay
acting in the same direction as the effect arising from the mass. For systems
with the central rotating black hole the time lag can be of the order of
minutes \cite{Goicoechea92}. 

One more interesting effect was noted by Carini et al \cite{carini92}. They
showed  that angular momentum of a Kerr black hole gives rise to a rotation of
the plane of polarization of a linearly polarized light. They noted that
the rotation of the polarization plane induced by the angular momentum
of the central gravitational source is analogous to Faraday effect of
the electromagnetic wave propagating through magnetized plasma and thus may
be referred to as the "gravitational Faraday rotation".

Still more effects may also possibly exist---in particular, it may be 
interesting to study the differential frequency shift between the beams 
propagating in the equatorial plane (where this effect would be maximal)
on the opposite sides of the black hole. Secondly, for a source 
of pulsed radiation (e.g. pulsars), it would be interesting to determine 
if a mechanism exists which can produce the difference in the time period 
between pulses in two images. In that case it could be possible that some (?)
of the pulsars in certain globular clusters are gravitationally lensed images
of the same source. This is the main motivation with which the present work
has started.      

\section{Photons in a gravitational field}

It was Galileo's famous experiment that asserted that big balls fall at the 
same rate as small balls. By the more general statement, known as EP (Einstein
equivalence principle) all particles follow the same geodesic path,
independent of the mass or internal structure. Photons move along null geodesics
in curved spacetime, and all photons are deflected in a gravitational
potential by the same angle independent of their energy or polarization.
This is mostly used in studies of the gravitational lensing around black
holes. However, some special situations are reported, where gravitational
birefringence and dispersive effects are described. For example, in quantum
electrodynamics, virtual electron loops induce curvature couplings in the
effective action of low-frequency photons \cite{gravrainbow}. In such a case,
photon propagation can be influenced by tidal effects coming from local
spacetime curvature. Drummonds and Hathrell \cite{drummonds80} found that the
propagation of photons in spatially anisotropic background (Schwarzchild and
gravitational-wave) was polarization-dependent. Similarly, gravitational
birefringence was also found to occur in Reissner-Norstr\"om spacetimes
\cite{Daniels94}. The energy-dependent deflection of light was shown in the
context of string theory \cite{Mende} and higher derivative gravity, which
arises from QED radiative corrections \cite{gravrainbow}. There is some
controversy in literature. For example, Masshoon firmly states
that birefringence is possible only if the central body is a rotating body
and it is a consequence of the coupling of the intrinsic spin of the photon
with the rotation of the central body. That is, for a \Schw black hole the
amplitudes for the  scattering of right circular polarized (RCP) and left
circular polarized (LCP) waves are equal and, hence, the spherical symmetry
of this field preserves the polarization of the incident radiation in the
scattered waves. However, for a Kerr black hole RCP and LCP radiations are
scattered differently and the deflection of the radiation by a rotating mass
becomes polarization dependent  \cite{mashhoon}. 

\subsection{Spinning particles}

We would like to pay special attention here to the fact that photon is an
elementary particle with spin. Spin angular momentum of matter, occurring in nature
in units of $\hbar /2$, is the physical notion which seems necessary for a
successful extension of GR to microphysics. All elementary particles can be 
classified by means of irreducible unitary representations of the Poincar\`e
group and can be labeled by mass $m$ and spin $s$. Mass is connected with the
translational part of the Poincar\`e group and spin with the rotational part.
Mass and spin are elementary notions, each with an analogous standing not
reducible to that of the other. 

The problem of the motion of a particle with internal angular momentum (spin) in
external gravitational field consists of two parts: spin precession and spin influence
on the trajectory of motion. Gyroscope precession in a centrally symmetric
gravitational field was considered more than 70 years ago \cite{Fokker21} and
is now a well-described effect, since it is extremely important in the
satellite industry. The situation is different with the second part of the
problem, which refers to the spin influence on the trajectory. According to EP
a spinning mass should fall at the same rate as non-spinning one. However, in
a seminal paper, Papapetrou \cite{Papapetrou} showed that trajectories of
spinning particles (in the limit of a vanishing mass) deviate from geodesics.
His derivation was based upon taking a macroscopic mass, expanding it in
moments about the center, and deriving the equations of motion for each
moment. The basic result was that the deviation is proportional to the spin of
the particle (vanishing for a zero spin) and the curvature of the space. It is
presumed that the results are valid for fundamental particles, but there is no
classical theory that yields the result \cite{Theodore}. The difficulty is
that there has not been a satisfactory theory which derives these equations
from a variational principle. There have been numerous attempts, though; a
good review on equations for particles with spin is recently presented by
Frydryszak \cite{spin-review}.  Most recent claims, involving a generalized
Clifford algebras, are presented by Pezzaglia \cite{Pezzaglia}. The problem of
the influence of spin on the trajectory of a particle in an external
gravitational field is not merely of a theoretical interest. This problem is
closely related to the gravitational waves field, where one has to calculate
the gravitational waves radiation from the inspiralling black (Kerr) holes. In
gravitational lensing this problem can be of the importance. It could lead to
the deviation of the photon trajectories from the geodesics and to the
gravitational birefringence in an arbitrary gravitational field. In the next
section, the modified geodesic equations of photon in an external
gravitational field are investigated. The results obtained in the
Section~\ref{sec:Prasanna} are used in the subsequent section
(Section~\ref{sec:epsilon}) in the context of recent claims
\cite{drummonds80,Daniels94,Khriplovich} that photons acquire superluminal
velocities in the \Schw, FRW and gravitational wave background. We extend the
results obtained by \cite{prasanna-main} to the Kerr background and find, in
concordance to previous results for other metrics, that in Kerr metric this
does not happen. In the last section we derive the Papapetrou equations for
the photon directly.

\section{Modified geodesic equation for photons}\label{sec:Prasanna}
\subsection{Theoretical background}

The action of the electromagnetic field in the curved spacetime is given by:
\begin{equation}
S = \int \, d^4 x \, \sqrt{-g} F_{\mu\nu} F^{\mu\nu} ,
\end{equation}
where $F_{\mu\nu}$ has its usual meaning$ F_{\mu\nu}=A_{\mu;\nu} -
A_{\nu;\mu}$. Varying the action with respect to $A_{\mu}$ we obtain the
source equations for a free electromagnetic field in curved space-time:
\begin{equation}
\nabla^{\nu}F_{\mu\nu} = 0\,\,.
\label{eq:2nd pair}
\end{equation}
These equations together with the property of the field:
\begin{equation}
\nabla_{[\mu}F_{\l\nu]} =0
\label{eq:1st pair}
\end{equation}
constitute Maxwell equations in an empty 4-space. To obtain a wave equation
for a photon, we will apply $\nabla^{\mu}$ to the Eq.~\ref{eq:1st pair}:
\begin{equation}
\nabla^{\mu}\left(\n_{\mu}F_{\nu\l}\right) + \n^{\mu}\left(\n_{\nu}
F_{\l\mu}\right) + \nabla^{\mu}\left(\n_{\l}F_{\mu\nu}\right) =0\,\,.
\end{equation}  
Using the properties of a commutator of the covariant derivatives 
$$
\left[\n^m,\n_n\right]A_{ik}=\n^m\left(\n_n A_{ik}\right) - 
\n_m\left(\n_n A_{ik}\right)\,\,,
$$
and from the field equations (\ref{eq:1st pair}) we have  
\be 
\square F_{\nu\l} +
\underbrace{\left[\n^{\mu},\n_{\nu}\right]F_{\l\mu}}_{\mbox{\ding{172}}} + 
\underbrace{\left[\n^{\mu},\n_{\l}\right]F_{\mu\nu}}_{\mbox{\ding{173}}} = 0 \,\,.  
\label{eq:6}
\end{equation}
The operator $\square$ will be frequently referred to. In general coordinates it is to
be taken as defined by
\begin{equation}
\square A_{\mu\nu...} \stackrel{\mathrm{def}}{\equiv}g^{\a\b}
\left(A_{\mu\nu...}\right)_{;\a;\b}
\end{equation}
It is also sometimes written in the form (see \cite{eddington}, p. 64) 
$\square = \left(\left(...\right)_{;\a}\right)^{;\a}$,  that is, we perform a
covariant and a contravariant differentiation and contract them. In some texts
it means also $\square = \n^{\mu}\n_{\mu}$, where $\nabla$ is a covariant
derivative. \\
Following the definition of a commutator of covariant derivatives
\begin{equation}
\left[\n_c,\n_d\right] X^a_{\;\;b} \stackrel{\mathrm{def}}{=}
R^a_{\;\;\l cd}X^{\l}_{\;\;b}-R^{\l}_{\;\;bcd}X^a_{\;\;\l}\,\,, 
\end{equation}
we expand the numbered terms in the equation (\ref{eq:6}):
\berrno
\mbox{\ding{172}} & \hskip 0.6in & \left[\n^{\mu},\n_{\nu}\right] F_{\l\mu} \equiv
\left[\n_{\mu},\n_{\nu}\right] F_{\l}^{\;\;\mu} =
-\left[\n_{\mu},\n_{\nu}\right] F^{\mu}_{\;\;\l}= 
R^{\rho}_{\;\;\l\mu\nu}F^{\mu}_{\;\;\r} - \hskip  4in  \\[0.05in]
&& - R^{\mu}_{\;\;\r\mu\nu}F^{\r}_{\;\;\l}=
R^{\rho}_{\;\;\l\mu\nu}F^{\mu}_{\;\;\r} - R_{\rho\nu}F^{\r}_{\;\;\l}\,\,.   
\errno
\berrno
\mbox{\ding{173}} & \hskip 0.6in &\left[\n^{\mu},\n_{\l}\right] F_{\mu\nu} \equiv
\left[\n_{\mu},\n_{\l}\right] F^{\mu}_{\;\;\nu} = 
R^{\mu}_{\;\;\r\mu\l}F^{\r}_{\;\;\nu} -   
R^{\r}_{\;\;\nu\mu\l}F^{\mu}_{\;\;\r}= \hskip  10in  \\[0.05in]
&& =R_{\r\l} F^{\r}_{\;\;\nu} - R^{\rho}_{\;\;\nu\mu\l}F^{\mu}_{\;\;\r}\,\,.     
\errno
Using symmetry properties of a Riemann tensor
$$
 R^{\r}_{\;\;\l\mu\nu}  = - R^{\r}_{\;\;\l\nu\mu} 
\stackrel{\&}{\Longrightarrow} R^{\r}_{\;\;\nu\mu\l} =-R^{\r}_{\;\;\l\nu\mu}-
R^{\r}_{\;\;\mu\l\nu} 
$$ 
we obtain the wave equation for a photon in a curved space-time:
\begin{equation}
\square F_{\nu\l} - R_{\r\nu}F^{\r}_{\;\;\l}+R_{\r\l}F^{\r}_{\;\;\nu} +
R^{\r}_{\;\;\mu\l\nu}F^{\mu}_{\;\;\r} = 0\,\,.
\label{eq:wave} 
\end{equation}
In empty space this equation becomes
\begin{equation}
\square F_{\nu\l} + R^{\r}_{\;\;\mu\l\nu}F^{\mu}_{\q\r} = 0\,\,.
\end{equation}
This equation was first derived by Eddington \cite{eddington}, though not from
the action. This equation is quite unusual, therefore it deserves some more
explanations. Let us consider the 4-potential $A_{\mu}$. Since it is not
gauge invariant, we will add the standard Lorentz gauge:
\begin{equation}
\left(A^{\mu}\right)_{;\mu} = 0\,\,.
\label{eq:Lorentz gauge}
\end{equation}
With the boundary condition at the infinity, the value of $A_{\mu}$ 
becomes completely determined. 

From Maxwell equations in the presence of a source, we get
\begin{eqnarray*}
J^{\mu} & = & \left(F^{\mu\nu}\right)_{;\nu} = \left(g^{\a\b}F_{\mu\b} \right)_{;\a}=
g^{\a\b}\left(F_{\mu\b} \right)_{;\a}=g^{\a\b}\left(A_{\mu;\b ;\a} -
A_{\b ;\mu ;\a}\right) = \\
& & g^{\a\b}\left(A_{\mu ;\b ;\a} - A_{\b ;\a ;\mu} +
R^{\e}_{\;\;\b\a\mu}A_{\e} \right) = g^{\a\b}\left(A_{\mu}\right)_{;\b ;\a} -
\left(A^{\a}_{;\a}\right)_{;\mu} + R^{\e}_{\;\mu} A_{\e}\,\,.
\end{eqnarray*}
In the Lorentz gauge, we have
$$
\square A_{\mu} = J_{\mu} - R^{\e}_{\mu}A_{\e}\,\,.
$$
In empty space this becomes $\square A_{\mu}=0$, showing that $A_{\mu}$ is
propagated with the fundamental velocity. We see that the velocity of
propagation of electromagnetic potential and of electromagnetic force is not
the same. It should not be surprising, since $A_{\mu}$ is not physically
important, being fixed by arbitraty convention (Eq.~\ref{eq:Lorentz gauge}).
Bu the result (\ref{eq:wave}) is very interesting, showing that Ricci and
Riemann curvature coupling terms give rise to the polarization dependent
deviation of photon orbits from null geodesics. The derivation of the result
(\ref{eq:wave}) from the action principle was first given in
\cite{prasanna-main}.

In order to derive equation for the characteristics of photon propagation we
make the geometrical optics plane-wave approximation. In geometrical
optics, one assumes that the electromagnetic waves have a wavelength which is
much smaller than the radius of curvature of the background geometry, or the
scale of variation of the amplitude of the wave front. 
We also assume in our analysis that the influence of the electromagnetic field
on the metric field can be neglected. Thus, ``locally plane" waves can be
represented by approximate solutions of Maxwell's equations of the form 
\be F_{\mu\nu}=e^{is(x)} f_{\mu\nu}\,\,,
\label{eq:anzatz} \end{equation} 
where $f_{\mu\nu}$ is a lowly varying amplitude, and $s$, a rapidly
oscillating phase. (It is understood that one takes the real part of the right
hand side of Eq.~\ref{eq:anzatz}). We should emphasize that although this is
an approximation for the actual solution for the photon, it exactly determines
the characteristics of propagation and hence the casual structure of the
solution, at least in domains where the gradient of the $s$ does not vanish
\cite{The-book}. The wave vector $k_{\mu}$ is given by the gradient of the
phase \be k_{\mu}=\n_{\mu} s\,\,. \end{equation} Operating by $\n_{\mu}$ on
(\ref{eq:anzatz}) and neglecting the derivative of a slow changing amplitude,
we obtain \be \n_{\mu}F_{\a\b} = i \,F_{\a\b}\n_{\mu}\, s 
=ik_{\mu}F_{\a\b}\,\,. \label{eq:eikonal} \end{equation} Using
(\ref{eq:anzatz}), Eq.~\ref{eq:wave} becomes $$
\n^{\mu}\n_{\mu}\left(e^{is}f_{\nu\l}\right) - R^{\r}_{\;\nu} F_{\r\l} +
R^{\r}_{\;\l}F_{\r\nu} + R^{\r\mu}_{\;\;\;\;\l\nu}F_{\mu\r} =0\,\,. $$
Considering $k_{\mu}$ a constant vector, after some algebra we obtain the
wave equation (\ref{eq:wave}) gives
\be
-k_{\mu}k^{\mu}f_{\nu\l} +R^{\r}_{\;\l}f_{\r\nu} +R^{\r\mu}_{\;\;\;\;\l\nu}
f_{\mu\r}-R^{\r}_{\;\nu}f_{\r\l} =0\,\,.
\label{eq:k-wave}
\end{equation}
Since, by definition, wave vector $k_{\mu}$ is a gradient, $\n^{\a}k_{\mu}=
\n_{\mu}k^{\a}$, and 
$$
\n^{\a}\left(k_{\mu}k^{\mu}\right)= 2k^{\mu}\n_{\mu}k^{\a}\,\,.
$$
Light rays are defined as the integral curves $x^{\a}(s)$ of the vector
field $k^{\a}$, that is, $ k^{\a}=dx^{\a}/ds$. Therefore, 
\berr
&& \hspace{-.4in}\n^{\a}\left(k^{\mu}k_{\mu} \right)=2 \fr{dx^{\mu}}{ds}
\left(k^{\a}\right)_{;\mu} = 2 \fr{dx^{\mu}}{ds}\left(k^{\a}_{\;\;,\mu} +
\G^{\a}_{\;\;\mu\s} k^{\s}\right)  = \no \\
& =& 2 \fr{dx^{\mu}}{ds}\left(\fr{\del}{\del x^{\mu}} \fr{dx^{\a}}{ds} +
\G^{\a}_{\;\;\mu\s} \fr{dx^{\s}}{ds}\right)=2\left(\fr{d^2x^{\a}}{ds^2} +
\G^{\a}_{\;\;\mu\s} \fr{dx^{\mu}}{ds}\fr{dx^{\s}}{ds} \right)\,\,.
\label{eq:LHS-geodesic}
\err
We can write the Eq.~\ref{eq:k-wave} in the form:
$$
k^2f_{\nu\l} = R^{\r}_{\;\l}f_{\r\nu} + R^{\r\mu}_{\;\;\;\;\l\nu}f_{\mu\r} -
R^{\r}_{\;\nu}f_{\r\l} \,\,,
$$
and obtain the dispersion relation for the wave vector:
\be
k^2 = \left(R^{\r\mu}_{\;\;\;\;\l\nu}f_{\mu\r} + R^{\r}_{\;\l}f_{\r\nu} -
R^{\r}_{\;\nu}f_{\r\l} \right) \cdot \fr{f^{\nu\l}}{(f_{\a\b}f^{\a\b})}\,\,.
\label{eq:dispersion}
\end{equation}
Combining (\ref{eq:LHS-geodesic}) and (\ref{eq:dispersion}), we obtain the
modified geodesic equation for photons:
\be
\fr{d^2x^{\a}}{ds^2} + \G^{\a}_{\mu\s} \fr{dx^{\mu}}{ds}\fr{dx^{\s}}{ds} =
\fr{1}{2} \n^{\a}\left[\left(R^{\r\mu}_{\;\;\;\;\l\nu}f_{\mu\r} +
R^{\r}_{\;\l}f_{\r\nu} - R^{\r}_{\;\nu}f_{\r\l} \right)
\fr{f^{\nu\l}}{(f_{\a\b}f^{\a\b})}\right]\,\,.
\label{eq:modified}
\end{equation}
The nonzero right hand side of the modified geodesic equation leads to a polarization
dependance in the gravitational redshift, deflection angle and Shapiro time
delay of light. In order to investigate these equations we consider the
\Schw \/ space-time.

\subsection{\Schw metric}
 
\Schw metric is represented by:
\be
d\tau^2=\left(1-\fr{2m}{r}\right)\,dt^2 - \left(1-\fr{2m}{r}\right)^{-1} 
\,dr^2 - r^2 \,d\theta^2-r^2\sin^2{\theta} \,d\phi^2\,\,.
\label{eq:metric}
\end{equation}
Equations of motion (Eq.~\ref{eq:modified}) in an empty space are:
\be
\frac{d^2 x^{\a}}{ds^2} + \Gamma^{\a}_{\mu\nu} 
\frac{dx^{\mu}}{ds} \frac{dx^{\nu}}{ds} = \frac{1}{2} \underbrace{\nabla^{\a}
\left( R_{\rho \mu \nu \l} f^{\rho \mu} \right)}_{\circledast}
\frac{f^{\nu \l}}{|f^2|}\,\,.
\label{eq:geodesic}
\end{equation}
Covariant derivative of a Riemann tensor:
\be
R_{\a \beta \mu \nu;\l} = R_{\a \beta \mu \nu,\l} - \G^k_{\a\l} 
R_{k \beta \mu \nu} - \G^k_{\b\l} R_{\a k \mu \nu} - \G^k_{\mu\l} 
R_{\a \beta k \mu} -\G^k_{\nu\l} R_{\a \beta \mu k}\,\,.
\end{equation}
After expanding $\circledast$, the right hand side of the Eq.~\ref{eq:geodesic}
is: 
\begin{align} 
&\mbox{R.H.S.} =  \frac{1}{2 |f^2|} \left[ f^{\rho \mu}
\left(\del_{\a} R_{\rho \mu \nu \l} - \G^k_{\rho \a} 
R_{k \mu \nu \l} - \G^k_{\mu \a} R_{\rho k \nu \l} -\G^k_{\q\nu \a} 
R_{\rho \mu k \l} -  \right. \right. \no \\[0.07in]
&  -\left. \left. \G^k_{\l \a} R_{\rho \mu \nu k} \right) 
+ R_{\rho \mu \nu \l} \left(\del_{\a} f^{\rho \mu}
+ \G^{\rho}_{\a k} f^{k \mu} +\G^{\mu}_{\a k} f^{\rho k} \right)
\right] f^{\nu \l}\,\,.
\label{eq:RHS}
\end{align} 
We will write this equation in components.
\renewcommand{\descriptionlabel}[1]%
{\hspace{\labelsep}\textsf{#1}}

\begin{description}

\item[A.] Let $\a=\v$. \\
Due to the axisymmetry of the metric,
$$
\del_{\v}R_{\rho \mu \nu \l}=0\,\,,
$$ 
$$
\del_{\v}f^{\rho\mu}=0\,\,.
$$
After expanding $\nabla^{\v}$, the right hand side (Eq.~\ref{eq:RHS}) is:
$$
\hspace{-0.5in}\mbox{RHS} = 
\fr{2}{|f^2|}\left[f^{tr}f^{t\v}\,\G^{\v}_{r\v}R_{t\v t\v} + 
f^{t\t}f^{t\v}\,\G^{\v}_{\t\v}R_{t\v t\v} +
f^{tr}f^{t\v}\,\G^{r}_{\v\v}R_{tr tr} +
f^{t\v}f^{t\t}\,\G^{\t}_{\v\v}R_{t\t t\t}\right]\,\,.
$$
Using the non-vanishing components of the affine connection and Riemann
tensor, we find:  
$$
\mbox{RHS}=\frac{6 m (2m-r)\sin^2{\theta}}{r^3}\frac{f^{tr}f^{t\v}}{f^2}\,\,.
$$
\item[B.] $\a=t$ 
$$
\del_tR_{\rho \mu \nu \l}=0\,\,,
$$
and we use the fact that in the absence of charges $\del_{\a}f^{\a\beta}=0$.
After expanding $\nabla^{t}$, the RHS for the time component $t$ is: 
$$ 
\mbox{RHS}=0\,\,.
$$

\item[C.] $\a=r$\\
\noindent
After expanding $\nabla^{r}$, the RHS for the $r$ component is:
\begin{align*}
&\hspace{-0.5in}\mbox{RHS}=
\frac{2}{|f^2|} \left[ \left(f^{tr}\right)^2 \del_r R_{trtr} +
\left(f^{t\t}\right)^2 \del_r R_{t\t t\t} +
\left(f^{t\v}\right)^2 \del_r R_{t\v t\v} +  
+f^{t\t}R_{t\t t\t}\, \del_r f^{t\t}+\right. \\[0.07in] 
& + \left. f^{t\v}R_{t\v t\v}\,\del_r f^{t\v} - 
 -\left(f^{t\t}\right)^2 R_{t\t t\t}\left( \G^t_{tr} + 
\G^{\t}_{r\t} \right) - 
\left(f^{t\v}\right)^2 R_{t\v t\v}\left( \G^t_{tr} + 
\G^{\v}_{r\v} \right) \right]\,\,.
\end{align*}
Using non-vanishing components of the affine connection and Riemann tensor, we
find: \begin{align*}
&\hspace{-0.8in}\mbox{RHS}=\frac{2m}{r^2}\frac{1}{f^2} \left[-\left(f^{rt}\right)^2 
\frac{6}{r^2} + \left(f^{\theta t}\right)^2 \frac{(2r-5m)}{r} +
+\left(f^{\v t}\right)^2 \fr{(2r-5m)\sin^2{\t}}{r} + \right.\\[0.07in]
&+ \left.  f^{\theta t}\,\del_rf^{\theta t}\,(2m-r) +
f^{\v t}\,\del_rf^{\v t} \,(2m-r)\sin^2{\theta}\right]\,\,.
\end{align*}

\item[D.] $\a=\t$\\
After expanding $\nabla^{\t}$, the RHS for $\theta$ component is:
\begin{align*}
& \hspace{-0.5in}\mbox{RHS}=
\fr{2}{|f^2|} \left[ \left(f^{rt}\right)^2 \del_{\t} R_{trtr} +
\left(f^{t\t}\right)^2 \del_{\t} R_{t\t t\t} + 
\left(f^{t\v}\right)^2 \del_{\t} R_{t\v t\v}+
+ f^{tr} R_{trtr} \,\del_{\t} f^{tr} + \right.\\[0.07in]
+ & \left. f^{t\v} R_{t\v t\v}\,\del_{\t} f^{t\v}- 
f^{tr} f^{t\t}\left(\G^{r}_{\t\t} R_{trtr}+\G^{\t}_{r\t} 
R_{t\t t\t}\right) -\left(f^{t\v}\right)^2 
\G^{\v}_{\t\v}R_{t\v t\v} \right]\,\,.
\end{align*}
Using non-vanishing components of the affine connection and Riemann tensor, we
find: \begin{align*}
&\hspace{-0.5in}\mbox{RHS}=
\frac{2m}{r^2}\frac{1}{f^2} \left[ \left(f^{\v t}\right)^2 
3 (r-2m)\cos{\t}\sin{\t}+f^{t\t}f^{tr}\,\frac{3(r-2m)}{r}+\right.\\[0.07in] 
&+\left.f^{tr}\,\del_{\t}f^{tr}\,\frac{2}{r} +f^{t\v}\,\del_{\t}f^{t\v}\,
(2m-r)\sin^2{\t} \right]\,\,.
\end{align*}
\end{description}

For the equatorial plane, $\t=\pi /2$, the equations are simplified:\\
RHS for $r$
\begin{align*}
&\mbox{RHS} = \frac{2m}{r^2}\frac{1}{f^2}\left[-\left(f^{rt}\right)^2 
\frac{6}{r^2} + \left(f^{\theta t}\right)^2 \frac{(2r-5m)}{r}\, + \right. \\[0.07in]
& \left. \left(f^{\v t}\right)^2 \frac{(2r-5m)}{r} +f^{\theta t}\,
\del_r f^{\theta t}(2m-r) + f^{\v t}\,\del_r f^{\v t}\,(2m-r)\right]\,\,,
\end{align*}
RHS for $\v$
$$
\mbox{RHS}=\frac{6m(2m-r)}{r^3} \frac{f^{tr}f^{t\v}}{f^2}\,\,,
$$
RHS for $\t$:
$$
\mbox{RHS}=\frac{6m(r-2m)}{r^3} \frac{f^{t \theta}f^{tr}}{f^2}\,\,,
$$
RHS for $t$:
$$
\mbox{RHS}=0\,\,.
$$

\subsection{The final set of equations}

Combining together left hand sides and right hand sides and noticing that
we considered the amplitude to be nearly constant, the equations of motion
are: 

\noindent 
Equation for $r$:
\begin{align}
&\hspace{-0.5in}\fr{d^2r}{ds^2}+ \fr{m(r-2m)}{r^3} \left(\fr{dt}{ds}\right)^2 -
\fr{m}{r(r-2m)}\left(\fr{dr}{ds}\right)^2 - 
(r-2m)\left(\fr{d\t}{ds}\right)^2 - 
(r-2m)\sin^2{\t}\left(\fr{d\v}{ds}\right)^2 =  \no \\[0.07in]
&\frac{2m}{r^2}\frac{1}{|f^2|}\left[- \frac{6}{r^2}\left(f^{rt}\right)^2 +
\frac{(2r-5m)}{r}\left(f^{\t t}\right)^2 + 
\frac{(2r-5m)\sin^2{\t}}{r}\left(f^{\v t}\right)^2 \right]\,\,.
\label{eq:r}
\end{align}
Equation for $\t$:
\begin{align}
&\hspace{-0.2in}\frac{d^2\t}{ds^2} + \fr{2}{r}\fr{d\t}{ds}\fr{dr}{ds} -
\sin{\t}\cos{\t}\left(\fr{d\v}{ds}\right)^2 = \fr{2m}{r^2}
\fr{1}{|f^2|}\left[ \fr{3(r-2m)}{r}f^{tr}f^{t\t}
+3(r-2m)\cos{\t}\sin{\t}\left(f^{t\v}\right)^2 \right]\,\,. \no\\
& 
\label{eq:theta}
\end{align}
Equation for $\v$:
\be
\frac{d^2\v}{ds^2}+\frac{2}{r}\frac{dr}{ds}\frac{d\v}{ds}
+2\cot{\t}\,\fr{d\t}{ds}\fr{d\v}{ds} = 
\frac{6m(2m-r)\sin^2{\t}}{r^3}\frac{f^{tr}f^{t\v}}{|f^2|}\,\,.
\label{eq:phi}
\end{equation}
Equation for $t$:
\be
\fr{d^2t}{ds^2}+\fr{2m}{r(r-2m)}\fr{dt}{ds}\fr{dr}{ds}=0\,\,.
\end{equation}

\vspace{0.1in}
And for $\t=\pi /2$, they become
\begin{align}
&\mbox{[a]}\hskip 0.2in \frac{d^2\v}{ds^2}+\frac{2}{r}\frac{dr}{ds}\frac{d\v}{ds}
= \frac{6m(2m-r)}{r^3} \frac{f^{tr}f^{t\v}}{f^2}\,; \no \\[0.1in]
&\mbox{[b]} \hskip 0.2in\fr{d^2t}{ds^2}+\fr{2m}{r(r-2m)}\fr{dt}{ds}\fr{dr}{ds}=0\,; 
\no \\[0.1in]
&\mbox{[c]} \hskip 0.2in\frac{d^2\t}{ds^2}=\frac{6m(r-2m)}{r^3} \frac{f^{t \theta}
f^{tr}}{f^2}\,; \no \\[0.1in]
\begin{split}
&\mbox{[d]} \hskip 0.2in\fr{d^2r}{ds^2}-\fr{m}{r(r-2m)}\left(\fr{dr}{ds}\right)^2 - 
(r-2m) \left(\fr{d\v}{ds}\right)^2 + 
\fr{m(r-2m)}{r^3} \left(\fr{dt}{ds}\right)^2 =  \\[0.07in]
& =\fr{2m}{r^2}\frac{1}{f^2}\left[-\left(f^{rt}\right)^2 \frac{6}{r^2} +
\left(f^{\theta t}\right)^2 \frac{(2r-5m)}{r} + \left(f^{\v t}\right)^2 
\frac{(2r-5m)}{r} \right]\,\,.    
\end{split}
\label{eq:equatorial}
\end{align}

\vspace{0.1in}
Eqs.~\ref{eq:equatorial} give the following conserved quantity, the energy
$E$: \be
\fr{r-2m}{r}\fr{dt}{ds}=E\,\,.
\end{equation}
However, we notice from Eqs.~\ref{eq:theta} and~\ref{eq:phi} that the
direction of the angular momentum is not preserved; photon that started at,
say, $\t=\pi/2$, will not remain in the equatorial plane---the right hand side
of the Eq.~\ref{eq:equatorial}[c] gives the acceleration to a photon in the
$\t$ direction.

\section{Velocity of photons in Kerr metric}\label{sec:epsilon}

In this section we  come back to the results of the Section~\ref{sec:Prasanna}
in view of the claims \cite{drummonds80,Khriplovich} that in higher derivative
gravity which arises by QED radiative corrections, the photon velocity in
local inertial frame can exceed the velocity of light in the Minkowski
background. This result was extended by Daniels \& Shore to charged
\cite{Daniels94} and rotating black holes \cite{Daniels96} with the same
conclusions. However, Mohanty \& Prasanna \cite{prasanna-main} showed that the
claims of superluminal photon velocity are due to the neglect of the curvature
coupling terms in the wave equation derived from the minimal
$F_{\mu\nu}F^{\mu\nu}$ Lagrangian (see Section~\ref{sec:Prasanna},
Eq.~\ref{eq:wave}) and that the velocity of photons is less than $c$ in \Schw
and FRW metrics, as well as in a higher derivative gravity, (as long as the
SEC (see Section~\ref{sec:chapter3}) is satisfied). We extended these results
to a Kerr gravitational background.

\vspace{-0.2in}
\subsection{$\e$-matrix}
To obtain the photon velocity we will use the dispersion relation
(Eq.~\ref{eq:dispersion}). Due to the internal Maxwell equations
(\ref{eq:1st pair}), only three components of $F_{\mu\nu}$ are independent.
Choosing the components of the electric field vector $E_i=f_{oi}$ as the
independent components, we have from (\ref{eq:1st pair}) and
(\ref{eq:eikonal}
\be
k_0f_{ij} + k_if_{j0} + k_jf_{0i}=0\,\,.
\label{eq:electric}
\end{equation}
Using ({\ref{eq:electric}) to substitute for $f_{ij}$ in terms of electric
field components $f_{j0}$ and $f_{0i}$ in the wave equation (\ref{eq:k-wave}),
we obtain
\be
-k_{\mu}k^{\mu}f_{0i} -R^j_{\,\,i} f_{0j} - R^0_{\,\,0}f_{0i} + R^j_{\,\,0}
\fr{k_i}{k_0} f_{0j} - R^j_{\,\,0}\fr{k_j}{k_0}f_{0i} +
2R^{0j}_{\,\,\,\,0i}f_{0j} + 4R^{lj}_{\,\,\,\,0i}\fr{k_l}{k_0}f_{0j} =0\,\,.
\end{equation}
Rewriting it through only one component, $f_{0j}$, gives
\be
\left[\left(k_{\mu}k^{\mu} +R^0_{\,\,0} +R^l_{\,\,0}\fr{k_l}{k_0}\right)
\d^j_i +\left(R^j_{\,\,i} - R^j_{\,\,0}\fr{k_i}{k_0} - 2R^{0j}_{\,\,\,\,0i}-
4R^{lj}_{\,\,\,\,0i}\fr{k_l}{k_0}\right)\right]f_{0j}=0\,\,.
\label{eq:full}
\end{equation}
This is the wave equation obeyed by the three components of the electric field
vector. We can simplify this equation by defining the matrix $\e_i^{\,\,j}$:
\be
\e_i^{\,\,j} \equiv \left(R^0_{\,\,0} +R^l_{\,\,0}\fr{k_l}{k_0}\right)
\d^j_i +\left(R^j_{\,\,i} - R^j_{\,\,0}\fr{k_i}{k_0} - 2R^{0j}_{\,\,\,\,0i}-
4R^{lj}_{\,\,\,\,0i}\fr{k_l}{k_0}\right)\,\,,
\end{equation}
and thus,  (\ref{eq:full}) becomes
\be
\left(k^2\d^j_i +\e^j_i\right)f_{0j} = 0\,\,.
\label{eq:short}
\end{equation}
We can rewrite this equation as matrix equation and by diagonalising
the $\e_i^j$, we find the equations for normal modes:
$$
\e_i^j f_{0j} = -k^2 f_{0i}\,\,,
$$
and
\be
\left(k^2 + \e_i\right)f_{0i} = 0 \qq (i=1,2,3)\,\,,
\label{eq:normal-modes}
\end{equation}
where $\e_i$ are the eigenvalues of the $e_j^i$ matrix. By defining the
projection operator $P_{jk}=\d_{jk}-n_j n_k$, with $n^i=k^i/|{\bf n}|$, these
equations can be decomposed into a transverse part   
\be
f_{0j}^{(T)} = \left(\d_j^i - n^i n_j\right) f_{0i}\,\,,
\end{equation} 
and a longitudal part
\be
f_{0j}^{(L)} = n^i n_j f_{0i}\,\,.
\end{equation}
By substituting back into (\ref{eq:normal-modes}), we obtain the wave
equations for the transverse photons (polarization) 
\be
\left[k^2 + \e_i \right]\left(\d_i^j - n^jn_i \right) f_{0j} = 0\,\,,
\label{eq:transverse}
\end{equation}
and equation for the longitudal photons (polarization)
\be
\left[k^2 + \e_i \right]n^jn_i  f_{0j} = 0\,\,.
\end{equation}
\subsection{Kerr metric}
First we have to find the components of the $\e_i^j$ matrix. Kerr metric is
represented in the form
\be
ds^2=\rho^2\frac{\Delta}{\Sigma}(dt)^2 - \frac{\Sigma^2}{\rho^2}
\sin^2{\theta}\left(d\varphi - 
\frac{2 a M r}{\Sigma^2}\,dt\right)^2 - \frac{\rho^2}{\Delta}(dr)^2 -
\rho^2(d\theta)^2\,\,, 
\end{equation}
where 
\begin{align}
&\r^2=r^2 +a^2\cos^2{\t}\,\,;\no \\
&\D=r^2+a^2-  2Mr\,\,;\no\\
&\S=(r^2+a^2)^2-a^2\D\sin^2{\t}\,\,,
\end{align}
and parameters $a$ and $m$ have dimensions of length. Full form of
$\e$-matrix for the Kerr metric is presented in App.~C. Here we will work in
the linear in $a$ approximation. Then, in the coordinate frame
$\left\{t,r,\t,\v\right\}$ the $\e$-matrix is reduces to \begin{equation} 
\left( \barr{ccc}
\fr{4m}{r^5}\left(r^2+3ak_3\right)&-\fr{12amk_3(r-2m)}{r^5}\cot{\t}&
\fr{12amk_2(r-2m)}{r^5}\cos{\t}\sin{\t}\\
&&\\
-\fr{24amk_3}{r^6}\cot{\t}&-\fr{2m}{r^5}\left(r^2+3ak_3\right)&
-\fr{6amk_1(r-2m)}{r^5}\cos{\t}\sin{\t}\\
&&\\
-\fr{12am}{r^6}\left(rk_1 - 2k_2\cot{\t}\right)&\fr{6am}{r^5}\left(
k_2+k_1(r-2m)\cot{\t}\right)&-\fr{2m}{r^3}
\earr 
\right)\,\,.
\end{equation}

In the equatorial plane, $\t=\pi/2$, the deviation from the geodesic is
expected to be maximal. Thus,
\begin{equation} \e_i^j = \left( \barr{ccc}
\fr{4m}{r^5}\left(r^2+3ak_3\right)&0&0\\
&&\\
0&-\fr{2m}{r^5}\left(r^2+3ak_3\right)&0\\
-\fr{12amk_1}{r^6}&\fr{6amk_1}{r^5}&-\fr{2m}{r^3}
\earr 
\right)\,\,.
\end{equation}
Eigenvalues here are:
\begin{align}
&\l_1=\fr{4m}{r^5}\left(r^2+3ak_3\right)\,;\no\\
&\l_1=\fr{2m}{r^5}\left(r^2+6ak_3\right)\,;\no\\
&\l_3=-\fr{2m}{r^3}\,\,.
\end{align}
Considering {\bf radial trajectories} with ${\bf n} =\left(r,\t\v\right) =
\left(1,0,0\right)$, equation (\ref{eq:transverse}) yields the wave equation
for the transverse fields $E_2,\,E_3$: 
\be 
\left( \barr{cc} k^2-\fr{2m}{r^3}&0  \\
0&k^2-\fr{2m}{r^3}
\earr
\right)\left(\barr{c}
E_2\\E_3
\earr \right) = 0\,\,.
\end{equation}
The dispersion relation yields
\be
\o^2-k_1^2=\fr{2m}{r^3}\,\,.
\end{equation}
The velocity of propagation of the transverse photon is
\be
v^r=\pder{\o}{k_1}=\fr{k_1}{\sqrt{2m/r^3 + k_1^2}} = 
\fr{1}{\sqrt{1+2m/k_1^2 r^3}} <1\,\,.
\end{equation}
Thus, the photon velocity is subluminal. It becomes luminal (i.e. equal to
$c$), when either $m=0$ (no black hole), or at large distances $r\rightarrow
\infty$.

\noindent
{\bf For tangential trajectories}:

\vspace{-0.2in}
\begin{trivlist}
\i (i) ${\bf n}=\left(0,1,0\right)$, and the equations
for the transverse fields $E_1$ and $E_3$ yield:
\be 
\left( \barr{cc} k^2+\fr{4m}{r^3}&0  \\
0&k^2-\fr{2m}{r^3}
\earr
\right)\left(\barr{c}
E_1\\E_3
\earr \right) = 0\,\,.
\end{equation}
The dispersion relation is $\left(k^2+\fr{4m}{r^3}\right)
\left(k^2-\fr{2m}{r^3}\right)$ and the root, corresponding to the propagating
mode, is
\be
\o^2-k^2=\fr{2m}{r^3}\,\,.
\end{equation}
The velocity of propagation is
\be
v^{\t}= \fr{1}{\sqrt{1+2m/k_2^2 r^3}} <1\,\,.
\end{equation}
The velocity is subluminal again.
\i (ii) ${\bf n}=\left(0,0,1\right)$, and the equations
for the transverse fields $E_1$ and $E_2$ yield:
\be 
\left( \barr{cc} k^2+\fr{4m}{r^5}\left(r^2+3ak_3\right)&0  \\
0&k^2-\fr{2m}{r^5}\left(r^2+6ak_3\right)
\earr
\right)\left(\barr{c}
E_1\\E_2
\earr \right) = 0\,\,.
\end{equation}
The equation for the propagating mode is
\be
\o^2=k_3^2+\fr{2m}{r^5}\left(r^2+6ak_3\right)\,\,,
\end{equation}
and the velocity of wave propagation is given by
\be
v^{\v}=\pder{\o}{\k_3} = \half \left(2k_3 +\fr{12am}{r^5}\right)
\left[k_3^2 +\fr{2m}{r^5}\left(6ak_3+r^2\right)\right]^{-\half}=
\fr{1+\fr{6am}{k_3r^5}}{\left[1+\fr{2m}{k_3^2 r^3}
\left(1+\fr{6ak_3}{r^2}\right)\right]^{1/2}}\,\,.
\end{equation}
In this case, the condition for the velociy to be subuminal reduces to:
$$
\fr{18a^2m}{r^7} <1\,\,.
$$
For $a^2<m^2$, we know that the null horizon is at $r_{\pm}=
m\pm(m^2-a^2)^{1/2}$. In this case, for $r_{+}$, the condition goes to
$\simeq  18a^2m<(2m)^7$, which is satisfied. For the extreme case, $a=m$, the
nul lhorizon is at $r_{\pm}=m$, and we have $18/m^4 <1$. 

We conclude that velociy of photons is always subluminal in the field of the
Kerr black hole.  

\end{trivlist}

\section{Derivation of the Papapetrou equation for photons}\label{sec:Zafar}

According to the EP of the general theory of relativity, the motion of
structureless test particles in a gravitational background field is determined
only by the spacetime geometry: particle worldlines are the geodesics of the
spacetime. Things become more complicated for test particles which are not
structureless but carry, for example, a non-vanishing charge or spin. In such
cases, the worldline of the test particle, in general, is no longer a
geodesic, but is modified by electromagnetic and/or spin-gravity forces (see
\cite{Pomeranski,Balakin} and references therein). It is far from obvious
whether one can observe, in practice, the spin corrections to the equations of
motion of elementary particles. However, the problem of influence of the spin
on the trajectory of a particle in an external field is not only of pure
theoretical interest. Spin-dependent corrections certainly exist in
differential cross sections of scattering processes. It was proposed long ago
that it is posible to separate charged particles of different polarizations
through the spin interaction with external fields in a storage ring of an
accelerator \cite{Niinikoski}. Though this proposal is being discussed rather
actively (see review \cite{Heinemann}), it is not yet clear whether it is
feasible technically. The EP can be put to test in an astrophysical setting. A
recent proposal is based on the analysis of the differential time delay
between the arrival of left and right-handed circularly polarized signals from
the millisecond pulsar PSR 1937+214 \cite{Biplab}. The authors have reported
the measured difference. However, far few papers exist on the theoretical
foundations of possible deviations of the photon motion from geodesics (see,
for example \cite{Biplab,prasanna-main}). Here we report yet another approach
to  this long-standing problem, based on the Lagrangian approach. If an
appropriate Lagrangian density is taken into account, then the photon equation
of motion (modified geodesic equation) can be found from the Euler-Lagrange
equations.    

\subsection{Formalism}

The generalised concept exists that classical particles follow the path of
least distance between the endpoints, even in curved spacetime . Thus, in
(pseudo) Riemannian spaces geodesic equation is found by variation of an
action $S$ identified with the parameter $s$ of a curve, interpreted as its
length. The same method has been used to find the geodesic equation for light
where, however, one technical problem arises: photon's worldline is null. One
way to avoid this difficulty is to consider motion of a massive vector
particle, then, if the equation obtained does not contain mass explicitly,
simply put the mass to zero, with one further condition that the four velocity
of the particle be a null vector. This method has been widely used for the
scalar particles and is known to give the equation of geodesic regardless of
mass. In this work we apply this approach, which makes it possible to use the
length parameter in our considerations of the action principle. Usually in
order to describe the behaviour of a field in a given gravitational background
one solves the corresponding field equations for the given metric. If the goal
is to describe waves, one can take the corresponding wave solution. In a
spherically symmetric spacetime the solutions contain factors expressed in
terms of spherical harmonics. This method works well when the wavelength is
comparable to the scales under consideration. It is not so in the case of
light, propagating in the vicinity of a massive object and, to describe the
propagation of light as an electromagnetic field, one would have to employ the
spherical harmonics of a very high order. The corresponding solution would
look too complicated and tell little about the behaviour of light.

Many efforts have been spent in the last decades to work out a simple
approach to this problem. The idea is find a satisfactory approximation to
the wave as some curves, that could be called ``rays", and, at the same time,
would take into account the polarization of light. In this section we work out
a simple approximation of this type. Our idea consists of the following: we
consider a massive vector field obeying the Proca equation, describing the
propagation of this field in some restricted domain of spacetime. The
shape of the domain can be chosen as that of some world-tube transverse to the
wave, with the cross-section comparable to or no more than two orders of
magnitude greater than the wavelength. As this tube is timelike there must
exist a timelike curve $\L$ in its interior, which specifies the time axis of a
local coordinate system. If the tube is not too wide, this
coordiniate system  would cover the entire interior. If $s$ is the proper time
on the curve $\L$, the curve can be chosen in such a way that the field
equation reduces  to \be \fr{D^2A_i}{Ds^2} +m^2A_i = 0\,\,,
\end{equation}
the same way as it happens in standard Cartesian coordinates for
Minkowskian spacetime. Correspondingly, the field Lagrangian is
$-\dot{\bf A}^2 + m^2{\bf A}^2$, where dot stands for the covariant derivative
on $s$, if and only if the curve $\L$ is chosen properly. This
Lagrangian should contain one more term, responsible for the
shape of $\L$, such that we get the geodesic equation, when we switch off the
field.  The form of this term is well known: $1/2 m\dot {\bf x}^2$, thus 
our final Lagrangian is
\be
2L = m \dot {\bf x}^2 - \dot {\bf A}^2 + m^2 {\bf A}^2 \,\,,
\label{eq:lagrangian}
\end{equation}
and coupling between the field and the shape of the curve $\L$ is 
incorporated in the form of a covariant derivative $\dot A$, which contains 
the product of connection, velocity $\dot x$ and the field.\\
The derivation of the conservation laws is more convenient in orthonormal 
frames. In what follows, $e^a_i$ will denote the components of an 
orthonormal 1-form frame field,
\be
\t^a = e^a_i(x)dx^i\,\,,
\end{equation}
and $e_a^i$ the components of its dual vector frame field,
\be
{\bf e}_a=e^i_a(x)\dd{x^i}\,\,.
\label{eq:e^i_a}
\end{equation}
Here frame indices are always $a,b,c,\ldots$; coordinate indices are
$i,j,k,\ldots$. The metric tensor can be expressed as
\be
g=g_{ij}dx^i \otimes dx^j=\d_{ab}\t^a \otimes \t^b\,\,.
\end{equation}
The connection 1-form for these frames may be introduced through the first
structure equation (see App.~D):
\be
d\t^a=\o_b^{\;\;a} \wedge \t^b\,\,,
\end{equation}
and the connection coefficients $\g_{abc}$
are that of the expansion of this 1-form in the local frames
$\left\{\t^a\right\}$: \be
\o_b^{\;\;c} =\g_{ab}^{\;\;\;c}\t^a\,\,.
\end{equation} 
Thus,
\be
\dot A^a = \fr{d A^a}{ds} + \gamma_{bc}^{\;\;\;a}A^c\fr{dx^b}{ds} =0
\end{equation}
or $\dot A^a$ is a covariant derivative in orthonormal frame on a curve 
$x^i(s)$ with $\gamma_{bc}^{\;\;\;a}$ being a spin connection.

\subsection{Equations of motion}
Each generalized coordinate has its conjugate generalized momentum:
\berr
&&p_a \equiv \pder{L}{\dot x^a} = m \eta_{ab} \dot x^b - \dot A^b \eta_{ab} 
\dd{\dot x^c} \left[ \fr{dA^c}{ds} + \g_{ba}^{\;\;\;c}A^a\dot x^b \right] =
m \eta_{ab} \dot x^b - \eta_{db} \dot A^b A^c \g_{ac}^{\;\;\;d}\;,\\[0.1in]
\label{eq:p_a}
&&E_a \equiv \pder{L}{\dot A^a} = -\dot A^b \eta_{ab}\,\,.
\label{eq:E_a}
\err
The Euler-Lagrange equations are
\berr
&&\fr{d}{ds}p_a = \pder{L}{x^a}\;,\\[0.1in]
&&\fr{d}{ds}E_a = \pder{L}{A^a} \,\,.
\label{eq:eulerforE_a}
\err
Let us first consider equations for the $E_a$ (\ref{eq:eulerforE_a}). RHS of
this equation is: 
\berr
& & \hspace{-0.3in}\pder{L}{A^a} = \dd{A^a}\fr{1}{2} \left[\left(-\dot A^b\right)^2 +
m^2 \left(A^b\right)^2\right]= -\dot A^c \eta_{bc} \dd{A^a} 
\dot A^b + m^2 \dd{A^a} A^b = \no \\[0.1in]
& & \hspace{-0.3in}-\dot A^c \eta_{bc} \dd{A^a} \left[ \fr{d A^b}{ds} + 
\g_{cd}^{\;\;\;b}\dot x^c A^d \right] + m^2 A^c\eta_{ac} =
-\dot A^c \eta_{bc} \g_{ca}^{\;\;\;b}\dot x^c + m^2 A^c\eta_{ac}\,.
\err
And Euler-Lagrange equations for $E_a$ are therefore:
\be
\left(\fr{d}{ds}E_a + \dot A^c \eta_{bc} \g_{ca}^{\;\;\;b} \dot x^c \right)-
m^2A^c\eta_{ac} = 0\,\,.
\end{equation}
With the help of Eq.~\ref{eq:E_a},
\be
\left(\fr{d}{ds}E_a -E_a \g_{ca}^{\;\;\;b} \dot x^c \right)-
m^2 A^c\eta_{ac} = 0\,\,.
\end{equation}
Expression in the brackets is a covariant derivative for $E_a$, thus, we obtain:
\be
\dot E_a - m^2 A^c\eta_{ac} = 0\,\,.
\end{equation}
Again using Eq.~\ref{eq:E_a}, we finally obtain the Euler-Lagrange equations:
\be
\ddot A^b \eta_{ab} + m^2 A^b \eta_{ab} = 0 \,\,,
\end{equation}
which, as we can see, reduces to Proca equation for the four-vector 
field $A_{\mu}(x)$ \cite{QFTbook}:
\be
\left(\Box +m^2\right) A_{\r} =0\,\,.
\label{eq:proca}
\end{equation}
We describe spin of the particle in this model directly by a tensor of spin
$S_{ab}$:
\be
mS_{ab} \equiv \fr{1}{2}\left(\dot A_a A_b - \dot A_b A_a \right)\,\,.
\label{eq:spin_def}
\end{equation} 
Thus, spin enters the expression for the generalized momentum (Eq.~\ref{eq:p_a}):
\be
p_a=\eta_{ab}m \dot x^b - \g_{ab}^{\;\;\;c} S_{c}^{\;\;b}\,\,,
\end{equation}
To obtain the equation of motion for spin, we write
\be
\fr{d S_{bc}}{ds} = \fr{1}{2}\fr{d}{ds}\left[\dot A_b A_c - 
\dot A_cA_b \right] = \fr{1}{2} \dot x^a \dd{x^a} \left[\dot A_b A_c - 
\dot A_cA_b \right]\,\,.
\end{equation}
Due to the wave equation (\ref{eq:proca}) all derivatives of $A$ vanish and
we are left with: 
\berr
& & \hspace{-1.in}\fr{d S_{db}}{ds} = \fr{1}{2} \dot x^a \left[\dot A_i A_j - 
\dot A_j A_i \right] \dd{x^a} \left(e^i_{d} e^j_{b}\right) = \no \\
& & \fr{1}{2} \dot x^a \left[\dot A_i A_j 
- \dot A_j A_i \right]\left(e^i_{d}\left(e^j_{b}\right)_{,a} +
e^j_{b}\left(e^i_{d}\right)_{,a}\right) = \no \\
& & \fr{1}{2} \dot x^c \left[\dot A_i A_j - \dot A_j A_i \right]
\left(e^i_{d}\g_{cb}^{\;\;\;a}  e^j_{a} + e^j_{\;b} \g_{cd}^{\;\;\;a} 
e^i_{\;a} \right) = \no \\
& & \fr{1}{2} \dot x^c\left(\g_{cd}^{\;\;\;a} \left[\dot A_a A_b-
\dot A_b A_a\right] + \g_{cb}^{\;\;\;a} \left[\dot A_d A_a -
\dot A_ba A_d\right]\right)\,\,.
\err
Here $e^i_a$ is the matrix introduced in the equation (\ref{eq:e^i_a}) and
derivatives $\left(e^i_{b}\right)_{,a}$ are obtained from the first structure
equation (see App.~D for the derivation). Using our definition of a spin
tensor (Eq.~\ref{eq:spin_def}), we obtain: \be \fr{d S_{db}}{ds} = \dot x^c
\left(\g_{cd}^{\;\;\;a} S_{ab} +  \g_{cb}^{\;\;\;a} S_{ad} \right)\,\,,
\end{equation}
or 
\be
\fr{DS_{db}}{Ds} = 0 \,\,.
\label{eq:spin0}
\end{equation}
Thus, spin is transported parallel to itself along the worldline. \newline
To derive Euler-Lagrange equations for the generalized momentum, we return
to the coordinate basis. In this case Lagrangian (Eq.~\ref{eq:lagrangian})
takes the form: 
\be 
2L=mg_{ij}\dot x^i \dot x^j - g_{ij}\left(\fr{dA^i}{ds} +
\G^i_{kl} \dot x^k  A^l\right) \left(\fr{dA^j}{ds} + \G^j_{kl} \dot x^k
A^l\right) +  m^2g_{ij}A^iA^j\,\,,
\end{equation}
where $\G^j_{kl}$ are the Christoffel symbols. This gives the generalized
momentum: \be
p_i \equiv \pder{L}{\dot x^i} = m g_{ij}\dot x^j - \G^m_{il} A^l \dot A_m\,\,.
\end{equation}
The Euler-Lagrange equations: 
\be
\fr{d}{ds}p_i = \pder{L}{x^i} \,\,.
\end{equation}
LHS of this equation is:
\be
\fr{dp_i}{ds} = m g_{ij}\fr{d \dot x^j }{ds} + m \dot x^j \dot x^k \del_k g_{ij}
- \dot x^j A^l \dot A_m \del_j \G^m_{il} - 
\G^m_{il}\left(\fr{dA^l}{ds}\dot A_m +  A^l \fr{d\dot A_m}{ds}\right) \;\;,
\label{eq:5_3-LHS}
\end{equation}
where derivatives $\del_i$ are defined as $\del_i=\dd{x^i}$. RHS is:
\be
\pder{L}{x^i}= \fr{1}{2} m \del_i g_{jk} \dot x^j \dot x^k + 
\del_ig_{mn} \left(-\dot A^m \dot A^n + m^2 A^m A^n \right) - 
g_{mn} \dot x^k A^l \dot A^n \del_i \G^m_{kl}  \,\,.
\label{eq:5_3-RHS}
\end{equation}
From the definition of the Christoffel symbols \be
\del_i g_{mn}=\fr{1}{2} \left(g_{kn}\G^k_{mi} +g_{km}\G^k_{in}\right)\,\,.
\label{eq:del_g}
\end{equation}
Using  
\be
\dot x^i \dot x^j \del_j g_{ik} = \fr{1}{2}\dot x^i \dot x^j 
\left(\del_j g_{ik} + \del_i g_{jk}\right)\,\,,
\end{equation}
we obtain:
\berr
& & \hspace {-0.3in}m g_{ij}\fr{d\dot x^j}{ds} + \fr{1}{2} \dot x^i \dot x^j 
\left(\del_j g_{ik} + \del_i g_{jk} -\del_k g_{ij} \right) - 
m^2 \G^k_{mi} A^m A_n-\G^n_{il} A^l\fr{d\dot A_m}{ds}+\no \\[0.1in]
& & \G^n_{mi} \dot A^m \dot A_n - 
\G^m_{il} \dot A_m\fr{dA^l}{ds} - \dot x^j A^l \dot A_m \del_j \G^m_{il} +
g_{mn}\dot x^k A^l \dot A^n \del_i \G^{m}_{kl} = 0 \,\,.
\err
First two terms of this equation are the covariant derivative of $m\dot x^i$,
and using the Proca equation (\ref{eq:proca}) after some algebra we get:
\be
m \fr{D \dot x^i}{Ds}= -\dot x^j A^l\dot A_k \left(\del_j\G^k_{il} -
\del_i\G^k_{jl} + \G^k_{im}\G^m_{jl} - \G^k_{mj}\G^m_{il} \right)\,\,.
\end{equation}
With the definition of the curvature tensor and spin tensor
(\ref{eq:spin_def}), the equation takes the evidently covariant form:
\be
g_{ij}\fr{D\dot x^j}{Ds} = R^{k}_{\;\;jil} \dot x^j S_k^{\;\;l} \,\,.
\label{eq:pap1}
\end{equation} 
This equation coincides with the Papapetrou equation in the linear-in-spin
approximation \cite{Papapetrou}. It must be pointed out that in case of a
zero spin this equation becomes a geodesic. If we reparametrise the curve
with some new parameter $\l$ in such a way that 
\be
g_{ij}\fr{dx^i}{d\l}\fr{dx^j}{d\l} =0\,\,,
\end{equation}
we can rewrite Eq.~\ref{eq:pap1} as
\be
g_{ij}\fr{D\dot x^j}{D\l} = R^{k}_{\;\;jil} \dot x^j S_k^{\;\;l} \,\,.
\end{equation} 
This equation is valid for the massless particles with spin as well, since mass
does not enter the equation explicitely.

\section{Conclusions}

In this chapter we considered the propagation of photons in the strong
gravitational field. Strong field can have additional effects on the light
propagation which do not exist (or are of vanishingly small order) in the
weak-field approximation. Gravitational lensing of light in the field of black
holes (and, may be, neutron stars) is strongly influenced by these effects and
the basic theoretical formulation underlying it is necessary. However, contrary
to previous claims, we confirmed that the photon's speed does not become
superluminal even in the field of the Kerr black hole. If, in fact, photons do
not follow geodesics in the fields of Kerr and \Schw black holes, as can be
followed from the Papapetrou equation, or they do experience gravitational
birefringence, all the calculations for the gravitational lensing in these
metrics done so far have to be revised. The main aim of our future work,
following the narration of this chapter is to solve the equations of motion of
photons in the fields of \Schw and Kerr black holes and obtain the trajectories
for RCP and LCP. We expect to prove that, even in the first order approximation
to the geodesic, the polarizations separate. We also plan to apply the results
of Sec.~\ref{sec:epsilon} to the phenomenon of gravitational Faraday rotation
in of pulsar radiation.

\chapter{Summary and Final Comments}
Over the last few decades the status of gravitational lensing has been raised
from that of a theoretical speculation to that of a precision measurement.
Indeed, gravitational lensing is now widely used to establish concordance
of a given cosmological model. In this thesis we have applied gravitational
lensing in the above spirit.

Gravitational lensing provides an exciting new probe for detecting exotic
objects in the universe that may be described by matter fields which are not
yet detected. Proposals have been made to look for signatures of cosmic
strings \cite{Vilenkin}, boson stars \cite{dabrowski} and neutralino stars
\cite{sazhin1-2} from their gravitational lensing effects. There is no
compelling evidence that any of the observed GL systems are due to these
objects, however, it is essential to develop new lens models with objects which
are not forbidden on the theoretical grounds. In chapter three we have
explored the consequences of the existence of matter violating the energy
conditions. We develop the formalism of gravitational lensing by a point
negative mass lens. Negative mass microlensing simulations are carried on,
showing the resulting shapes of the images, the intensity profiles, the time
gain function, the radial and tangential magnifications, and other features. We
provide an in-depth study of the theoretical peculiarities that arise in
negative mass microlensing, both for a point source and for extended source
situations for various source intensity profiles and discuss the differences
between them. We have also numerically analyzed the influence of a finite
size of the source on the negative-mass gravitational lensing event. 

Since the primary motivation for our theoretical work is observational, in
order to set the experimental context more explicitly, we take wormholes as a
useful and appealing theoretical scenario for the appearance of negative
masses. But it applies, {\it mutatis mutandis}, to any possible object
possessing a negative mass. Thus, we provide a useful comparison arena where
to test the possible existence of wormholes or other kinds of
negative mass objects. Our work gives a method of analyzing observational
predictions quantitatively. The next step would be to test these predictions
using archived, current, and forthcoming observational microlensing
experiments. The only search done up to now included the BATSE database of
$\g$-ray bursts \cite{doqui}. There is still much unexplored territory in the
gravitational microlensing archives. Following our suggestions, the MOA
(Microlensing Observations in Astrophysics) group is currently adapting their
systems to look for the effects of negative mass microlensing in the first
direct online search for exotic objects in the universe \cite{yock}.

It has been recently suggested that there can exist large classical violations
of the energy conditions \cite{ULT}. Disregarding the fundamental mechanism
by which the EC are violated (e.g. fundamental scalar fields, modified
gravitational theories, etc.), if large-scale localized violations of null
energy condition exist in our universe, we show that it would be possible  to
detect them through cosmological macrolensing. Contrary to the usual case,
where arc-like structures are expected, finger-like ``runaway" filaments and a
central void appear. Comparison of the effects of negative masses on
background fields with macrolensing effects produced by equal amounts of
positive mass, located at the same redshift, brings out notorious differences.
These results make the cosmological macrolensing produced by matter violating
the null energy condition observationally distinguishable from the standard
situation. Whether large-scale violations of the EC, resulting in spacetime
regions with average negative energy density indeed exist in the universe can
therefore be decided through observations.

Alternative cosmologies have been more and more actively discussed in the
literature following serious theoretical and observational difficulties with
standard cold dark matter FRW cosmology \cite{alternative}. An open FRW
model with linear evolution of the scale factor, $a(t) \propto t$, has
generated a lot of interest. Many of the problems of a standard model are
naturally resolved in such a cosmology. This model is consistent with
nucleosynthesis in the early universe \cite{annu} and presents a good fit to
the current SNE Ia data \cite{abha}. It is now well known that gravitational
lensing is a useful probe of the geometry of the universe. It has been pointed
out \cite{fukugita,turner90} that the expected frequency of multiple imaging
lensing events for high-redshift sources is sensitive to a cosmology. In
chapter four we used this test to constrain the power index of the scale
factor of a general power law cosmology, $a(t) \propto t^{\alpha}$. Expected
number of lens systems depends upon the index $\alpha$ through the angular
diameter distances. By varying $\alpha$, the number of lenses changes and on
comparison with the observations gives us constraints on $\alpha$. We
describe the general formulation of the gravitational lensing statistics and
use it to constraint the parameters for that cosmology. These results lead to
the conclusion that open linear coasting cosmology, $a(t)=t$, is consistent
with present observations.

Linear evolution of the scale factor is a generic feature in a class of models
that attempts to dynamically  solve the cosmological constant problem
\cite{models}. Such models have a scalar field non-minimally coupled to the
large scale curvature of the universe. With the evolution of time, the
non-minimal coupling diverges, the scale factor quickly approaches linearity
and the non-minimally coupled field acquires a stress energy that cancels the
vacuum energy in the theory. Such an evolution of the scale factor is also
possible in alternative effective gravity and higher gravity theories. It is
also possible in the ``toy" model \cite{dakshmeetu} that combines the Lee-Wick
construction of non-topological soliton (NTS) solutions in a variant of an
effective gravity model proposed by Zee \cite{zee}. The conditions of the
theory are sufficient for the existence of large NTS's with the size equal or
larger than typical halos of galaxies. The interior and exterior of such a
domains would be regions with effective gravitational constant $G_{\rm
eff}^{\rm in}=$ constant and $G_{\rm eff}^{\rm out}=0$, respectively. Such a
domain is characterized by an effective refraction index in its interior,
which would cause bending of light incident on it. In chapter four we examine
the possibility of a special kind of NTS, gravity ball, to be a gravitational
lens. We investigate its lensing properties, calculate the deflection angle of
a point source, derive the lens equation and plot the lens curve for the
gravity ball with specific parameters. We have shown that depending on the
parameters of the model there can be one, two or three images (two inside the
Einstein radius and one outside). For the extended sources we showed how a
large gravity ball can induce surprisingly large distortions of images of
distant galaxies. It is assumed that in order to produce large arcs a
cluster must have its surface-mass density approximately supercritical,
$\Sigma \ge \Sigma_{crit}$. We found that the empty balls alone are able to
produce arc-like features, including straight arcs, arclets and radial arcs.
We have modeled the gravitational lensing effect of our gravity ball on a
background source field and found that the ball produces distortion (`shear')
of that field, consistent with observations. We also obtain constraints on
the size of a large gravity ball from the existing observations of clusters
with arcs. Empty G-balls can show in observations as arcs without evident mass
distribution---a lens and/or through the distortion of a background random
galaxy field. Observationally empty G-balls have not yet been found, but we
conclude from our analysis that the existence of this special kind of NTS,
large G-balls, cannot be ruled out by gravitational lensing effects.

Most cosmological studies of gravitational lensing are performed in the weak
field approximation. However, weak field approximation becomes invalid in the
vicinity of compact objects like black holes and pulsars. Propagation of light
in such backgrounds must be described as occurring in strong-field regime.
Chapter five is dedicated to the studies of the propagation of light in the
fields of \Schw and Kerr black holes. Strong fields can exhibit additional
effects on the light propagation which do not exist (or are vanishingly
small) in the weak-field approximation. Gravitational lensing of light in the
field of black holes (and, may be, neutron stars) is strongly influenced by
these effects and it is necessary to develop the basic theoretical formulation
underlying it. Contrary to previous claims \cite{drummonds80,Khriplovich}, we
confirm that the photon's speed does not become superluminal even in the field
of the Kerr black hole. We derive the modified geodesic equation for photons
in \Schw background and present a way to solve it. Alternatively, we derive the
Papapetrou equation for photons in the first order approximation. If, in fact,
photons do not follow geodesics in the fields of Kerr and \Schw black holes,
as can be followed from the Papapetrou equation, or they do experience
gravitational birefringence, gravitational lensing calculations in these
metrics done so far have to be revised. This is a part of our ongoing research
work.

\begin{appendix}
\chapter{Wormhole solution to the Einstein equations}
Building of the wormhole solution of the \E equations was done in quite a 
non-traditional way, that is why we will describe it in details. The
traditional way of obtaining solutions to Einstein equations is to start
by picking one's favourite Lagrangian for the matter fields that could 
support the wormhole's spacetime. One then calculates the relevant
stress-energy tensor and solves the \E field equations. Finally, one checks for
the presence of curvature singularities in the solution so obtained---the goal
is to avoid any singularity whatsoever (like event horizon or naked
singularity), since the main idea is to obtain a traversable stable wormhole. 

Till the seminal paper of Morris and Thorne \cite{motho}, all attempts at
following the above described prescription failed. The resulting wormholes
were pathological in one or more manners. Morris and Thorne adopted a more
engineering oriented approach. First, assume the existence of a suitably
well-behaved geometry. Then calculate the Riemann tensor associated with this
geometry and use the \E field equations to deduce what the distribution of
stress-energy must be. Finally, decide whether the deduced distribution is 
physically reasonable. 

\section*{Desired properties of traversable wormholes}

It is interesting to note that wormholes as objects of study in mathematical
relativity predate black holes: Within one year after \E\!'s final
formulation of his field equations, the Viennese physicist Ludwig Flamm
recognized that the \Schw solution of \E\!\!\!\!'s field equations represents a
wormhole \cite{flamm}. Possible roles of the \Schw and other wormholes in
physics were speculated upon in the 1920s by Herman Weyl \cite{weyl}, in the
1930s by \E and Rosen \cite{e-rosen}, and in the 1950s-1960s by John Wheeler
\cite{wheeler}. However, all the above described configurations were unsuitable
for the use of travel. Here we will state the properties we would like the
wormholes to possess in order to be traversable---a Lorentzian wormhole that is
(quasi)-permanent and macroscopic so as to allow for traverse by human beings.
We will state them verbally without the mathematical details.
\vskip -0.1in
\begin{enumerate}
\itemsep=-0.1in 
\item[(i)] The metric should be both spherical symmetric and static. 
\item[(ii)] The solution must everywhere obey the \E field equations, since we
assume the correctness of GR theory.
\item[(iii)] To be a wormhole, the solution must have a throat that connects
two asymptotically flat regions of spacetime. 
\item[(iv)] There should be no horizon, and tidal gravitational forces
experienced by a traveler must be bearably small.
\item[(v)] A traveler must be able to cross through the wormholes in a finite
and reasonably small proper time as measured not only by a traveler, but also
by observers outside the wormhole.
\item[(vi)] The matter and fields that generate the wormhole's spacetime
curvature must have a physically reasonable stress-energy tensor. 
\item[(vii)] The solution should be perturbatively stable (especially as a 
traveler passes through).
\item[(viii)] It should be possible to assemble a wormhole. For example, the
assembly should both require much less than the mass of the universe and much
less time than the age of the universe. This is closely related to the
property (vi), in the sense that we shall make the wormhole construction
material as compatible as possible to our present prejudices about the forms
of matter allowed by physical laws.
\end{enumerate}

\begin{figure}
\centerline{
\includegraphics[width=0.9\textwidth]{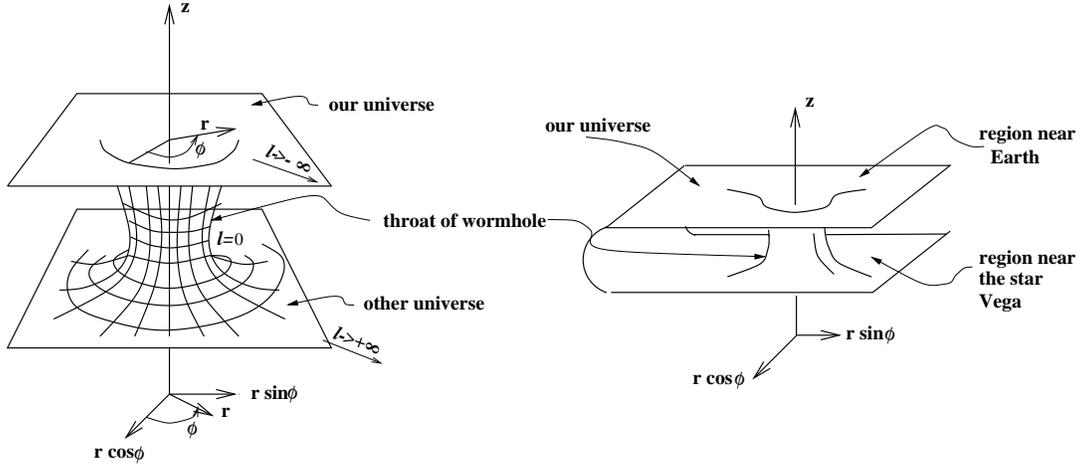}}
\caption[Embedding diagrammes for wormholes]{{\it Left}: Embedding diagramme
for a wormhole that connects two different universes. {\it Right}: Embedding
diagramme for a wormhole that connects two distant regions of our universe.
Each diagramme depicts the geometry of an equatorial slice ($\t =\pi/2$)
through space at a specific moment of time ($t=$ constant).}
\label{fig:embedding}  \end{figure}

Thus, adopting \Schw ($t,r,\t,\v$) coordinates, the metric can be put into the
form
\be
ds^2=-e^{2\phi_{\pm}(r)}dt^2 +\fr{dr^2}{1-b_{\pm}(r)/r} + 
r^2 \left[d\t^2 + \sin^2{\t}\,d\phi^2\right]\,\,.
\end{equation}
\begin{itemize}
\itemsep=-0.09in
\i Radial coordinates: $r$, such that $2\pi r=$ circumference; $l$---proper
radial distance from the throat. They are related by $l(r)=\pm \int_{r_0}^r
dr/(1-b_{\pm}(r')/r')^{-1/2}$. Two coordinate patches ($[r_0,+\infty]$),
each covering one universe, join at $r_0$, the throat of the wormhole. 
 \i Solution is determined by two freely specifiable functions of $r$: shape
function, $b(r)$, defined by $dl/dr=\pm(1-b/r)^{-1/2}$ and redshift function,
$\phi(r)$, defined by $g_{tt}=-e^{2\phi}$. The mass of the wormhole, as seen
from spatial infinity, is given by $b_{\pm}=2GM_{\pm}$, the two mouths of the
same wormhole can in general have different masses; besides, there is no {\it a
priory} requirement that $\phi_{+}(\infty)=\phi_{-}(\infty)$, which implies
that time can run at different rates in the two universes. Constraints on
$b(r)$ and $\phi(r)$ to produce a traversable wormhole: 
\begin{itemize} 
 \partopsep=-1.in
\itemsep=-0.1in 
\i[(a)]
Spatial geometry must have wormhole shape (see Fig.~\ref{fig:embedding}). No
horizons or singularities, $\phi$ is everywhere finite; $t$ measures proper
time in \ass flat regions $\Leftrightarrow \phi \rightarrow 0$, as $l
\rightarrow \pm \infty$. 
\i[(b)] Radial velocity of a traveler as she passes radius $r$, as measured
by a static observer there, in terms of distance traveled $dl$, radius
traveled $dr$, coordinate time lapse $dt$ and proper time lapse as seen by
the traveler, $d\tau$, is $v=dl/e^{\phi}dt = \mp dr/(1-b/r)^{1/2}e^{\phi}dt$.
Defining $\g\equiv[1-(v/c)^2]^{-1/2}$ we find \be
v\g = \mp \fr{dr}{(1-b/r)^{1/2}d\tau}\,\,.
\end{equation}
\end{itemize}
\i The material that generates the wormhole's \st curvature:
\begin{itemize}
\partopsep=-1.in
\itemsep=-0.2in
\i[(a)] Stress-energy tensor as measured by static observers in the orthonormal
frame:\berr
&&T_{\hat t \hat t}=\r c^2=(\mbox{density of mass-energy});\,\, T_{\hat r \hat
r}=-\tau=-(\mbox{radial tension});\,\,\no\\
&&T_{\hat \t \hat \t}=T_{\hat \v \hat \v}= p=(\mbox{lateral pressure})\,\,.
\label{eq:stress-tensor}
\err
\i[(b)] \E field equations:
\be
\r=\fr{b'}{8\pi G c^{-2}r^2}\,;\q \tau=\fr{b/r -
2(r-b)\phi'}{8\pi G c^{-2}r^2}\,;\q
p=\fr{r}{2}\left[(\r c^2-\tau)\phi'-\tau'\right]\tau\,\,. \end{equation} 
\i[(c)] The constraints placed on the shape function $b(r)$ give rise, via the
\E field equations, to constraints on $\r$, $\tau$ and $p$, that generate the
spacetime curvature. The most severe constraints occur in the wormhole's
throat. The requirement that the wormhole be connectible to \ass flat
spacetime leads to the flaring-out condition
$$
\fr{\tau_0 - \r_0c^2}{|\r_0c^2} > 0 \qq \mbox{at or near the throat}.
$$
The constraint $\tau_0 > \r_0c^2$ means that in the throat the
tension of the wormhole's material must be so large as to exceed the
total density of mass-energy $\r_0c^2$. Material with this property, $\tau >
\r c^2 > 0$, is called ``exotic". An observer moving through the throat sees an
energy density (projection of the stress-energy tensor
(Eq.~\ref{eq:stress-tensor}) on traveler's time basis vector ${\bf e}_{\hat
0'}= \g {\bf e}_{\hat t} \mp \g (v/c){\bf e}_{\hat r}$) ) given by  \be
T_{\hat 0' \hat 0'}=\g^2(\r_0c^2-\tau_0)+\tau_0\,\,.
\end{equation}
If such an observer moves sufficiently fast, with a radial velocity close to
the speed of light, $\g \gg 1$, he will see a negative density of
mass-energy. A rough explanation  of th negative mass-energy density seen by
some observers is that bundles of light rays (null geodesics) that enter a
wormhole at one mouth and emerge from the other must have cross-section at
areas that initially decrease and then increase. The conversion from
decreasing to increasing can only be produced by gravitational repulsion of
matter through which the rays pass, a repulsion that requires negative energy
density.  \end{itemize} \end{itemize}
The above material was compiled using Refs.~\cite{motho,visser}.

\chapter{Lensing clusters sample}
\vspace{-0.4in}
\begin{table}[ht!]
\caption[Lensing cluster sample]{Lensing clusters sample}
\vskip 0.05in
\begin{center}
\begin{tabular}{|l|c|l|c|c|c|} \hline \hline
\emph{cluster} & \emph{$z_{\rm cl}$} & \mbox{$arc^{\,a}$} & \emph{$z_{\rm
arc}$}  & \emph{$r\,\, ({\rm arcsecs})^{\,b}$}& \mbox{ref.$^{\,c}$} \\ \hline
A2280         &  0.326  &   large arc      & $\approx 0.7$   &                       &    \\
              &         &                  & or 1.0?         &  $14''$               & 1  \\
MS 0440+0204  &  0.190  &   1 GLA +        &                 &                       &    \\
              &         &   3 bright arcs  & 0.53?           &  $22''$               & 2  \\
A370          &  0.374  &   arclet A5      & 1.306?          &  $63.7''$             & 3  \\
PKS 0745-191  &  0.1028 &   bright arc     & 0.433           &  $18.2''$             & 4  \\
A2390         &  0.231  &   GLA; straight  &                 &                       &    \\
              &         &   image          & 0.913           &  $37''$               & 5  \\
A963          &  0.206  &   GLA+counter    &                 &  image separation=    &    \\
              &         &   arc            & 0.77            &
$30'';\,\,\te=15''$  & 6  \\ Cl 2244-02    &  0.329  &   GLA            & 2.237
          &  $14''$               & 7  \\ Cl 0024+1654  &  0.391  &   five
arcs      & 1.39?           &  $ 35''$              & 8  \\ Cl 0500-24    &
0.316  &   GLA            & 0.913           &  $\sim 26''$          & 9 \\
S295          &  0.299  &   GLA            & 0.93            &  $\sim 20''$
      & 10  \\ \hline
\end{tabular}\label{tab:clusters}
\end{center}
\end{table}
\vspace{-0.1in}
\noindent
$ ^a$  Lensing phenomenon. GLA---giant luminous arc.\\
$ ^b$  Distance from the centre of cluster where arc(s) are detected.\\
$ ^c$  Only one reference is provided for lensing feature. References are
listed in \cite{table}.

\chapter{Full $\e_i^j$ matrix in {$t,\v,r,\t$} components}
\vspace{-0.6in}
\begin{align*}
&\hspace{-0.5in}\e_{\v}^{\v}=-\fr{8Mr\left(-3a^2+2r^2-3a^2\cos{2\t}\right)}
{\left(a^2+2r^2+a^2\cos{2\t}\right)^3}\,\,;\\[0.1in]
&\hspace{-0.5in}\e_r^r= \fr{8 M r \left(-3 a^2 + 2 r^2 -  3 a^2 \cos{2
\theta}\right) \left(5 a^2 + 6 a k_1 + 4 r^2 - a^2 \cos{2 \t}\right)}
{\left(a^2 + 2 r^2 + a^2\cos{2 \t}\right)^4}\,\,;\no\\[0.1in]
&\hspace{-0.5in}\e_{\t}^{\t}= -\fr{16 M r \left(-3 a^2 + 2 r^2 -3 a^2
\cos{2\t}\right)\left(2a^2 + 3ak_1 + r^2 - a^2\cos{2\t}\right)}
{\left(a^2 + 2 r^2 +a^2\cos{2\t}\right)^4} \,\,;\\[0.1in]
&\hspace{-0.5in}\e_{\v}^r=-\fr{8ak_3M\left(a^2 + r (-2 M + r)\right)
\left(a^2-6 r^2 + a^2 \cos{2 \t}\right) \sin{2 \t}}
{\left(a^2 + 2 r^2 + a^2 \cos{2 \t}\right)^4}\,\,;\\[0.1in]
&\hspace{-0.5in}\e_r^{\v}= -\fr{8 a M \left(6 k_2 r \left(-3 a^2+ 2
r^2 - 3 a^2\cos{2\t} \right) - \fr{k_3 \left(-5 a^2 + 4(2M-r)r +
a^2 \cos{2\t} \right) \left(a^2 - 6 r^2 + a^2 \cos{2\t}\right)\cot{\t}}
{a^2-2Mr+r^2}\right)}{\left(a^2 +2r^2 +a^2\cos{2\t}\right)^4}\,\,;\\[0.1in]
&\hspace{-0.5in}\e_{\v}^{\t}= \fr{8 a k_2M \left(a^2 + r(-2 M + r)\right) 
\left(a^2 -6 r^2 + a^2 \cos{2\t}\right)\sin{2\t}}{\left(a^2 + 2 r^2 + 
a^2 \cos{2\t}\right)^4}\,\,;\\[0.1in]
&\hspace{-0.5in}\e_{\t}^{\v}=\fr{16 a M \left(3 k_3 r(3 a^2-2 r^2 + 
3a^2 \cos{2\t})+k_2 \left(-2 a^2 + (2M-r)r + a^2 \cos{2\t}\right) 
\left(a^2 - 6 r^2+ a^2 \cos{2 \t}\right)\cot{\t}\right)}
{\left(a^2 + 2 r^2 +  a^2 \cos{2 \t}\right)^4}\,\,;\\[0.1in]
&\hspace{-0.5in}\e_r^{\t}= \fr{8 a M \left(6 r^2 -a^2- a^2 \cos{2 \t}\right) 
\left(-3a^3 - 5 a^2 k_1 + a(4M-3r)r + 4 k_1(2M-r)r+a(3a^2+ak_1+3r^2-4Mr) 
\cos{2\t}\right)\cot{\t}}{\left(a^2 - 2 M r + r^2\right)\left(a^2 + 2 r^2
+ a^2 \cos{2\t}\right)^4}\,\,;\\[0.1in]
&\hspace{-0.5in}\e_{\t}^r= \fr{8 a M \left(6 r^2-a^2 -a^2 \cos{2 \t}\right) 
\left(-3a^3 - 4 a^2 k_1 + a(2M-3r)r + 2k_1(2M-r)r + a(3a^2+2ak_1+3r^2-2 Mr) 
\cos{2\t}\right) \cot{\t}}{\left(a^2 + 2r^2 + a^2\cos{2\t}\right)^4}\,\,.
\end{align*}

\chapter{Structure equations}
\vspace{-0.5in}
\section{Connections}
\vspace{-0.1in}
Consider a Riemann space $M$ endowed with a coordinate system
$\{\,x^i\}$ and let 1-forms $\nu^a$
\begin{equation}
\nu^a=e_i^a(x)dx^i
\end{equation}
constitute an orthonormal frame in every point of the space $M$, such that
its metric tensor may be expressed through the Kronecker delta:
$$
g=g_{ij}dx^i \otimes dx^j=
\delta_{ab} \nu^a \otimes \nu^b.
$$
Any given connection is characterized by two tensors, torsion and curvature.
The torsion can be expressed as a vector-valued 2-form $T^a$ (we
usually assume zero torsion) and the curvature as a (1,1)-tensor-valued
2-form $\Omega_a^b$. The defining relations for these two tensors are
known as Maurer-Cartan structure equations: \be
\hspace*{-7.5cm}\mbox{Ist Structure Equation:}\qq\qq   \qq
T^a=d\nu^a + \omega_b^{\;a} \wedge \nu^b\,\,,
\label{eq:Ist}
\end{equation}
and
\be
\hspace*{-7.cm}\mbox{IInd Structure Equation:}\qq \qq     \qq
\Omega_a^{\;b} =d\omega_a^{\;b} + \omega_a^{\;c}
\wedge \omega_c^{\;b}\,\,,
\label{eq:IInd}
\end{equation}
where $\omega_a^{\;b}$ is called the 1-form of connection. Due to the
orthonormality of the frame, the connection 1-form satisfies the condition
\be
\omega_{a b} +\omega_{ba} = 0\,\,,
\end{equation}
and the connection coefficients $\g_{abc}$ are that of the expansion of
this 1-form in the local frames $\left\{\t^a\right\}$:
\be
\omega_{b}^{\;c}=\gamma_{ab}^{\;\;\;c}\nu^a\,\,.
\label{eq:omega}
\end{equation}
The Riemann curvature is represented by the 2-form
$\Omega_{a}^{\;b}$, defined by the second structure equation,
and coefficients of its expansion in the local frame
of 2-forms $\left\{\nu^a \wedge \nu^b \right\}$ constitute the components
of the Riemann curvature tensor
\be
\Omega_a^{\;b} = R_{a\,\,cd}^{\,\,b} \nu^c \wedge \nu^d\,\,.
\end{equation}
\vspace{-0.5in}
\section{Derivatives of the components of the frame}
Suppose, we have an orthonormal frame of 1-forms
\be
\nu^a=e^a_i dx^i\,\,,
\label{eq:nu}
\end{equation}
and dual to it vector frame
\be
dx^i = e^i_a \nu^a\,\,.
\label{eq:dx}
\end{equation}
From Eqs.~\ref{eq:Ist} and~\ref{eq:omega}, we write Ist structure
equations as (torsion is zero)
\be
d\nu^a +\g_{cb}^{\,\,\,\,\,a} \nu^c \wedge \nu^b = 0\,\,.
\label{eq:intermediate}
\end{equation}
Using (\ref{eq:nu}), we write 
$$
d\nu^a=de^a_i \wedge dx^i \,\,,
$$
where the derivatives of the tetrad components can be written by analogue
with a gradient of a function, $df=f_idx^i= \left(\dd{x^i}f\right)dx^i=
\left(\vec e_c \circ f\right)\nu^c\equiv f_{,c}\nu^c$, as
\be
de^a_i =\left(\vec e_c \circ e^a_i\right) \nu^c \equiv
\left(e^a_i\right)_{,c} \nu^c\,\,.
\end{equation}
Thus, with the use of (\ref{eq:dx}),
$$
d\nu^a=\left(e^a_i\right)_{,c}\nu^c \wedge dx^i =  
\left(e^a_i\right)_{,c} e_b^i \nu^c \wedge \nu^b\,\,.
$$
Then, we can write (\ref{eq:intermediate}) as 
\be
\left(e^a_i\right)_{,c} e_b^i \nu^c \wedge \nu^b =
-\g_{cb}^{\,\,\,\,\,a} \nu^c \wedge \nu^b\,\,.
\end{equation}
The Ist structure equation can then be written as
\be
\left(e^a_i\right)_{,c} = -\g_{cb}^{\,\,\,\,\,a} e_i^b\,\,.
\end{equation}

\end{appendix}

\end{document}